\documentclass[aps,epsf,amsmath,superscriptaddress,citeautoscript,twocolumn,
showpacs,floatfix,prd]{revtex4-2}
\usepackage{bm}
\usepackage{amssymb}
\usepackage{graphicx}
\usepackage{amsmath, amsthm, amssymb, graphicx}
\usepackage[usenames]{color}
\graphicspath{{figures/}} 
\usepackage{ulem}

\usepackage{amsmath,amsthm,amsfonts,amscd,amssymb}
\usepackage{eucal} 	 	
\usepackage{verbatim}      	
\usepackage{makeidx}       	
\usepackage{braket}

\usepackage{bibentry}
\usepackage{hyperref}

\usepackage{mathrsfs}
\usepackage{graphicx}
\usepackage{dcolumn}
\usepackage{bm}

\begin{document}
	
	\preprint{}
	
	\title{Propagators in the Correlated Worldline Theory of Quantum Gravity}
	
	\author{Jordan Wilson-Gerow}
	\affiliation{Department of Physics and Astronomy, University
		of British Columbia, 6224 Agricultural Rd., Vancouver, B.C., Canada
		V6T 1Z1}
	\affiliation{Theoretical Astrophysics, Cahill,
		California Institute of Technology, 1200 E. California Boulevard,
		MC 350-17, Pasadena CA 91125, USA}
	\author{ P.C.E. Stamp}
	\affiliation{Department of Physics and Astronomy, University
		of British Columbia, 6224 Agricultural Rd., Vancouver, B.C., Canada
		V6T 1Z1}
	
	\affiliation{Theoretical Astrophysics, Cahill,
		California Institute of Technology, 1200 E. California Boulevard,
		MC 350-17, Pasadena CA 91125, USA}
	\affiliation{Pacific
		Institute of Theoretical Physics, University of British Columbia,
		6224 Agricultural Rd., Vancouver, B.C., Canada V6T 1Z1}

	\begin{abstract}
		
		Starting from a formulation of Correlated Worldline (CWL) theory in terms of functional integrals over paths, we define propagators for particles and matter fields  in this theory. We show that the most natural formulation of CWL theory involves a rescaling of the generating functional for the theory; correlation functions then simplify, and all loops containing gravitons disappear from perturbative expansions. The spacetime metric obeys the Einstein equation, sourced by all of the interacting CWL paths. The matter paths are correlated by gravitation, thereby violating quantum mechanics for large masses. We derive exact results for the generating functional and the matter propagator, and for linearized weak field theory. For the example of a two-path experiment, we derive the CWL matter propagator, and show how the results compare with conventional quantum theory and with semiclassical gravity. We also exhibit the structure of low-order perturbation theory for the CWL matter propagator.

	\end{abstract}
	
	\maketitle

	
	\section{Introduction}
	\label{sec:intro}


	\subsection{Background}
	\label{Ssec:backG}

	Efforts have been made for decades to marry quantum mechanics (QM) and General Relativity (GR) in a consistent theory of quantum gravity \cite{QGrav}. Most theoretical efforts tend to focus on the very high energy regime, at Planck energy scales $\sim E_p = M_p c^2 \equiv (\hbar c^5/G)^{1/2} \sim 1.22 \times 10^{19}$ GeV (the Planck mass $M_p = (\hbar c/G)^{1/2}  \sim  2.18 \times 10^{-8}$ kg), and/or at length scales $\sim \ell_P = \hbar/cM_p \sim 1.64 \times 10^{-35}$ m. This work assumes the validity of QM at all energies, and addresses problems like UV renormalizability, sums over different topologies, the breakdown of GR near singularities, quantum black holes, etc.  
	
	There are however alternative scenarios, wherein one assumes QM to fail because of gravity even at {\it low energies}, because of a perceived incompatibility between GR and QM . Theoretical discussion of this possibility began over 60 years ago \cite{chapelH57,feynmanGR,karolhazy,kibble1,kibble2,penrose96}.  In this case one can expect departures from QM when rest masses approach $M_P$, ie., for mesoscopic objects. Theories of this kind are also motivated by widespread reservations over the validity of QM for macroscopic systems \cite{ajl,PCES09}. These motivating factors are reviewed in section II.A.
	
	In these low-energy scenarios, high-energy questions are put to one side as being premature. One instead starts with low-energy gravity \cite{donoghue-lowE,carney19}, an effective field theory with well-established foundations (see sec. 2.A below). One then looks for deviations from QM within this framework.  
	
	The focus of the present paper is the ``Correlated Worldline" (CWL) theory of quantum gravity \cite{stamp15,BCS18,CWL2}. This is an internally consistent field theory which does predict departures from QM at rest mass scales $\sim O(M_p)$, even for slowly moving masses. The parameters entering the theory are $G_N, \hbar$ and $c$, plus any parameters required to deal with the underlying physics of matter fields (higher-order curvature terms in the action are not excluded, although we will be employing the simple Einstein action in this paper).  
	
	CWL theory is a quantum field theory (QFT), which still has all the usual fields of conventional QFT, including the gravitational metric field $g^{\mu\nu}(x)$. These fields are still `quantized': we define factors $\sim e^{iS/\hbar}$ to be attached to paths, and these factors involve Planck's constant. 
	
	However, CWL theory violates a key assumption of conventional QM or QFT. Instead of the usual independent QM sum over all possible paths (including paths for both matter fields and for $g^{\mu\nu}(x)$), correlations between all paths are mediated by the gravitational field $g^{\mu\nu}(x)$. Note that {\it only the gravitational field is involved in these correlations}. The form of the correlations is uniquely determined by an extension of the equivalence principle. In section 3 we describe the structure of CWL theory in more detail.  
	
\begin{figure}
		\includegraphics[width=3.2in]{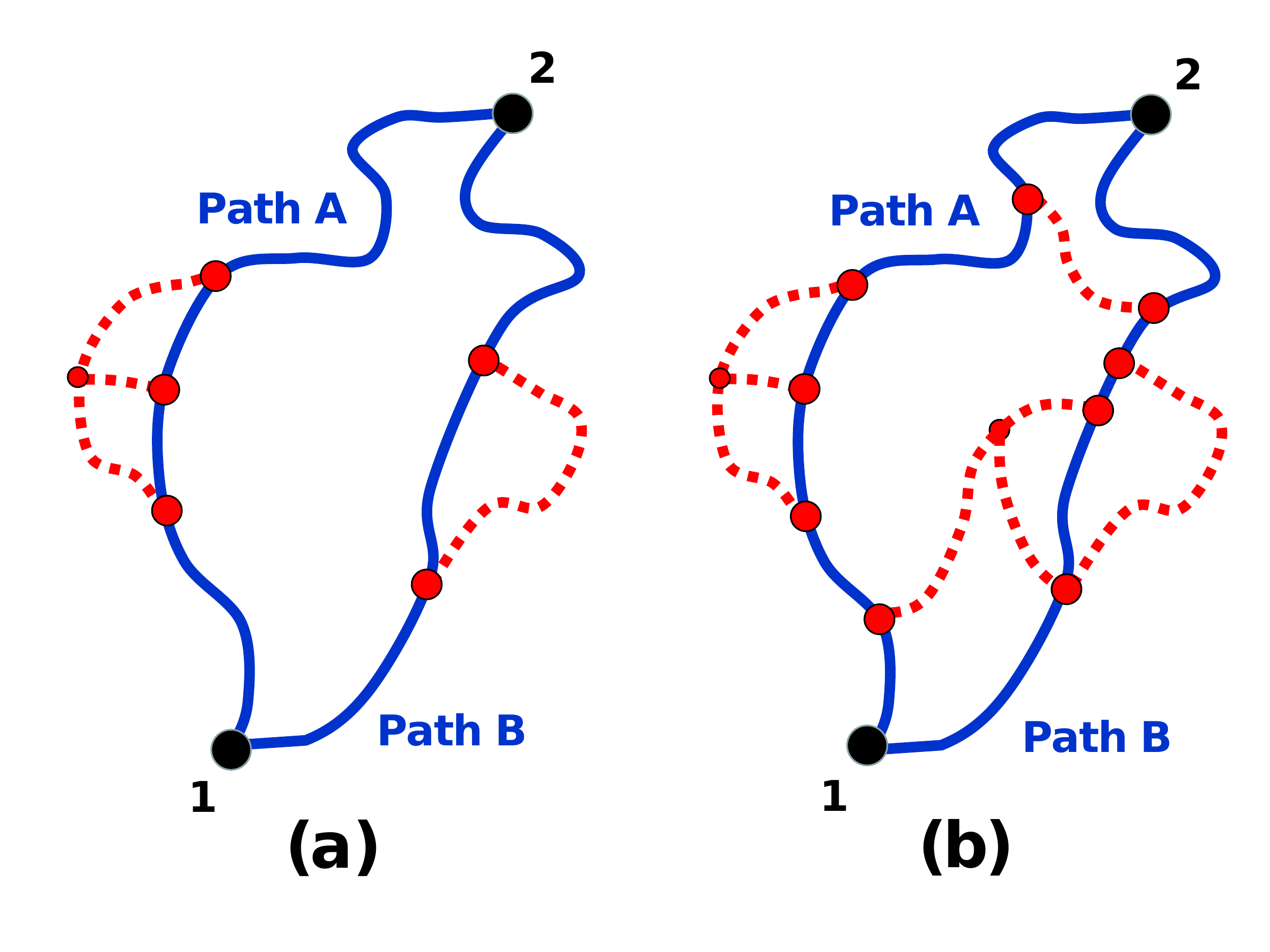}
		\caption{\label{fig:2path-CWL} A comparison between 2-path processes in ordinary QM or QFT and CWL theory, where a single particle is coupled to gravitation. The particle path is shown as a solid line, gravitons as hatched lines. In (a) we see a typical process in conventional QFT - to get the QM amplitude for the process we sum the contributions from path A and path B. In (b) we see a process in CWL theory, in which one cannot separate or sum over the contributions from paths A and B, because gravity has coupled them via ``CWL correlations". }
	\end{figure}

	To see the difference between conventional theory and CWL theory, consider a typical ``2-path" or a ``2-slit" experiment. In conventional QM or QFT (\ref{fig:2path-CWL}(a)) the two different paths are summed over independently to give the quantum transition amplitude between 2 states \cite{feynmanIII,feynH65}. In CWL theory this superposition rule is violated:  processes like that shown in Fig. \ref{fig:2path-CWL}(b) exist, in which gravitational interactions occur between different paths for a {\it single quantum system}.
	
	Even the lowest order perturbative processes involving gravity then give results different from conventional QM \cite{stamp15,BCS18,CWL2,stamp12,carney19}. However, these CWL corrections are expected to be immeasurably small until the mass of the objects involved approaches $\sim M_p$; for microscopic masses they are far too weak (see, eg., the numbers given in ref. \cite{stamp15}, and in much more detail in section 7.B below). For large masses, 2nd-order perturbation theory suggests \cite{stamp15} that the key physical process is one of ``path-bunching", in which the CWL interaction between different paths, for a {\it single particle}, ultimately cause the paths of that particle to bunch together.
	
	At first glance it would seem very hard to find a consistent non-perturbative theory of this kind. In 2 recent papers \cite{BCS18,CWL2} we have described various consistency checks, and the theory has passed all of these. 
	
	However, one would like to address another kind of consistency for CWL theory, viz., consistency with experiment. With an eye on `2-path' interference experiments in, eg., optomechanical systems \cite{bouwm03,QG-interf,aspelm19}, at rest mass scales approaching $M_P$, we here focus on (i) the 2-path experiment (section 6), and on low-order perturbation theory (section 7). 
	
	Some of our previous work has discussed the motivation for CWL theory  \cite{stamp15,BCS18,CWL2}; we briefly recall this rationale in section 1.B below. One can also ask what CWL theory is good for, ie., what does it do better than other theories of quantum gravity, and what new approaches and new experiments it suggests. We give a preliminary answer to this question at the end of the paper, in section VIII.

	
	\subsection{CWL Theory: Physical Discussion \& Motivation}
	\label{Ssec:CWL-phys}
	
	The rationale for CWL theory is largely based in physical arguments. On the one hand one has the strong suspicion that QM must break down in some way at the macroscopic scale, and on the other hand many questions have been raised about the compatibility of QM and GR at low energies (where GR is supposed to work very well). The problems can be summarized as follows:
	
	(a) {\it Macroscopic QM}: Doubts about QM at the macroscopic scale \cite{ajl,PCES09} have led to many tests of quantum superposition, quantum interference and coherence, and of Bell and Leggett-Garg inequalities, at the nanoscopic scale. Examples include ``mass superpositions" (ie., superpositions with a mass in 2 different positions) of large molecules in 2-slit or similar systems \cite{arndt19} and of large masses in optomechanical systems \cite{optoM}, flux superpositions for SQUID devices \cite{SQUID}, and spin superpositions for magnetic systems \cite{polzik,Taka11}. 
	
	Although there is some dispute over how to measure the `macroscopicity' of these states \cite{birgitta,ajl16,arndt14}, the largest `2-path' mass superpositions (in which the paths actually separate) that have been found so far \cite{arndt19} involve masses $< 10^5$D, ie., $< 10^{-14} M_p$, (note that $M_p = 1.311 \times 10^{19}$D). Such masses are far too small for one to see gravitational effects. 
	
	(b) {\it Low-energy GR}: Doubts about the low-energy compatibility of QM and GR rest on several arguments, including (i) the problem of the mis-match between spacetimes derived from mass superpositions \cite{wald84,penrose96,penroseCone}, and the consequent inability to define causal relations for quantum fields \cite{wald84}, (ii) paradoxes such as the black hole information paradox \cite{RMW-WGU}, which involves low-energy excitations; and (iii) incompatibilities between orthodox theory quantum measurement theory and standard GR, again at low energy \cite{kibble2,unruh84}. 
	
	Clearly, the theoretical assumption that QM must work at `macroscopic' rest mass scales $\sim O(M_P)$ (let alone at cosmological scales), involves a large extrapolation beyond current laboratory experiments; and it poses clear theoretical problems. Note that this extrapolation is quite different in character from the enormous extrapolation of QFT made in, eg., string theory, up to the Planck energy (an energy $\sim 10^{16}$ higher than that in current particle accelerators).
	
	The idea that gravity could play a role in a low-energy breakdown of QM stems essentially from (a) the problems just noted with macroscopic mass superpositions (b) the idea that gravity is different from the other fields in nature, in that it sees all fields (including itself) in the same way, and provides causal relations \cite{wald84,penroseCone} for all fields (including itself); and (c) that it is the only obvious known physical mechanism that might lead to a breakdown in QM. 
	
	The difficulty is of course to find a low-energy theory of this kind, which is both theoretically consistent and consistent with experiment. This subject has an interesting history. In several remarkable papers, Kibble et al. \cite{kibble1,kibble2} sketched a theory wherein intrinsic non-linearity led to a breakdown of the superposition principle; and they sought this non-linearity in gravitation.  They concluded that such a non-linear theory was unworkable, and also argued that semiclassical gravity was internally inconsistent (see also refs. \cite{unruh84,carney19}). In parallel work, Weinberg \cite{weinberg79} set up a framework for non-linear generalizations of QM (while again keeping the operators, Hilbert space, and measurements of QM). It was shown very quickly \cite{polchinski79} that even this loose framework violated causality, and entailed superluminal communication.
	
	The moral we take from this story is that one needs to drop at least part of the formal structure of QM to make progress (and Kibble tried to dispense with Hilbert space, even for ordinary QM \cite{kibbleQM79}). The idea of CWL theory \cite{stamp15,BCS18} is that one starts from path integrals, and generalizes these beyond the usual QM framework. The idea that one start from path integrals is of course not new \cite{hartle88,hawking-book,hartleJ,oeckl,CTC}). However the CWL framework also drops the linearity inherent to QM, QFT, and conventional quantum gravity, since in CWL theory paths are coupled. 
	
	In CWL theory, notions like ``measurement" are secondary \cite{stamp15}; measurements are just another physical process. Instead, the difference between the microscopic and macroscopic worlds arises from within the theory; for sufficiently large masses, the usual quantum dynamics of the system fails as CWL correlations between paths set in.  
	
	In previous papers our study of CWL theory focused on formal questions. It was found that (i) when $G_N \rightarrow 0$, we get back conventional QM or QFT, and letting $\hbar \rightarrow 0$ gives GR; (ii) one may formulate consistent expansions about the classical limit $\hbar \rightarrow 0$ and the non-gravitational limit $G_N \rightarrow 0$, and (iii) calculate correlation functions. Finally (iv) it was shown that the theory was gauge and diffeomorphism invariant, and obeyed all relevant Ward identities \cite{BCS18,CWL2}. 
	
	In the present paper we focus more on the physics of CWL dynamics. After a theoretical preamble, in section 2, our new results fall into 3 main categories:
	
	(a) First, we rescale the generating functional of the theory to better organize various prefactors.  It simplifies the expressions for correlation functions, and leads to a massive simplification in the perturbative structure of the theory - in the interaction between CWL lines shown in Fig. (\ref{fig:2path-CWL}(b)), no loops containing gravitons survive. This is done in sections 3 and 4. 
	
	(b) We calculate matter propagators between `boundary data' defined on 2 different hypersurfaces. The dynamics of the matter field is different from standard QM. In section 5, we establish key exact results for the matter propagator and for the connected generating functional $\mathbb{W}$. We also derive the weak-field linearized form of CWL theory. 
	
	(c) To study the CWL dynamics in more detail, section 6 looks at the 2-path experiment, and gives explicit results in the linearized regime for CWL theory, for conventional linearized gravity, and for semiclassical gravity. The three results all differ from each other. Then in section 7 we look at the dynamics of a single particle at lowest non-trivial order in $G_N$ (ie., $\sim O(\ell_P^2)$). This calculation shows clearly how CWL theory departs from standard QM for large rest masses. Finally, in section 8 we summarize the lessons learned from these calculations. 
	
	There are several things we do not address here. We do not discuss quantum measurement theory in any detail - this is a large topic requiring discussion of real measuring systems. We also ignore questions of renormalizability - this is the subject of a separate investigation. Finally, we assume a simple structure for spacetime - no attempt is made to discuss horizons, achronal regions, or singularities. For the explicit calculations in the paper, of relevance to potential experiments (in sections 6 and 7), we assume a background flat spacetime.
	
	Finally, a notational point - for most of the paper we will put $\hbar = 1$, except when we wish to emphasize its role in the theory.

	
	\section{Theoretical Preliminaries}
	\label{sec:theoP}
	
	
	Let us first recall some key features of conventional theory, and also of the formal structure of CWL theory. This will also allow us to establish notation.
	
	In section 2.A we define `ring paths' for the generating functional ${\cal Z}$ in both conventional QFT and in conventional quantum gravity; we then show how to define matter field propagators and field correlators in these theories. To make all of this clearer we give more detail, in Appendix A, on how this works for ordinary QM, for scalar field theory, and for conventional quantum gravity. 
	
	In section 2.B, we briefly recall the form of the generating functional $\mathbb{Q}$ for what we call the `unscaled version' of CWL theory \cite{BCS18,CWL2}, and the $n$-point matter field correlators it leads to.  In Appendix B we also deal with a technical question in this unscaled theory, which was left unresolved in previous papers, viz., the form of the regulator $c_l$.

	\subsection{Conventional Theory}
	\label{sec:conv-QGr}
	
	Here we give a summary of the ring path definition of ${\cal Z}$, and definition of the propagator in terms of it, for a particle and a scalar field, both on a flat spacetime. We then look at the same two quantities for a scalar field coupled to gravity. Again, we let $\hbar = 1$.

	\subsubsection{Ring Paths and Propagators}
	\label{sec:ringP}
	
	In conventional QFT one defines a generating functional ${\cal Z}[J]$ for some matter field (eg., for a scalar field $\phi(x)$) as a functional of some external current $J(x)$ coupling to $\phi(x)$. Here we define the generating functional in terms of `ring paths' (see also our previous papers \cite{stamp15,BCS18,CWL2}). Our goal is to then define propagators, starting directly from ${\cal Z}$.

	\vspace{3mm}
	
\begin{figure}
		\includegraphics[width=3.2in]{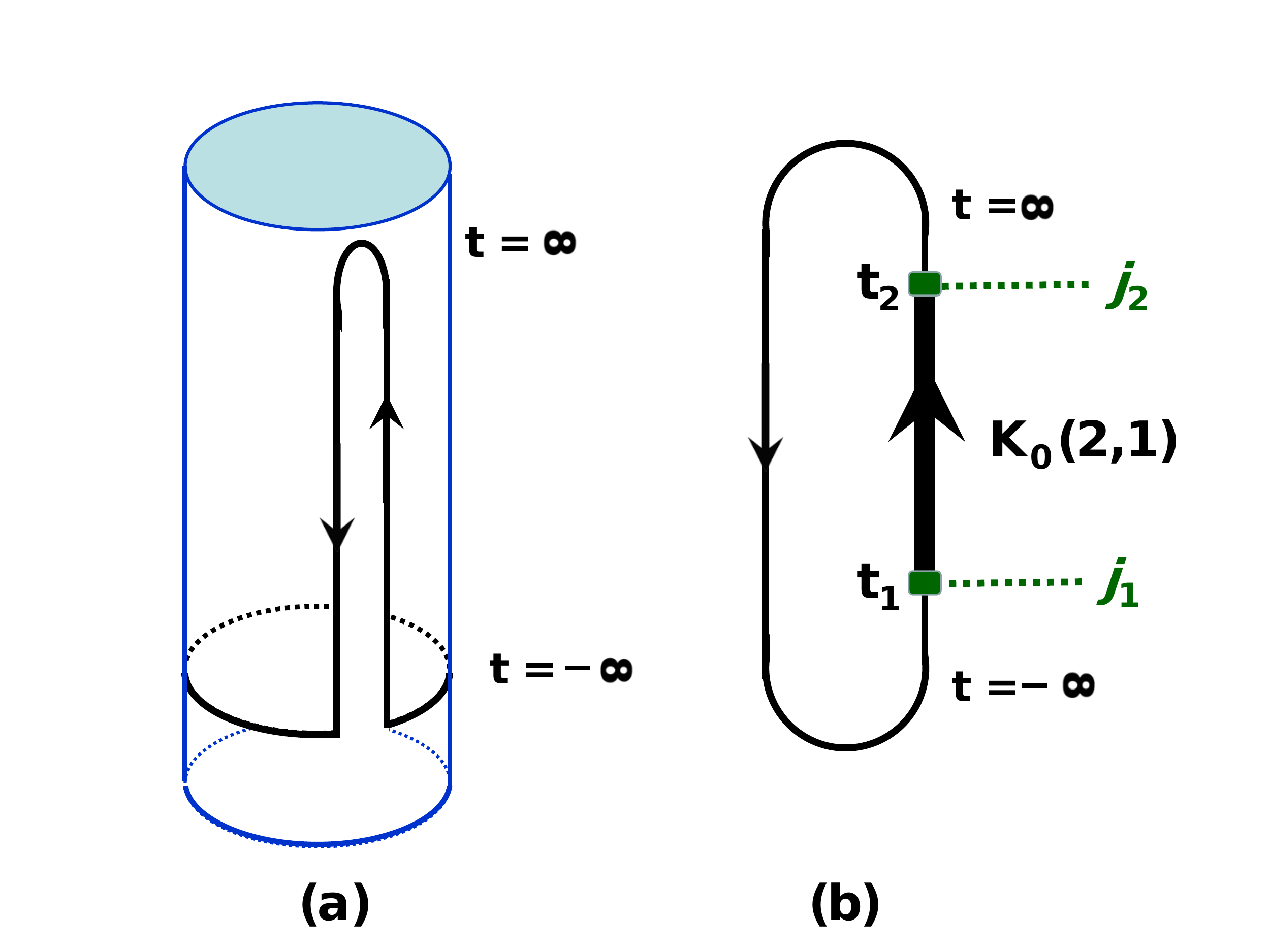}
		\caption{\label{fig:ring+K1} In (a) we show the contour of the ``ring" diagram for the generating functional of the single particle discussed in the text. This extends from proper time $t = -\infty$ up to $t = \infty$ and back again; it is then closed at $t = -\infty$ around the ``temperature cylinder" of circumference $2\pi/kT$. In (b) the contour is represented by a ring, and we show how the propagator $K_o(2,1)$ defined in the text is produced by injecting external currents ${\bf j}_1,{\bf j}_2$ at times $t_1, t_2$ on the upward section of the contour from $t= -\infty$ to $t= \infty$ (and then integrating over ${\bf j}_1$ and ${\bf j}_2$.   }
	\end{figure}

	{\bf (i) Particle Dynamics}: Consider a non-relativistic particle with action $S_o[{\bf r}, {\bf \dot{r}}]$; and a particle coordinate ${\bf r}(t)$ coupling to some external current ${\bf j}(t)$, giving a generating functional
	\begin{equation}
		{\cal Z}_o[{\bf j}] \;=\; \oint {\cal D} {\bf r}(t) \; e^{i(S_o[{\bf r}, {\bf \dot{r}}] + \int {\bf j}\cdot {\bf r})}
		\label{Zo-QM}
	\end{equation}
	where the path integration $\oint {\cal D} {\bf r}(t)$ is taken over a set of closed ``ring paths". The usual way this is done is by extending a Schwinger-Keldysh contour \cite{schwingerK} from $t = -\infty$ to $t = + \infty$ and then back again, and then closing this with a path in imaginary proper time (Fig. \ref{fig:ring+K1}(a)). We will adopt this procedure here.
	
	We can then define the propagator for the particle, starting directly from this generating functional. To simplify the discussion here, we assume a simple propagator between 2 times $t_1$ and $t_2$ on the `upward' path of the ring; this defines one of the 4 Keldysh propagators (for more details, see Appendix A).  We thus introduce two `cuts' at the  times $t_1$ and $t_2$ in the ring path, by writing ${\bf j}(t) = {\bf j}_1 \delta(t-t_1) + {\bf j}_2 \delta(t-t_2)$, so that
	\begin{eqnarray}
		{\cal Z}_o[{\bf j}] & \rightarrow & {\cal Z}_o[{\bf j}_1, {\bf j}_2] \; \equiv \; {\cal Z}_o[{\bf j}_1 \delta(t-t_1) + {\bf j}_2 \delta(t-t_2)]  \nonumber \\
		&=& \; \oint {\cal D} {\bf r}(t)   e^{i(S_o[{\bf r}]}\; e^{i({\bf j}_1 \cdot {\bf r}(t_1) + {\bf j}_2 \cdot {\bf r}(t_2))} \quad
		\label{Zo-j1j2}
	\end{eqnarray}
	in which the cuts have vector magnitudes ${\bf j}_1$ and ${\bf j}_2$ respectively. 
	
	We now integrate over both ${\bf j}_1$ and ${\bf j}_2$ between these 2 cuts, which defines the function 
	\begin{equation}
		\aleph(2,1) = \int d{\bf j}_1  d{\bf j}_2 \; e^{-i ({\bf j}_1 \cdot {\bf x}_1 + {\bf j}_2 \cdot {\bf x}_2)}  {\cal Z}_o[{\bf j}_1, {\bf j}_2]
		\label{Ko-Zo}
	\end{equation}
	which is shown in App. A to be equivalent to the product
	\begin{equation}
		\aleph(2,1) \; = \; K_o(2,1) \, f (2,1)
		\label{aleph}
	\end{equation}
	depicted in Fig. \ref{fig:ring+K1}(b), in which the two terms are

	(i) the usual Feynman propagator $K_o(2,1)$ between states $|1 \rangle \equiv |{\bf x}_1 \rangle$ and $|2 \rangle \equiv |{\bf x}_2 \rangle$ at times $t_1$ and $t_2$, which we write as
	\begin{equation}
		K_o(2,1) \;=\; \int_1^2 {\cal D} {\bf r}(t) e^{iS_{o}[{\bf r}]}
		\label{Kqm-21}
	\end{equation}
	ie., the heavy line shown in Fig. \ref{fig:ring+K1}(b); and  
	
	(ii) the light line shown in Fig. \ref{fig:ring+K1}(b) which completes the ring, and which is given by 
	\begin{equation}
		f(2,1) \;=\; \langle {\bf x}_1 | e^{-i H(t_1 - t_{in})} \, \hat{\rho}_{in} \,  e^{i H(t_2 - t_{in})} | {\bf x}_2 \rangle
		\label{f-21-rho}
	\end{equation}
	where we let $t_{in} \rightarrow -\infty$, and $\hat{\rho}_{in}$ is the density matrix defined on the contour around the cylinder defined at $t = -\infty$, which here is a thermal density matrix defined at temperature $T$.

	
	One can in the same way define the propagator for a relativistic particle, and for a density matrix, and define conditional propagators in which other conditions are prescribed in addition to the boundary information about the end-points (see Appendix A).

	\vspace{3mm}
	
	{\bf (ii) Scalar Field Dynamics}:  Consider a scalar field $\phi$ with action $S[\phi]$ and generating functional $Z_{\phi}[J]$ in the presence of an external field $J(x)$, defined on a spacetime in which a hypersurface $\Sigma$ bounds a `bulk' spacetime region ${\cal M}$. The surface $\Sigma$ is divided into spacelike past and future surfaces $\Sigma_1$ and $\Sigma_2$, along with a region $\Sigma_B$ at spatial infinity.
	
	Starting from $Z_{\phi}[J]$, and using the same methods as before (now imposing cuts at $\Sigma_1$ and $\Sigma_2$), we get a propagator between scalar field configurations $\Phi_1(x)$ and $\Phi_2(x)$, localized on $\Sigma_1$ and $\Sigma_2$, given by
	\begin{equation}
		K(2,1) \; \equiv \; K(\Phi_2,\Phi_1)  \;=\;  \int^{\Phi_2}_{\Phi_1} {\cal D}\phi\, e^{iS_{\phi}[\phi]}
		\label{K2-QFT-g0}
	\end{equation}
	
	The analogy with the particle derivation just given is clearest when the surfaces $\Sigma_1$ and $\Sigma_2$ are simple time slices at times $t_1$ and $t_2$. If they are not, then the discussion becomes a lot more technical (compare refs. \cite{jordan20} - \cite{jordanPhD}), but the basic principles are still the same. 
	
	We can also, again in analogy with the discussion for a particle, define a conditional propagator for the field $\phi(x)$, on a spacelike hypersurface $\bar{\Sigma}$  located between $\Sigma_1$ and $\Sigma_2$; and one can generalize these derivations to gauge field theories (see refs. \cite{jordan20,jordanPhD} for the case of QED).

	\subsubsection{Conventional Quantum Gravity}
	\label{sec:conv-QGr}
	
	By `conventional quantum gravity' we mean a low-energy theory of gravity, framed in terms of the Einstein action, which we will define using path integrals. Without the restriction to low energies, one expects severe problems: the theory is then non-renormalizable, and involves quantum sums over different spacetime topologies. Although there do exist procedures to define such sums \cite{loll12}, it is not clear whether topology-changing transitions are physically meaningful \cite{dewitt86}. 
	
	In a low-energy effective theory, one expands about a background metric configuration $g_0$. This is done either by expanding perturbatively in $G_N$, or in metric fluctuations. A UV cutoff is implicit, and there is no sum over different spacetime topologies. Here we will define a generating functional and matter field propagators for this theory, and establish our notation. 
	
	Consider again a scalar field  $\phi(x)$, now with action
	\begin{equation}
		S_{\phi}[\phi,g] \;=\; \tfrac{1}{2} \int d^4x g^{1/2}\;[g^{\mu \nu} \nabla_{\nu}\phi \nabla_{\mu}\phi - V(\phi)]
		\label{S-Phi}
	\end{equation}
	in the presence of the background metric $g \equiv g^{\mu\nu}(x)$ (here $\nabla$ denotes a covariant derivative). 
	
	Let us first write the matter generating functional with a fixed background spacetime $\bar{g}$, as 
	\begin{equation}
		Z_{\phi}[\bar{g}, J] \;=\;\oint D\phi\;e^{i(S_{\phi}[\phi,\bar{g}]\;+\;\int J\phi)} \;\; \equiv \;\; e^{i\,W_o[\bar{g}, J]}
		\label{Z-gj}
	\end{equation}
	so that $W_o[\bar{g}, J] = -i \ln Z_{\phi}[\bar{g}, J]$ is the generating functional for connected diagrams for the scalar field on the background $\bar{g}$. We assume, as before, a spacetime ${\cal M}$ bounded by the hypersurface $\Sigma$; and we assume all fields vanish fast enough at $\Sigma_B$ that we can integrate by parts freely on spatial derivatives, without picking up surface terms at $\Sigma_B$.

	We now unfreeze the metric $g^{\mu\nu}(x)$. The pure gravitational action is written as
	\begin{equation}
		S_{G}[g]=\ell_{P}^{-2}\left(I^{o}_{G}+I^{YGH}_{G}\right),
	\end{equation}
	where $l_P^2 = 16\pi G$ is the square of the Planck length, and we have put $\hbar =1, c = 1$. Here we include the bulk Einstein action $I_G^o = \int d^{4}x\sqrt{g} R$, in which $R$ is the Ricci scalar, defined in ${\cal M}$, and $I_{YGH}$ is a York-Gibbons-Hawking boundary term \cite{YGH}, given by
	\begin{equation}
		I^{YGH}_G \;=\; 2 M_{P}^{2}\int_{\Sigma}d^{3}y \; \epsilon(\Sigma) \,\sqrt{|\mathfrak{h}|}\,K
		\label{YGH-A}
	\end{equation}
	in which $\mathfrak{h}$ is the determinant of the induced metric on $\Sigma$, $K$ is the trace of the extrinsic curvature $K_{ab}$ of $\Sigma$, and
	$\epsilon(\Sigma)=\pm1$, depending on whether the relevant piece of $\Sigma$ is timelike or spacelike.
	
	Finally we include a gauge-fixing function $\chi^{\mu}(g(x))$, to get rid of the gauge redundancy in path integrals under diffeomorphisms $x^{\mu}\rightarrow x^{\mu}+\xi^\mu(x)$. With this term we write the total gravitational action as $I[g]/\ell_{P}^{2}$,  with
	\begin{equation}
		I[g] \;= \;  I_G^o + I_G^{YGH} + \tfrac{1}{2}\chi^{\mu} c_{\mu\nu}
		\chi^{\nu}.
		\label{S-EH}
	\end{equation}
	We've written eqtn. (\ref{S-EH}) in the compact DeWitt notation, in which the coordinates are folded in with the tensor indices and repeated indices imply a spacetime integration over these coordinates; thus
	\begin{equation}
		\chi^{\mu} c_{\mu\nu}
		\chi^{\nu} \;\; \equiv \;\;  \int d^{4}xd^{4}x'\, \chi^{\mu}(x) c_{\mu\nu}(x,x')\chi^{\nu}(x')
	\end{equation}
	
	To completely specify the path integral we define the Faddeev-Popov ghost operator \cite{FP67} as
	\begin{equation}
		\Xi^{\mu}_{\nu}(x,x'|g) \;=\;\left.
		\frac{\delta\chi^{\mu}(g^{\xi}(x))}{\delta\xi^{\nu}(x')}
		\right|_{\,\xi = 0}\, .
		\label{FPghost}
	\end{equation}
	
	Both the ghost operator $\Xi^\mu_\nu$ and the matrix $c_{\mu\nu}$ need to be invertible; we write the inverse of $\Xi^{\mu\nu}$ as
	\begin{equation}
		\Xi^\mu_\nu\,\mathfrak{G}^\nu_\lambda \;=\; \delta^\mu_\lambda
	\end{equation}
	which defines the ``ghost propagator"  $\mathfrak{G}^\nu_\lambda (x,x')$.  We also assume that  $c_{\mu\nu}(x,x')\propto \delta(x,x')$, for otherwise an extra ghost contribution $\sim {\rm Det}\, c_{\mu\nu}$ will be needed.
	
	Note that we are describing here a conventional theory with minimal coupling to the matter field. In reality quantum fluctuations generate non-minimal couplings in the action, in any background curved spacetime. In this paper we will ignore such terms, because we are interested in applications to low-energy laboratory experiments, where we expect non-minimal corrections to be unimportant.

	Consider now the generating functional ${\cal Z}[J]$ for this full theory. Naively this is written as
	\cite{dewitt67c,mandelstam68,fradkinV73,FP74}
	\begin{align}
		{\cal Z}[J] \;&=\; \oint Dg \, e^{i (I[g]/\ell_{P}^{2}- i {\rm Tr} \ln \Xi) }
		\; Z_{\phi}[g, J].
		\label{Zconv2}
	\end{align}
	One can also define the generating functional with the gauge-fixing represented explicitly as a constraint, in contrast with the ``Gaussian-smeared'' version above. We write this below in terms of the Faddeev-Popov functional determinant $\Delta(g) = \rm {Det}\,\Xi = e^{Tr \ln \Xi}$, , viz.,
	\begin{eqnarray}
		{\cal Z}[J] &=&  \oint\mathcal{D}g\,e^{iS_G[g]}\Delta[g]\delta(\chi^{\mu}(g)) \oint D\phi\;e^{i(S_{\phi}[\phi,g]\;+\;\int J\phi)}\nonumber \\
		&=&\oint\mathcal{D}g\,e^{iS_G[g]}\Delta[g]\delta(\chi^{\mu}(g)) \; Z_{\phi}[g, J].
		\label{Zconv3}
	\end{eqnarray}
	For $J=0$ these two definitions coincide, but not for general $J$. However, both expressions yield the same results when used to compute gauge-invariant quantities.

	Pursuing this approach, one then defines the propagator between two different field configurations $\Phi_1(x)$,$\Phi_2(x)$, and two induced metric configurations $\mathfrak{h}^{ab}_1, \mathfrak{h}^{ab}_2$, specified on $\Sigma_1$ and $\Sigma_2$ respectively. We get \cite{hartleJ,hawking-book}
	\begin{eqnarray}
		K(2,1) & \equiv & K(\Phi_2, \Phi_1; \mathfrak{h}^{ab}_2, \mathfrak{h}^{ab}_1) \nonumber \\
		&=&  \int^{\mathfrak{h}_2}_{\mathfrak{h}_1} {\cal D}g\,  e^{i S_G[g]} \Delta(g) \,\delta(\chi^{\mu})   \int_{\Phi_1}^{\Phi_2} {\cal D}\phi \; e^{iS_{\phi}[\phi, g]} \nonumber \\ &=& \int^{\mathfrak{h}_2}_{\mathfrak{h}_1} {\cal D}g \,  e^{i S_G[g]} \Delta(g) \,\delta(\chi^{\mu})  \; K_0(2,1|g)
		\label{K2-QFT-int-g}
	\end{eqnarray}
	where the function $K_{0}(2,1|g) \equiv  K_0(\Phi_2,\Phi_1|g)$ is the propagator for $\phi(x)$ when  $g^{\mu\nu}(x)$ is `frozen' in one particular configuration $g$.

	
	\begin{figure}
		\includegraphics[width=3.2in]{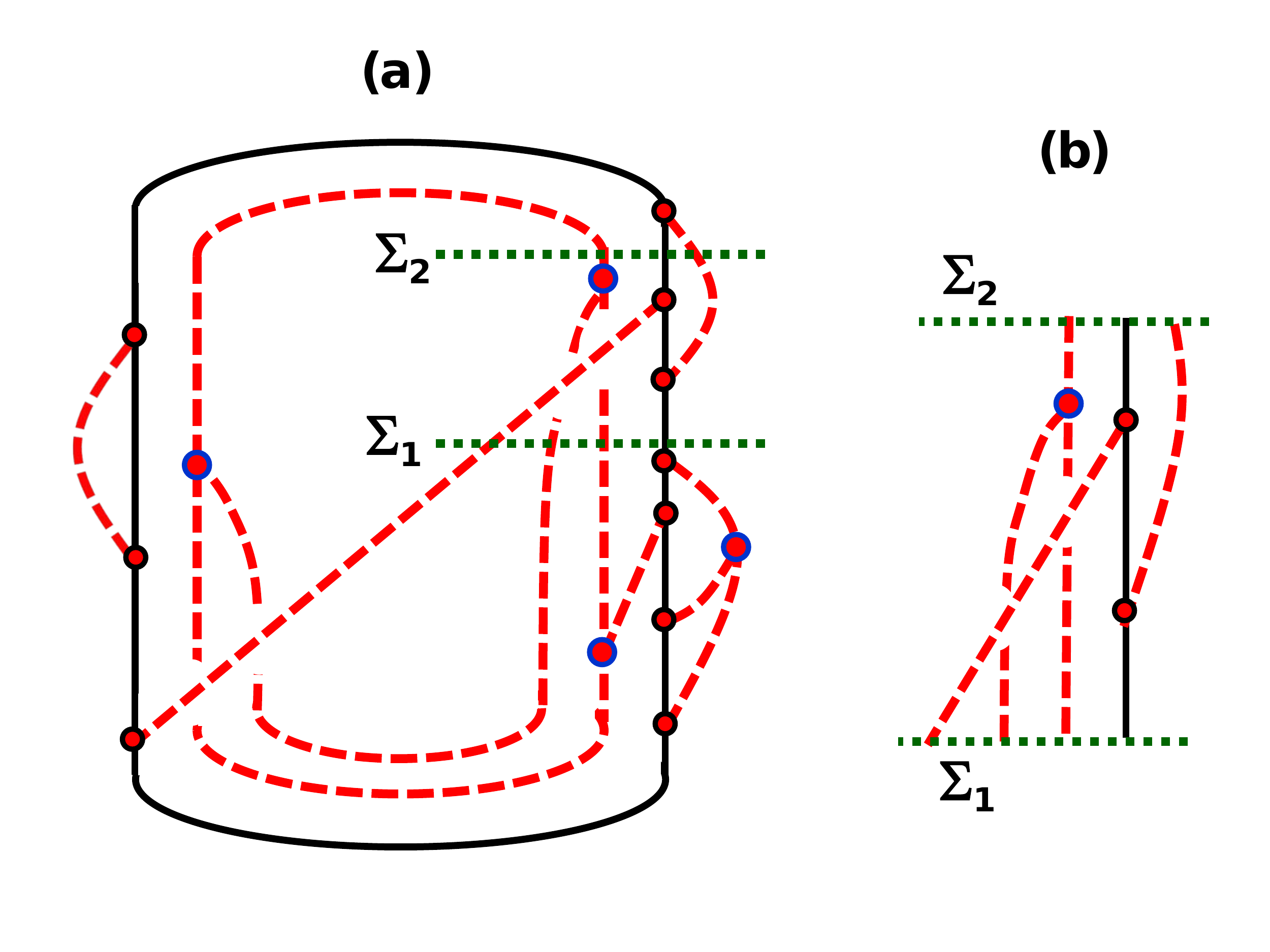}
		\caption{\label{fig:ring-G+M} In (a) we show a typical graph for a ``ring" contribution to the generating functional ${\cal Z}$ for a scalar field coupled to gravitons discussed, as in the text. Again, this extends from proper time $t = -\infty$ up to $t = \infty$ and back again, and is around the temperature cylinder. The dashed lines represent graviton propagators, and the solid line is the matter field. In (b) we show the propagator $K_0(2,1)$ generated by imposing cuts on the surfaces $\Sigma_1$ and $\Sigma_2$; the initial and final states (obtained from where the lines crossing the cuts on $\Sigma_1$ and $\Sigma_2$, in the graph in (a)) involve multiple gravitons as well as the scalar matter field.  }
	\end{figure}
	

	However, the problem with both the ring path functional (\ref{Zconv3}) and the propagator (\ref{K2-QFT-int-g}) is that if we impose no restrictions on the allowed configurations of the metric field $g^{\mu\nu}(x)$, then it is completely unclear what is meant by the integrations $\oint {\cal D}g$ and $\int_1^2 {\cal D}g$ in these formulas.  If all the spacetimes included in the integration were constrained to be compact (compare ref. \cite{hawking79-15}, pp. 749-52), or at least spatially closed \cite{hawking-book}, then one might attempt a rigourous definition of these path integrals; but of course there is no reason to make such restrictions.  
	
	This is where our restriction to low energies comes in. We now assume a slowly-varying background spacetime $g_0$; and we adopt the view, standard in QFT, that the path integration now defines a perturbative expansion about $g_0$, ie., some sort of graviton expansion. Again, we ignore non-minimal couplings. 
	
	The ring path diagrams then involve both matter and graviton states - a typical example is shown in Fig. \ref{fig:ring-G+M}(a), involving multiple gravitons. The cuts in this ring diagram required to produce the propagator in Fig. \ref{fig:ring-G+M}(b), on the surfaces $\Sigma_1$ and $\Sigma_2$, now involve external currents coupling to both the matter and graviton fields (see Appendix A). We get a propagator $K(2,1)$ in which the initial state $|1 \rangle \equiv |h_1, h'_1, h''_1; \Phi_1 \rangle$ has 3 incoming gravitons and a scalar field state $|\Phi_1 \rangle$, and the final state $|2 \rangle \equiv |h_2, h'_2; \Phi_2 \rangle$ has 2 outgoing gravitons and a final state $|\Phi_2 \rangle$ for the scalar field.  
	
	We can also generalize the above work to cover propagators for the density matrix (see Appendix A). The techniques for doing this were described in ref. \cite{BCS18}, and worked out in detail for linearized gravity in ref. \cite{jordan-CQG18}. Explicit expressions for eqtns. (\ref{K2-QFT-int-g}) and its particle analogue can be found in linearized gravity, in a way analogous to that for QED \cite{jordanMSc,jordan20}; we will not need these in the present paper.

	\subsection{Unscaled CWL Theory}
	\label{sec:conv-QGr}
	
	The unscaled version of CWL theory was described in detail in refs. \cite{BCS18,CWL2}, and we summarize it here.  Again, to be specific, we consider a scalar matter field. One starts by replacing the single scalar field $\phi(x)$ appearing in
	conventional QFT by a ``tower", ie., a set $\{\phi_{k}^{(n)}\}$ of multiple versions of $\phi(x)$, with $k=1,2,...n$, coupled to a set $\{ g_n \}$ of metric fields. One then writes a generating functional
	\begin{eqnarray}
		&&\tilde{\mathbb{Q}}[J]=\prod\limits_{n=1}^\infty {\cal Q}_n[\,J\,], \nonumber\\
		&&{\cal Q}_n[\,J\,]=
		\!\oint\! Dg_{n}\,e^{inS_G[\,g_{n}\,]} \, \Big(Z_{\phi}\Big[\,g_{n},
		\frac{J}{c_n}\Big]\Big)^n
		\label{bbQ-J1'}
	\end{eqnarray}
	in which we take the product over all $n$, ie., we take the product over {\it all} the towers of different $n$. The number $c_n$ is a regulating factor, whose form is derived in Appendix B. Here, and in what follows, we suppress all reference to gauge-fixing and Faddeev-Popov determinants; they will be absorbed into the path integral measure $\int {\cal D}g_n$, and only written explicitly when necessary.

	In previous papers we have sometimes referred to the $n$ different members $\{ \phi^{(n)}_k(x) \}$ of the tower as `copies' or `replicas' of the basic field $\phi(x)$ (or of some particle path $q^{\mu}(\tau)$). However this language is misleading, because it implies that each field has an independent existence, and that the permutations of the field labels can be treated as a symmetry under which the states can be organized into representations.
	
	In CWL theory, however, these `replicas' are simply a mathematical device used to represent different paths (or configurations) {\it of a single object}. In contrast with conventional QFT, `replica permutation' (ie., path permutation) in CWL inside some given tower should be treated as the analogue of a discrete gauge symmetry - the paths are indistinguishable and refer to a single physical system. As a matter of principle one should never try to physically distinguish one `replica', or path, from another. The `towers' are thus simply collections of $n$ different paths for the same object. 
	
	Notice that the gravitational action in the $n$-th tower (ie., for the $n$-path term) is rescaled by a factor $n$. This rescaling of $S_G[g_{n}]$ to $nS_G[g_{n}]$ implies a coupling constant scaling $G \rightarrow G/n$ for the metric $g_{n}$ in this tower, which apparently reduces the effect of metric fluctuations at high $n$. Note, however, that the stress-energy tensor $T_{\mu\nu}$ rescales in the opposite way, to $n T_{\mu\nu}$. Thus, as we will see, the classical Einstein equations still hold in the classical limit of CWL theory (and in this paper we will discover that they hold quite generally, even when the matter fields are in the quantum regime). 
	
	The generating functional for connected diagrams is given in unscaled CWL theory from (\ref{bbQ-J1'}), as
	\begin{eqnarray}
		\tilde{\mathbb{W}}[J] &=& -i \, \ln \tilde{\mathbb{Q}}[J]  \nonumber \\
		&=& -i  \lim_{N\rightarrow\infty} \; \sum_{n=1}^N \, \ln \, {\cal Q}_n[\,J\,]
		\label{bbW-J}
	\end{eqnarray}
	which is additive over the different towers. We immediately derive the connected correlation functions of the theory upon functional differentiation with respect to $J(x)$, to give \cite{BCS18}
	\begin{eqnarray}
		{\cal G}_l(\{ x_k \}) &=& \langle\,\phi(x_1)...\phi(x_l)\,
		\rangle^{CWL}_{\rm c} \nonumber \\
		&=&(-i)^l {1 \over \sum\limits_{n=1}^\infty\,
			n c^{-l}_n}
		\left.{ \delta^l
			\ln\tilde{\mathbb{Q}}[J] \over \delta J(x_1) .. \delta J(x_l)} \,\right|_{\,J = 0}
		\label{corrCWL-poss}
	\end{eqnarray}
	where the correlator is calculated for some state of the system; for the vacuum state $|\Phi_o\rangle$ we would have
	\begin{equation}
		\langle\,\phi(x_1)...\phi(x_l)\,\rangle \;\equiv \; \langle \Phi_o |\,\phi(x_1)...\phi(x_l)\,|\Phi_o\rangle
		\label{vacCorr}
	\end{equation}
	
	The result (\ref{corrCWL-poss}) contains the regulating factor $c_n$. In Appendix B we show that $c_n = 1$, for all $n$, so that (\ref{corrCWL-poss}) becomes
	\begin{equation}
		{\cal G}_l(\{ x_k \})
		\; =\;  {(-i)^l \over \sum\limits_{n=1}^\infty\,
			n }
		\left.{ \delta^l
			\ln\mathbb{Q}[J] \over \delta J(x_1) .. \delta J(x_l)} \,\right|_{\,J = 0}
		\label{corrCWL-unsc}
	\end{equation} 
	
	Since we expect the correlators $\{ {\cal G}_l(\{ x_k \}) \}$ to be finite, we then see that the divergent denominator in (\ref{corrCWL-unsc}) is exactly cancelled by the divergent numerator coming from (\ref{bbW-J}). This situation is mathematically unsatisfactory, and suggests that we rescale the original form for $\tilde{\mathbb{Q}}[J]$ in (\ref{bbQ-J1'}). As we now see, this rescaling, although not changing the theory in any fundamental way, does make it much simpler to work with.

	
	\section{Rescaled CWL Theory}
	\label{sec:CWL}
	
	
	We now turn to the rescaled version of CWL theory we shall use from now on. In section 3.A we describe the rescaled theory, and show how it leads to a much simpler form for the correlation functions. Then, in section 3.B, we show how the both the classical limit, and the decoupled limit (where $G_N = 0$) simplify in the rescaled CWL theory. Finally, in section 3.C we set out the diagrammatic rules for the calculation of perturbative expansions in $G_N$, for the connected generating functional $\mathbb{W}$.

	\subsection{Rescaled CWL Theory}
	\label{sec:CWLgenF-reS}
	
	One always has some liberty in how the generating functional $\mathbb{Q}$  defined, because it is  $\ln(\mathbb{Q})$ that is of importance in determining physical quantities. Thus, eg., multiplication of $\mathbb{Q}$ by some factor simply adds an irrelevant constant to $\ln(\mathbb{Q})$, and raising $\mathbb{Q}$ to some power amounts to a rescaling of $\ln(\mathbb{Q})$. In what follows we employ a very natural rescaling which greatly simplifies the theory.

	\subsubsection{Form of Rescaling}
	\label{sec:CWL-reS-form}

	Suppose we transform the unscaled generating functional $\tilde{\mathbb{Q}}[J]$ given in the last section, so that $\tilde{\mathbb{Q}}[J] \rightarrow \mathbb{Q}[J] = \tilde{\mathbb{Q}}^{\alpha}[J]$. Then the connected generating functional rescales as $\tilde{\mathbb{W}}[J] \rightarrow \mathbb{W}[J] = \alpha \tilde{\mathbb{W}}[J]$. This rescaling then multiplies the correlation functions, etc., by a factor $\alpha$.
	
	Here we rescale the generating functional to
	\begin{equation}
		\mathbb{Q}[J] \;=\; \lim_{N\rightarrow\infty} \left( \prod_{n=1}^N {\cal Q}_n[J] \right)^{\alpha_N}
		\label{bbQ-alpha}
	\end{equation}
	so that the rescaled connected generating functional is
	\begin{equation}
		\mathbb{W}[J] \;=\; -i \lim_{N\rightarrow\infty} \alpha_N \sum_{n=1}^N \ln \,{\cal Q}_n[J]
		\label{bbW-alpha}
	\end{equation}
	ie., we write the scaling factor $\alpha_N$ as a function of the number $N$ of towers, and then take the limit $N \rightarrow \infty$. 
	
	We now choose $\alpha_N$ to be
	\begin{equation}
		\alpha_N \;=\; \left(\sum_{n=1}^{N}n\right)^{-1} \;\;=\;\; {2 \over N(N-1)}
		\label{alphaN}
	\end{equation}
	and all of the subsequent theory in this paper will start from the rescaled versions of $\mathbb{Q}[J]$ and $\mathbb{W}[J]$ in eqtns. (\ref{bbQ-alpha})-(\ref{alphaN}). The $n$-th tower functional ${\cal Q}_n$ will be given by 
	\begin{equation}
		{\cal Q}_n[\,J\,]=
		\!\oint\! Dg_{n}\,e^{inS_G[\,g_{n}\,]} \, \Big(Z_\phi[\,g_{n},
		J \,]\Big)^n, 
		\label{bbQ-J1new}
	\end{equation}
	obtained by putting $c_n = 1$ in eqtn. (\ref{bbQ-J1'}).
	
	Because the generating functional factorizes, we see that $\mathbb{W}[J]$  is just a sum over single $g$ integrals, and we do not have correlations between $g_n$ and $g_m$ unless $n=m$, ie. the different towers do not `talk' to each other. We can thus also write (\ref{bbQ-J1new}) as \cite{BCS18}
	\begin{equation}
		{\cal Q}_n[\,J\,]=
		\!\oint\! Dg\,e^{inS_G[\,g\,]} \, \Big(Z_\phi[\,g,
		J \,]\Big)^n, 
		\label{bbQ-J1new'}
	\end{equation}
	with only one metric field.

	Let us write out $\mathbb{Q}[J]$ for the rescaled CWL theory in full, for future reference, always bearing in mind that it is the logarithm of this, ie., the connected generating $\mathbb{W}[J]$, which is the physical object. To be specific we assume a theory with a scalar matter field coupled to gravity. We then have
	\begin{widetext}
		\begin{equation}
			\label{eq:genfunc3}
			\mathbb{Q}[J]  \;\;=\;\;  \lim_{N\rightarrow\infty}\left[\prod_{n=1}^{N}\oint \mathcal{D}g_{n}\,e^{inS_{G}[g_{n}]}\prod_{k=1}^{n}\oint\mathcal{D}\phi^{(n)}_{k}\,
			e^{iS_{\phi}[\phi^{(n)}_{k},g_{n}]+J\phi^{(n)}_{k}}\right]^{\alpha_N} \;\;\;\equiv\;\;\;
			\lim_{N\rightarrow\infty} \left( \prod_{n=1}^N {\cal Q}_n[J] \right)^{\alpha_N}
		\end{equation}
	\end{widetext}
	with the exponent $\alpha_N$ given by (\ref{alphaN}). There are suppressed DeWitt indices in (\ref{eq:genfunc3}); thus $J$, $g$ and $\phi$ are all functions of spacetime coordinates, and the product $J\phi \equiv \int d^4 x J(x) \phi(x)$ is integrated over spacetime. We also omit Faddeev-Popov gauge fixing factors - these will be restored when needed. 
	
	We emphasize again that we will never use the functional $\mathbb{Q}[J]$ except for formal manipulations - it is $\mathbb{W}[J]$ which is physically significant.  As one expects, $\mathbb{Q}[J]$ is essentially a geometric mean of the individual tower generating functionals ${\cal Q}_n$, whereas  $\mathbb{W}[J]$ is a normalized sum over the different $W_n[J]$, where $W_{n}[J]=-i \hbar \log Q_{n}[J]$.

	\subsubsection{Correlators}
	\label{Ssec:CWL-corr}
	
	In the unscaled version of the theory we found that the prescription for computing the correlation functions was quite peculiar - one obtained an awkward formula in which a divergent sum in the main expression was supposed to be cancelled by the prefactor.
	
	In the rescaled version of CWL theory this problem disappears; the prefactor is finite, and the rescaling factor $\alpha_{N}$ removes the divergence. We then immediately find that the correlators are given from $\mathbb{W}[J]$ by straightforward differentiation, to get:
	\begin{equation}
		\label{eq:greensfunc3}
		\mathcal{G}(x_{1},..,x_{l}) \;=\; \frac{(-i)^{l+1}\delta^{l}}{\delta J(x_{1})...\delta J(x_{l})} \mathbb{W}[J]\bigg|_{J=0}.
	\end{equation}
	ie., the same formula as that in ordinary QFT.
	
	One sees explicitly what has happened if we return to the unscaled theory by simply setting  $\alpha_N = 1$ in (\ref{eq:genfunc3}). Then we get, instead of (\ref{eq:greensfunc3}), the result
	\begin{equation}
		\label{eq:greensfunc}
		\lim\limits_{\alpha_N \rightarrow 1}\mathcal{G}_m( \{ x_k \}) \;=\;  \mathfrak{C}\; \frac{(-i)^{l}\delta^{l}}{\delta J(x_{1})...\delta J(x_{l})}\log \mathbb{Q}[J]\bigg|^{J=0}
	\end{equation}
	where the normalizing factor $\mathfrak{C}$ is given by
	\begin{equation}
		\mathfrak{C} \;=\; \lim_{N\rightarrow\infty} \left(\sum_{n=1}^{N}n \right)^{-1}  \;\; \equiv \;\; \lim_{N\rightarrow\infty} \alpha_N
		\label{mfrakC}
	\end{equation}
	ie., the normalizing factor $\mathfrak{C}$ in the unscaled theory is exactly cancelled in the rescaled version by the factor $\alpha_N$, when $N\rightarrow\infty$, to give eqtn. (\ref{eq:greensfunc3}).

	\subsection{Two Limiting Cases}
	\label{Ssec:CWL-CWL-limits}

	Before continuing, we check that the rescaled theory reduces to sensible results in two limiting cases, viz., (i) the ``decoupled limit", where $G_N = 0$, so that the metric field $g^{\mu\nu}(x)$ decouples from any matter field; and (ii) the classical limit $\hbar \rightarrow 0$, where the theory has to reduce to classical Einstein gravity.

	\subsubsection{Decoupled Limit}
	\label{Ssec:CWL-GN-0}

	We wish to show that the rescaled generating functional has the correct limit when $G_N \rightarrow 0$;  we then want the theory to reduce to a conventional QFT defined in flat spacetime (ie, $g \rightarrow \eta$), with no gravitation at all.
	
	Starting from eqtn. (\ref{eq:genfunc3}), we get
	\begin{align}
		\mathbb{Q}[J]\bigg|_{G_{N}=0}&=\lim_{N\rightarrow\infty}\left[\prod_{n=1}^{N}\prod_{k=1}^{n}
		\oint\mathcal{D}\phi^{(n)}_{k}\,e^{iS_{\phi}[\phi^{(n)}_{k},\eta]+iJ\phi^{(n)}_{k}}\right]^{\alpha_N}
		\nonumber \\
		&=\lim_{N\rightarrow\infty}\left[\prod_{n=1}^{N}\left(\oint\mathcal{D}\phi\,e^{iS_{\phi}[\phi,\eta]
			+iJ\phi}\right)^{n}\right]^{\alpha_N} \nonumber \\
		&=\lim_{N\rightarrow\infty}\left[\left(\oint\mathcal{D}\phi\,
		e^{iS_{\phi}[\phi,\eta]+iJ\phi}\right)^{\sum_{n=1}^{N}n}\right]^{\alpha_N} \nonumber \\
		&=Z_{\phi}[J]
		\label{GNto0}
	\end{align}
	where $Z_{\phi}[J] \;=\; \int\mathcal{D}\phi\,e^{iS[\phi]+iJ\phi}$ is the conventional generating functional for a scalar field in the absence of gravity, ie., it is the function $Z_{\phi}[g;J]$ defined previously (in eqtn. (\ref{Z-gj})), but with $g = \eta$. This is precisely the desired result; it holds for any other matter field, or for particles.

	\subsubsection{Classical Limit}
	\label{Ssec:CWL-class}

	The actions for both the metric and the set of $n$ paths are unchanged by the overall rescaling factor $\alpha_{N}$. This means that the original discussion \cite{BCS18} of the saddle point for the unscaled version of CWL still applies. We see this as follows.
	
	The saddle point equations, now written in terms of the set $\{ g_n \}$ of metric field configurations, are
	\begin{eqnarray}
		&&n\,\frac{\delta S_G[\,g_n\,]}{\delta g_n}
		+\sum\limits_{k=1}^n
		\frac{\delta S_{\phi}[\,\phi_k^{(n)},g_n\,]}{\delta g_n} \;=\; 0 \nonumber \\
		&&\frac{\delta S_{\phi}[\,\phi_k^{(n)},g_n\,]}{\delta\phi_k^{(n)}} \,-\, J \;\;=\;\;0
		\label{saddleP}
	\end{eqnarray}
	in which the rescaling factor $\alpha_N$ does not appear. We now impose the same boundary conditions on all the different paths $\phi_k^{(n)}$ of the matter field, so that we have $\phi_k^{(n)} \rightarrow \phi^{(n)}$, ie., both the matter fields and the stress energy tensors in the different saddle point equations must also be the same.  The coefficient $n$
	in (\ref{saddleP}) then cancels out, and we get the Einstein equation for each of the metric fields:
	\begin{eqnarray}
		\frac{\delta S_G[\,g_n\,]}{\delta g_n}
		+\frac{\delta S_{\phi}[\,\phi^{(n)},g_n\,]}{\delta g_n}=0,
	\end{eqnarray}
	with source field $\phi^{(n)}$. Moreover, in contrast with ref. \cite{CWL2}, since the regulators $c_{n}$ are all taken equal to $1$, all reference to the tower index $n$ disappears, and thus $g_n$ and
	$\phi^{(n)}$ satisfy the same set of equations for all $n$, ie., we have
	\begin{eqnarray}
		&&\frac{\delta S_G[\, \bar{g}_c\,]}{\delta \bar{g}_c}
		+\sum\limits_{k=1}^n
		\frac{\delta S_{\phi}[\,\phi_c,\bar{g}_c\, ]}{\delta \bar{g}_c}=0 \nonumber \\
		&&\frac{\delta S_{\phi}[\,\phi_c,\bar{g}_c\,]}{\delta\phi_c}=0  \label{saddleP-E}
	\end{eqnarray}
	in which $\phi^{(n)}=\phi_c$, and $g_n= \bar{g}_c$, the classical solutions.
	
	At first glance the fact that the classical limit turns out to be Einstein theory seems a bit surprising, given that the gravitational coupling $G_N$ has  effectively become $G_N/n$. Why doesn't the theory then have a complete decoupling between gravity and the matter fields in the large-$n$ limit? The answer, already noted at the beginning of this section, is seen explicitly in eqtn. (\ref{saddleP-E}), in the sum over $k$ in the first equation. Because of this sum, $T_{\mu\nu}$ now effectively becomes $n T_{\mu\nu}$, so that the factors of $n$ cancel between the new effective gravitational coupling and the new effective stress-energy tensor. Thus we recover the usual coupling in the Einstein equation. 
	
	At this point our next step would normally be to set up a semiclassical expansion. However this is not so easy, even in standard QFT, because of the now well-established result that semiclassical expansions are not equivalent to loop expansions \cite{donoghue04}. This result invalidates the usual association between powers of $\hbar$ and numbers of loops, even in standard QED \cite{loop-comment}. In conventional quantum gravity, where loops contribute even to low-order calculations of, eg., classical perihelion precession \cite{mercury}, this point is particularly pertinent.
	
	In section 5 we return to the classical limit of CWL theory. Using a combination of diagrammatic and exact results, we will give a complete characterization of it.

	\subsection{Diagrammar for $\mathbb{W}$}
	\label{Ssec:CWL-W-Diag}
	
	We now turn to an analysis of the physical function $\mathbb{W}[J]$. We will develop a perturbative diagrammatic calculus for $\mathbb{W}[J]$, up to the point where one can see the general structure of the diagrammatic expansion.
	
	We then find a rather startling result, viz., that in CWL theory, the contribution of loop diagrams containing gravitons is exactly zero. We will not deploy rigorous proofs here - a more formal discussion, along with the implications for the renormalizability of CWL theory, will appear in a paper devoted to this topic \cite{jordan-largeN}.
	
	This result creates apparent paradoxes, since graviton loops are normally considered to be essential in the derivation of classical GR from conventional quantum gravity. Using results derived in in section 5, we we will return to these paradoxes in section 7.

	
	\begin{figure}
		\includegraphics[width=3.2in]{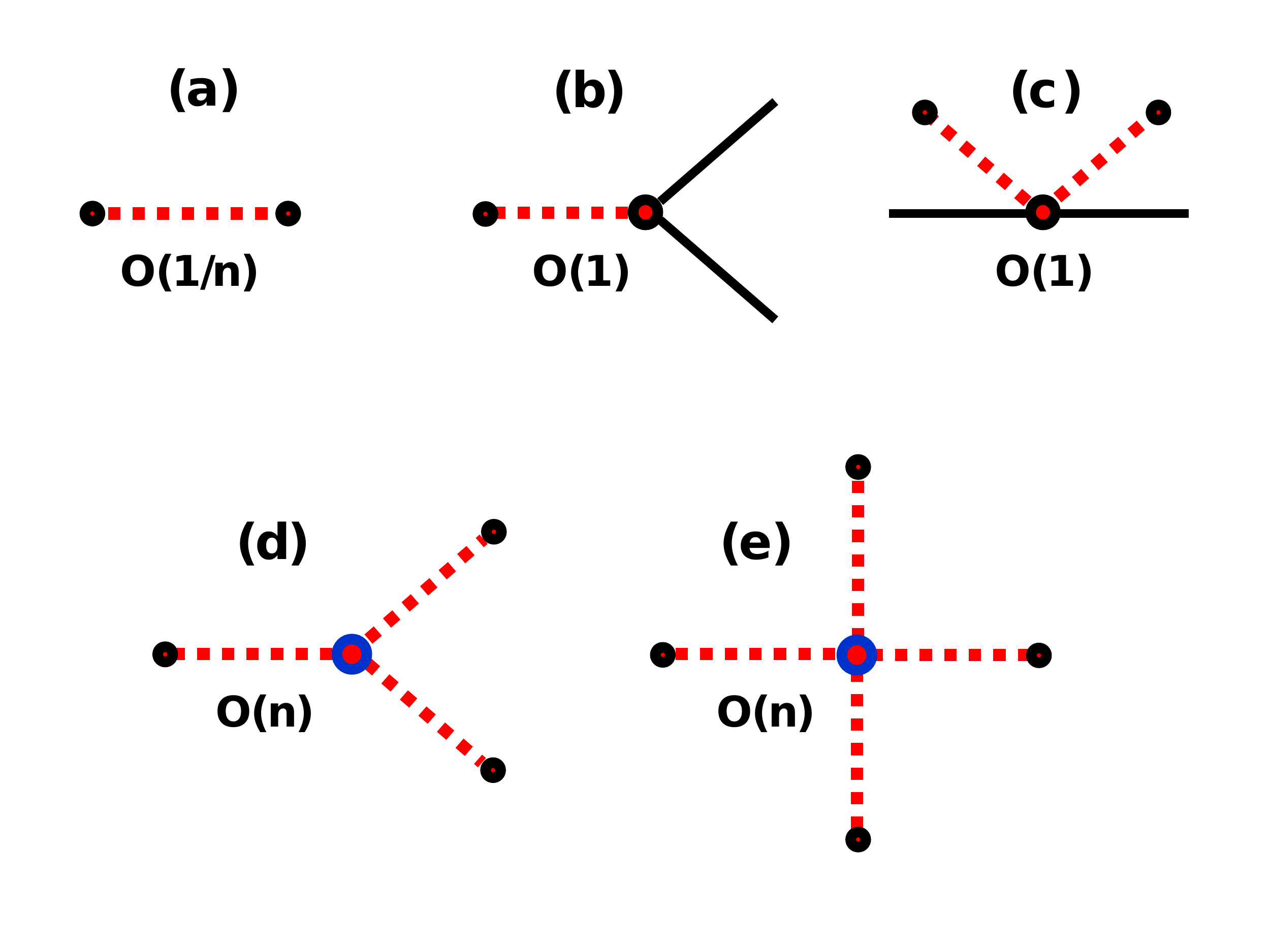}
		\caption{\label{fig:CWL-vertex5}  Order of the contribution to graphs in the $n$-th tower for different vertices. The graviton graph in (a) is $\sim O(1/n)$. In (b) and (c) we have interactions between the matter field and either one or two gravitons; these vertices are both $\sim O(1)$. In (d) and (e) we show 3-graviton and 4-graviton interaction vertices, which are both $\sim O(n)$.    }
	\end{figure}
	

	\subsubsection{Diagrammatic Rules}
	\label{Ssec:CWL-W-DiagR}
	
	To set up perturbation theory we proceed as in our discussion of conventional quantum gravity in section 2.A.2. Thus we again expand the metric about a flat background as $g_{\mu\nu}=\eta_{\mu\nu}+h_{\mu\nu}$, and expand both the Einstein and matter actions in powers of $h_{\mu\nu}$. We can then read off the diagram rules from the form of the action.  Since the matter action is independent of $n$, each of the matter-graviton vertices will be the same as conventional quantum gravity (ie., $\sim O(n^0))$; this is seen in the graph in Figs. \ref{fig:CWL-vertex5}(b) and (c).
	
	The Einstein action $S_G[g]$ appears in CWL theory multiplied by $n$, so each graviton-graviton vertex will come with a factor of $n$, and the graviton propagator (which is the inverse of the quadratic form in the action) will come with a factor $n^{-1}$. These results are illustrated in Figs. \ref{fig:CWL-vertex5}(a), (d), and (e).
	
	Now let us recall the rescaled CWL expression for the connected generating functional $\mathbb{W}[J]$, in equation (\ref{bbW-alpha}); note again that the rescaling factor $\alpha_N \propto 1/N^2$ in the limit $N \rightarrow \infty$. Again, we write 
	\begin{equation}
		W_{n}[J]=-i \hbar \log Q_{n}[J], 
		\label{Wn-J}
	\end{equation}
	and now expand this functional in a power series in $n$, as
	\begin{equation}
		W_{n}[J] \;=\; nW^{(1)}[J]+ n^0 W^{(0)}[J]+\mathcal{O}(n^{-1}),
		\label{Wn-exp}
	\end{equation}
	
	If we now substitute this into the full connected generating functional in (\ref{bbW-alpha}), we obtain
	\begin{align}
		\label{eq:freeenergy2}
		\tilde{\mathbb{W}}[J]&=\lim_{N\rightarrow\infty}\left[\alpha_{N}\sum_{n=1}^{N}\left(nW^{(1)}[J]+\mathcal{O}(n^{0})\right)\right] \nonumber \\
		& \rightarrow W^{(1)}[J]\lim_{N\rightarrow\infty}\left[1+\mathcal{O}(N^{-1})\right] \nonumber \\
		&=W^{(1)}[J].
	\end{align}
	
	Eqtn. (\ref{eq:freeenergy2}) shows that when computing the CWL connected generating functional perturbatively, we need only retain those connected diagrams at each level \textit{n} which scale linearly with $n$. All other diagrams, scaling with $n$ sub-linearly will be cancelled by $\alpha_N$, and their contribution will be identically zero.
	
	When we come to insert these vertices into graphs for $W_n$ or for ${\cal K}_n(2,1)$, it will also be clear that we must sum over the independent path (``replica") indices $\{ k \}$ in the matter lines. Thus a factor of $n$ will appear for every different sum over these indices, in any diagram for $W_n$ or for ${\cal K}_n(2,1)$.

	\subsubsection{Results for $\mathbb{W}$}
	\label{Ssec:CWL-W-diagRes}
	
\begin{figure}
		\includegraphics[width=3.2in]{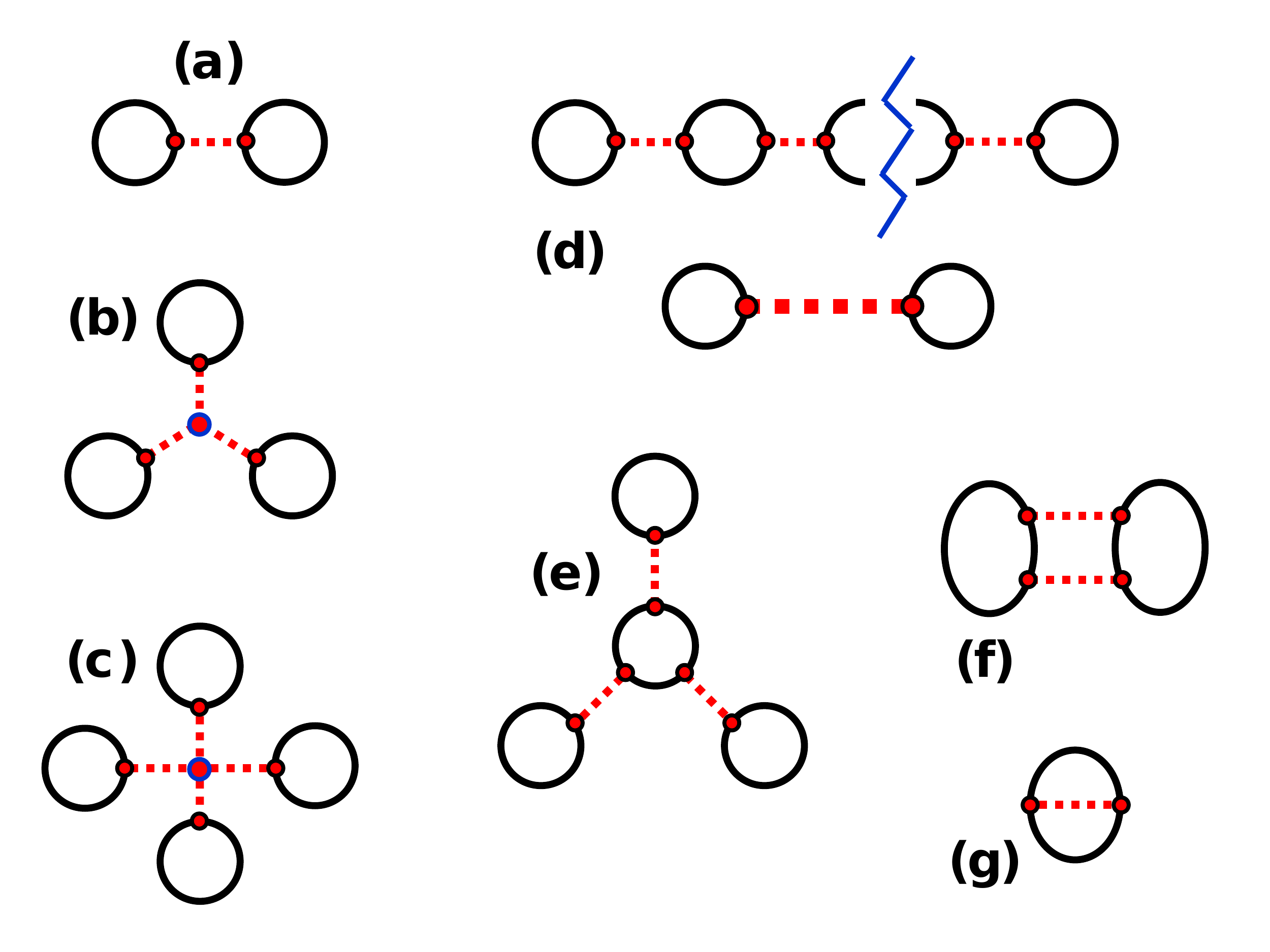}
		\caption{\label{fig:multiL-W} Graphs contributing to $W_{n}[J]$ in eqtn. (\ref{Wn-exp}). All graphs are $\sim O(n)$ except for graphs (f) and (g), which are $\sim O(1)$. The graphs in (a)-(c) are interactions between matter bubbles mediated by 2-point, 3-point, and 4-point graviton vertices respectively. Graph (d) illustrates how we can renormalize the graviton propagator by inserting an arbitrary number of bubbles into the graviton line - all these graphs are $\sim O(n)$. Graph (e) shows a renormalization of the 3-graviton vertex, also $\sim O(n)$. Finally graphs (f) and (g) have loops containing gravitons, and because they are $\sim O(1)$, they contribute nothing to $W_n[J]$.}
	\end{figure}

	To see how this works, let us now consider some typical diagrams for $\mathbb{W}$, with $J=0$. Fig. \ref{fig:multiL-W} shows some of the simpler ones. Thus, in the abbreviated DeWitt notation, Fig. \ref{fig:multiL-W}(a) can be written as $\tfrac{1}{2} G_a D^{ab} G_b$, where $G_a$ is the matter propagator, and $D^{ab}$ the graviton propagator. This diagram has sums over 2 different path replica indices $a$ and $b$ coming from the two matter loops, giving a factor $n^2$, with a factor $n^{-1}$ coming from $D^{ab}$. Thus this diagram is of order $n$. In the same way Fig. \ref{fig:multiL-W} (b) has a factor $n^3$ coming from the 3 matter loops, and a factor $n^{-3}$ from the three graviton lines; but there is also a factor $n$ from the 3-point graviton vertex, giving again an overall factor $n$.
	
	Figs. \ref{fig:multiL-W}(d) and (e) illustrate how vertices can be renormalized by the insertion of internal matter loops. Thus \ref{fig:multiL-W} (d) shows that we can renormalize the bare graviton propagator by insertion of an arbitrary number of matter bubbles; the sum of all these terms gives the full renormalized graviton propagator, since there are no other insertions that give terms $\sim O(n)$. In the same way we can insert a matter loop in place of the bare 3-graviton interaction, to give the result in Fig. \ref{fig:multiL-W}(e), which is still $\sim O(n)$.

	
	Figs. \ref{fig:multiL-W}(f) and (g) show how contributions of order lower that $\sim O(n)$ can arise. The first of these has two path replica sums, but the factor of $n^2$ is cancelled by a factor $n^{-2}$ coming from the two gravitons, so the resulting graph is $\sim O(1)$. The second has a similar problem - the single matter replica sum is cancelled by the single graviton contribution. Thus both these graphs are $\sim O(1)$, and do not contribute in the $N \rightarrow \infty$ limit.
	
	We can identify a simple underlying pattern determining the power of $n$ in each diagram. Suppose we first integrate out the matter fields, leaving us with an effective theory for the gravitons which has a new set of effective vertices. For example, the central matter bubble in Fig. \ref{fig:multiL-W}(e) would be considered as just one effective three-graviton vertex. Since the $n$ path replica's are symmetric, each such vertex is just $n$ times the result for a single matter field. Now we see that the effective diagram rules are: $n^{-1}$ for each graviton line and $n$ for each vertex, bare {\bf or} effective.
	
	With this counting, we see that any diagram with $I$ propagators and $V$ vertices must scale as $n^{V-I}$. For a connected graph every propagator comes with a 4-momentum integral, and every vertex comes with a momentum conserving delta function.  One of these delta functions conserves total momentum; the number of remaining 4-momentum loop integrals is then given by $L=I-(V-1)$. Thus a diagram with $I$ propagators and $V$ vertices is $\sim O(n^{1-L})$. Only diagrams with zero graviton loops are $\sim O(n)$ and able to contribute to $\mathcal{W}[J]$.
	
	Thus the following two simple rules apply here:
	
	\vspace{2mm}
	
	(i) If a graviton line forms any part of a closed loop in a diagram, then this is enough to kill the graph, ie., it will not contribute in the $N \rightarrow \infty$ limit. Fig. \ref{fig:multiL-W}(f) is a very simple example of this rule.
	
	\vspace{2mm}
	
	(ii) If in some graph, any matter line `self-connects' through a graviton line (ie., if a matter line with a given path/replica index interacts with itself via either a single graviton line or a sequence of graviton lines), then again this graph will not contribute in the $N \rightarrow \infty$ limit. Fig. \ref{fig:multiL-W}(g) is the simplest possible diagram illustrating this.
	
	\vspace{2mm}
	
	To summarize - only graviton tree diagrams are included in the theory, although matter loops still survive. Hence no self-interactions are allowed for paths. We see that CWL has a built-in ``large-\textit{N} limit'' which is  different from the large-\textit{N} limits considered in conventional QFT, since it refers here not to the number of matter fields but to the number of paths (recall again that one should think of CWL ``replicas" as distinct but indistinguishable paths). In this large-$N$ limit, all graviton loops are eliminated.
	
	As noted already, this seems to create two blatant paradoxes: it (a) apparently forbids obvious physical processes like gravitational self-energy or radiation-reaction effects, and (b) is in apparent contradiction with the classical limit - as already emphasized above, graviton loops in quantum gravity contribute to classical General Relativity. We discuss how to resolve these paradoxes in section 7.

	
	\section{Propagators in Correlated Worldline Theory}
	\label{sec:CWL-PropK}
	
	
	We now turn to one of the central questions of this paper, viz., the dynamics of matter fields or particles. In this section we define the CWL propagator in section 4.A,  and elucidate the structure of perturbation expansions for it, in powers of $G_N$, in section 4.B. This is done in the rescaled version of CWL theory, and we find that, just as for the correlation functions, the rescaling leads to a great simplification of the perturbative structure.

	\subsection{Propagators: Basic Definition}
	\label{Ssec:CWL-K-def}
	
\begin{figure}
		\includegraphics[width=3.2in]{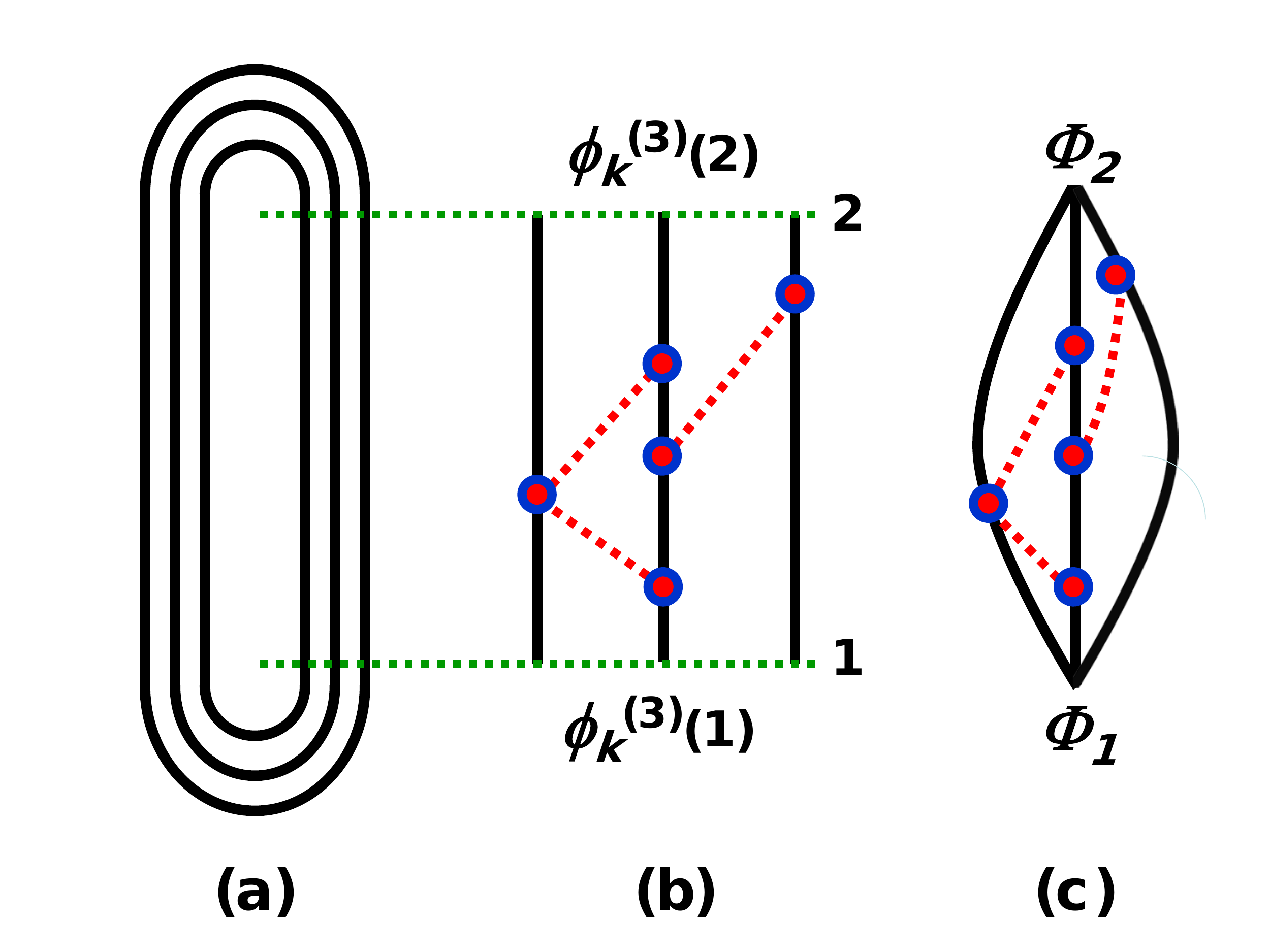}
		\caption{\label{fig:Qn-Kn}  Graphical definition of the CWL propagator for a field $\phi(x)$, starting from the generating functional $\tilde{\mathbb{Q}}$. In (a) we see a CWL graph for the $n$-th tower contribution ${\cal Q}_n$ to $\tilde{\mathbb{Q}}$, with $n=3$; only the matter lines are shown. In (b) we cut the matter lines and restore the CWL graviton interactions between the 3 different matter paths; the paths $\phi_k^{(n)}$, with $n=3$ and $k = 1, 2, 3$, terminate at states $\phi_k^{(n)}(1)$ and $\phi_k^{(n)}(2)$. Finally in (c) we tie the 3 matter lines together at the initial and final states, so that $\phi_k^{(n)}(1) \rightarrow \Phi_1$ and $\phi_k^{(n)}(2) \rightarrow \Phi_2$ for each of the $n=3$ matter lines. This gives a contribution to ${\cal K}_n(2,1)$, for $n=3$. }
	\end{figure}

	We will start from the CWL generating functional $\mathbb{Q}$, and just as was done in section 2 for conventional QFT, we define propagators using a cut procedure. To be definite, let us take a contribution to $\mathbb{Q}[J]$ from  ${\cal Q}_n$ (see eqtn. (\ref{eq:genfunc3}), and also Fig. \ref{fig:Qn-Kn}(a)). We now impose cuts on ${\cal Q}_n$, to get the situation shown in Fig. \ref{fig:Qn-Kn}(b). We take the product over $n$ later on.

	
	For this set (`tower') of $n$-path contributions we have a set of $n$ matter lines, each with different end-points. To define two specific end-point specific states $\Phi_1(x)$ and $\Phi_2(x)$ for the propagator ${\cal K}$, we must fix these states for each of the $n$ lines to be the same. Moreover, we must choose the same end states for the different towers - any different choice would make it impossible to reconcile the contributions from the different ${\cal Q}_n$.
	
	The resulting process of `tying together' the separate lines to get ${\cal K}(2,1)$ is shown in Fig. \ref{fig:Qn-Kn}(c). We denote by ${\cal K}_n$ the set of all contributions like that in Fig. \ref{fig:Qn-Kn}(c) to ${\cal K}(2,1)$, coming from $n$ matter lines - the full propagator ${\cal K}(2,1)$ will be given by a product over the ${\cal K}_n(2,1)$. We have
	\begin{eqnarray}
		{\cal K}_n(2,1)  &=& \int^2_1 \mathcal{D}g_{n}\,e^{inS_{G}[g_{n}]}
		\nonumber \\ && \qquad \times \;   \prod_{k=1}^{n} \int_{\Phi_1}^{\Phi_2}
		\mathcal{D}\phi_{k}^{(n)}\,e^{iS_{\phi}[\phi_{k}^{(n)},g_{n}]} \;\;\;\;
		\label{Kn-21}
	\end{eqnarray}

	This expression still needs to be properly normalized.  To fix this normalization we freeze the dynamics of the gravitational field to a particular configuration $\overline{g}$, so that it no longer plays any dynamic role in the theory. We then require that the propagator reduces to the conventional QFT expression for the scalar field, in the background field $\overline{g}$.
	
	Freezing the metric and carrying out the product we find
	\begin{align}
		\prod_{n=1}^{N} {\cal K}_n(2,1|\overline{g}) &= \prod_{n=1}^{N}\prod_{k=1}^{n}
		\int_{\Phi_1}^{\Phi_2} \mathcal{D}\phi_{k}^{(n)} \,e^{iS_{\phi}[\phi_{k}^{(n)},
			\overline{g}]} \nonumber \\
		&=\bigg(K_{o}(2,1|\overline{g})\bigg)^{C_N}
	\end{align}
	where $K_{\phi}(2,1|\overline{g})$ is just that function defined in eqtn. (\ref{K2-QFT-int-g}), and $C_N = \sum_{n=1}^{N}n$.
	
	If we are to match CWL propagators to conventional QFT when gravity is switched off, we must cancel the exponent $C_N$; moreover, $C_N$ is nothing but the inverse of the exponent $\alpha_N$ already introduced, ie., $C_N = \alpha_N^{-1}$. Thus, in the same way as with our treatment of $\mathbb{Q}[J]$, we must take the $\alpha_{N}\,^{th}$ root of the integral before taking the limit $N\rightarrow\infty$ limit (compare eqtn. (\ref{alphaN})). 
	
	After unfreezing the metric to restore functional integration over the metric field, we thus end up with the CWL propagator for the scalar field in the form
	\begin{widetext}
		\begin{equation}
			\label{eq:propDefn}
			{\cal K}(2,1) \; = \; \lim\limits_{N\rightarrow\infty}\;\left(\prod_{n=1}^{N} {\cal K}_n(2,1) \right)^{\alpha_N}  \;\;\; = \;\;\;\;  \lim\limits_{N\rightarrow\infty}\; \left[\prod_{n=1}^{N}\mathcal{N}_{n}^{-1}\int^2_1 \mathcal{D}g_{n}\,e^{inS_{G}[g_{n}]}  \prod_{k=1}^{n}\int_{\Phi_{1}}^{\Phi_{2}}\mathcal{D}\phi_{k}^{(n)}\,
			e^{iS_{\phi}[\phi_{k}^{(n)},\,g_{n}]}\right]^{\alpha_N}
		\end{equation}
	\end{widetext}
	
	We stress that this is so far a purely formal expression (as with all path integrals). One can alleviate the divergences somewhat by taking the logarithm of (\ref{eq:propDefn}), but we can also use it to generate perturbative expansions in $G_N$, which we do below. In the next section we will see that it can be evaluated exactly.
	
	As a check on (\ref{eq:propDefn}), we can refreeze the metric field $g^{\mu\nu}(x)$ in it to some fixed configuration $\overline{g}^{\mu\nu}(x)$;  it is clear that we will then recover the conventional QFT result, ie., we get ${\cal K}(2,1) \rightarrow K_{\phi}(2,1|\overline{g})$.
	
	It will also be clear from this derivation how to define a propagator between  initial and final position states for a particle. Thus, for a non-relativistic particle, the path integration $\int_{\Phi_{1}}^{\Phi_{2}}\mathcal{D}\phi_{k}
	^{(n)}$ for the field is replaced by $\int_{{\bf x}_{1}}^{{\bf x}_{2}}
	\mathcal{D}q_{k}^{(n)}$, where $q_k^{(n)}$ is the $k$-th path in the $n$-th 
	tower of paths, and ${\bf x}_1$ and ${\bf x}_2$ are the end points.  
	
	More generally, for both particles and matter fields, we can define propagation between two arbitrary states $|\alpha \rangle$ and $|\beta \rangle$. To do this, let's first note how one can write simple 1-particle QM in CWL language (without gravity). Recall that in ordinary QM, the propagator for a single non-relativistic particle propagating from state $|\psi_{\alpha} (t_1) \rangle \equiv |\alpha \rangle$ to state $|\psi_{\beta}(t_2) \rangle \equiv |\beta \rangle$ is
	\begin{equation}
		K_o(\beta, \alpha) \;=\; \int d^3{\bf x}_1 d^3{\bf x}_2 \; \langle \beta | {\bf x}_2 \rangle \, K_o(2,1) \, \langle {\bf x}_1 | \alpha \rangle
		\label{Kqm-ba}
	\end{equation}
	where $K_o(2,1) \equiv  K_o({\bf x}_2, {\bf x}_1; t_2, t_1)$ is just the 1-particle propagator between spatial positions ${\bf x}_1$ and ${\bf x}_2$ given in eqtn. (\ref{Kqm-21}) of section 2.
	
	To write this in CWL language one defines, for the $n$-th tower, a set of $n$ different spatial coordinates ${\bf x}_{k1}^{(n)}$ and ${\bf x}_{k2}^{(n)}$, these being the initial and final coordinates for the $k$-th particle line. We then integrate separately over each of the inner products $\psi_{\beta}({\bf x}_{k2}^{(n)}, t_2) =  \langle \beta | {\bf x}_{k2}^{(n)} \rangle$ and $\psi_{\alpha}({\bf x}_{k1}^{(n)}, t_1) = \langle {\bf x}_{k1}^{(n)} | \alpha \rangle$, for each of these $n$ lines, to get the final answer; ie., we write
	\begin{widetext}
		\begin{equation}
			K_o(\beta, \alpha) \;\;=\;\; \lim\limits_{N\rightarrow\infty}\left[\prod_{n=1}^{N}
			\left( \prod_{k=1}^n  \int d^3{\bf x}_{k1}^{(n)} \int d^3{\bf x}_{k2}^{(n)} \;\langle \beta | {\bf x}_{k2}^{(n)} \rangle \, K_o({\bf x}_{k2}^{(n)}, {\bf x}_{k1}^{(n)}; t_2, t_1) \, \langle {\bf x}_{k1}^{(n)} | \alpha \rangle \right) \right]^{\alpha_N}
			\label{Kqm-ba-CWL}
		\end{equation}
	\end{widetext}

	
	\begin{figure}
		\includegraphics[width=3.2in]{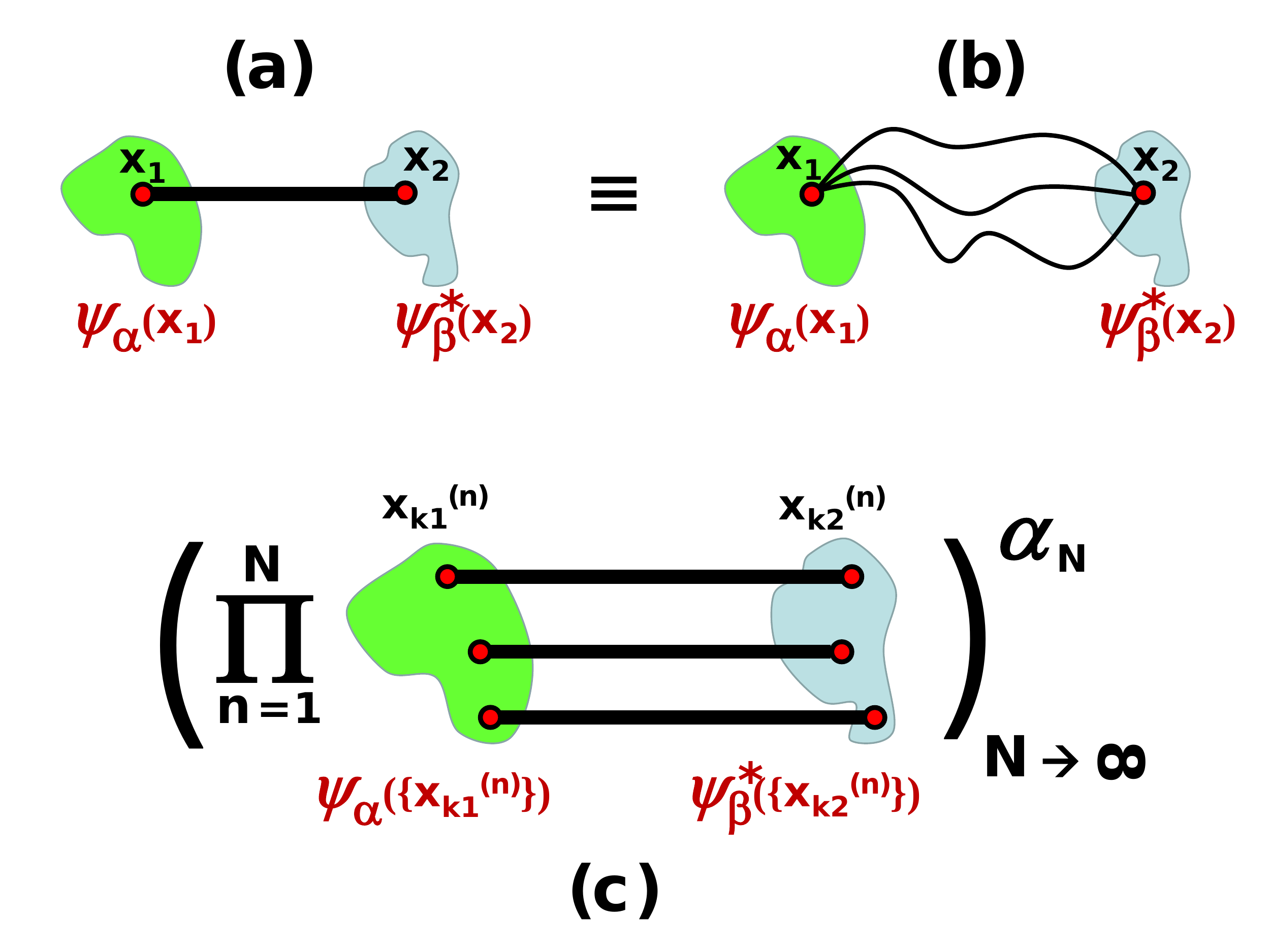}
		
		\caption{\label{fig:state-graph} Comparison two different ways of writing $K_o(\beta,\alpha)$ in CWL representation. In (a) the propagator $K_o(2,1)$ between ${\bf x}_1(t_1)$ and ${\bf x}_2(t_2)$ is shown as a heavy line on the left; this decomposes into the set of all paths (depicted as light lines) between ${\bf x}_1(t_1)$ and ${\bf x}_2(t_2)$, shown in (b) at right. The supports of the inner products $\langle {\bf x}_1|\alpha \rangle$ and $\langle \beta | {\bf x}_2 \rangle$ are shown as patches. In (c), which corresponds to eqtn. (\ref{Kqm-ba-CWL}), each different path contributing to $K_o(2,1)$ has a different set of end-points $\{ {\bf x}_{k1}^{(n)} \}$ and $\{ {\bf x}_{k2}^{(n)} \}$.
		}
		
	\end{figure}
	

	The formulas (\ref{Kqm-ba}) and (\ref{Kqm-ba-CWL}) for $K_o(\beta, \alpha)$ are of course identical (indeed, they are just the application of eqtn. (\ref{GNto0}) to the case of a non-relativistic particle).  However one can imagine two different graphical representations of this propagator, shown in Fig. \ref{fig:state-graph}.  On the one hand one collects the end points of all the paths into the same coordinates ${\bf x}_1(t_1)$ and ${\bf x}_2(t_2)$ (see Fig. \ref{fig:state-graph}(b)); whereas in the correct CWL treatment, the different paths have independent end-points (see Fig. \ref{fig:state-graph}(c)). 
	
	We see that it is important, in generalizing ordinary QM or QFT expressions for propagators to CWL theory, to keep the $2n$ end-points or end-fields in the $n$-th tower independent from each other.

	For completeness we give the complete expressions for CWL propagators for both a particle and scalar field, now including the functional integration over the metric. For ordinary particle propagation between states $|\alpha \rangle \equiv |\psi_{\alpha} \rangle$ and $|\beta \rangle \equiv |\psi_{\beta} \rangle$ we define
	\begin{eqnarray}
		&& \int_{|\alpha \rangle}^{|\beta \rangle}\mathcal{D}q_{k}^{(n)}  \;\;\; \equiv \;\;\;   \int d^3 x_{k1}^{(n)} \int d^3 x_{k1}^{(n)} \nonumber \\
		&& \qquad \qquad \qquad \qquad \times \;\langle \beta | x_{k1}^{(n)} \rangle  \langle x_{k2}^{(n)}|\alpha \rangle \;   \int_{x_{k1}^{(n)}}^{x_{k1}^{(n)}}\mathcal{D}q_{k}^{(n)} \qquad
		\label{pathI-CWL-q}
	\end{eqnarray}
	in which a set of $n$ different paths $\{ x_k^{(n)} \}$ propagates in 4-dimensional spacetime, in the $n$-th tower, between end-points $x_{k1}^{(n)}$ and $x_{k2}^{(n)}$ respectively. In the same way, for propagation between scalar field functionals $\Psi_{\alpha}$ and $\Psi_{\beta}$, we define
	\begin{eqnarray}
		&&\int_{\Psi_{\alpha}}^{\Psi_{\beta}}\mathcal{D}\phi_{k}^{(n)} \;\;\; \equiv  \;\;\;    \int {\cal D} \Phi_{k2}^{(n)} \int {\cal D} \Phi_{k1}^{(n)} \nonumber \\
		&&  \qquad \qquad \qquad \qquad \times \; \langle \beta | \Phi_{k2}^{(n)} \rangle  \langle \Phi_{k1}^{(n)}|\alpha \rangle \;   \int_{\Phi_{k1}^{(n)}}^{\Phi_{k2}^{(n)}}\mathcal{D}\phi_{k}^{(n)} \qquad
		\label{pathI-CWL-phi}
	\end{eqnarray}
	in terms of a set of ``end-fields" $\Phi_{k1}^{(n)}$ and $\Phi_{k2}^{(n)}$ for the scalar fields $\phi_k^{(n)}$ in the $n$-th tower. 
	
	The CWL propagator between states $|\alpha \rangle$ and $|\beta \rangle$, for either particle or a field, is then
	\begin{equation}
		{\cal K}(\beta, \alpha) \;=\; \lim\limits_{N\rightarrow\infty}\;\left(\prod_{n=1}^{N} {\cal K}_n(\beta,\alpha) \right)^{\alpha_N}
		\label{K-phi-ab}
	\end{equation}
	where ${\cal K}_n(\beta,\alpha)$ is produced from ${\cal K}_n(2,1)$ in (\ref{eq:propDefn}) by changing the integration limits according to either (\ref{pathI-CWL-q}) or (\ref{pathI-CWL-phi}), depending on whether we deal with a particle or a field.

	\subsection{Graphical Expansion of Propagator}
	\label{Ssec:CWL-K2}
	
	The structure of the CWL propagator ${\cal K}(\Phi_2, \Phi_1)$ is of course rather peculiar. However we can understand it better by using it to generate a perturbative expansion in $G_N$, in the same way that we did already for $\mathbb{W}$; we now outline this.

	\subsubsection{Graphical Rules}
	\label{Ssec:CWL-K2-rule}
	
	From eqtn. (\ref{eq:propDefn}), and from Fig. \ref{fig:Qn-Kn}, we see that a graphical construction of the perturbation expansion for  ${\cal K}(2,1)$ can be accomplished by 3 steps, as follows:
	
	\vspace{2mm}
	
	(i) for the contribution ${\cal K}_n$ to the propagator, draw a set of ``untethered" lines between start and end points $\phi_k^{(n)}(1)$ and $\phi_k^{(n)}(2)$ (see Fig. \ref{fig:Qn-Kn}(b)). These represent the $n$ different paths for the matter field (here a scalar field). At this point we have not yet identified the end points of the $n$ different lines.
	
	\vspace{2mm}
	
	(ii) Now insert all possible gravitational interactions between these $n$ lines. This is done in accordance with the usual Feynman rules for conventional quantum gravity, since we are working inside a specific ``tower", the $n$-th tower (ie., working with all diagrams involving $n$ paths for the $\phi$-field). Examples are shown in Fig. \ref{fig:Qn-Kn}(b).
	
	\vspace{2mm}
	
	(iii) Now tie together the end points of the $n$ untethered matter lines at their two end points, ie., let $\phi_k^{(n)}(1) \rightarrow \Phi_1$ and $\phi_k^{(n)}(2) \rightarrow \Phi_2, \forall n$. We then get graphs of the form shown in Fig. \ref{fig:Qn-Kn}(c), contributing to ${\cal K}_n(2,1)$. To get all graphs for ${\cal K}(2,1)$, we must then take the product over $n$, defined in eqtn. (\ref{eq:propDefn}).

	\vspace{2mm}
	
	This procedure again defines a set of diagrammatic rules, which we can use to represent high-order terms in a perturbation expansion. One should not think of these rules as producing conventional Feynman graphs; they do not represent the propagation of $n$ different fields, but instead correlations between $n$ paths, for a single field. Moreover, we still have to perform the product over $n$, which fundamentally changes the results, as we now see.

	
	\begin{figure}
		\includegraphics[width=3.2in]{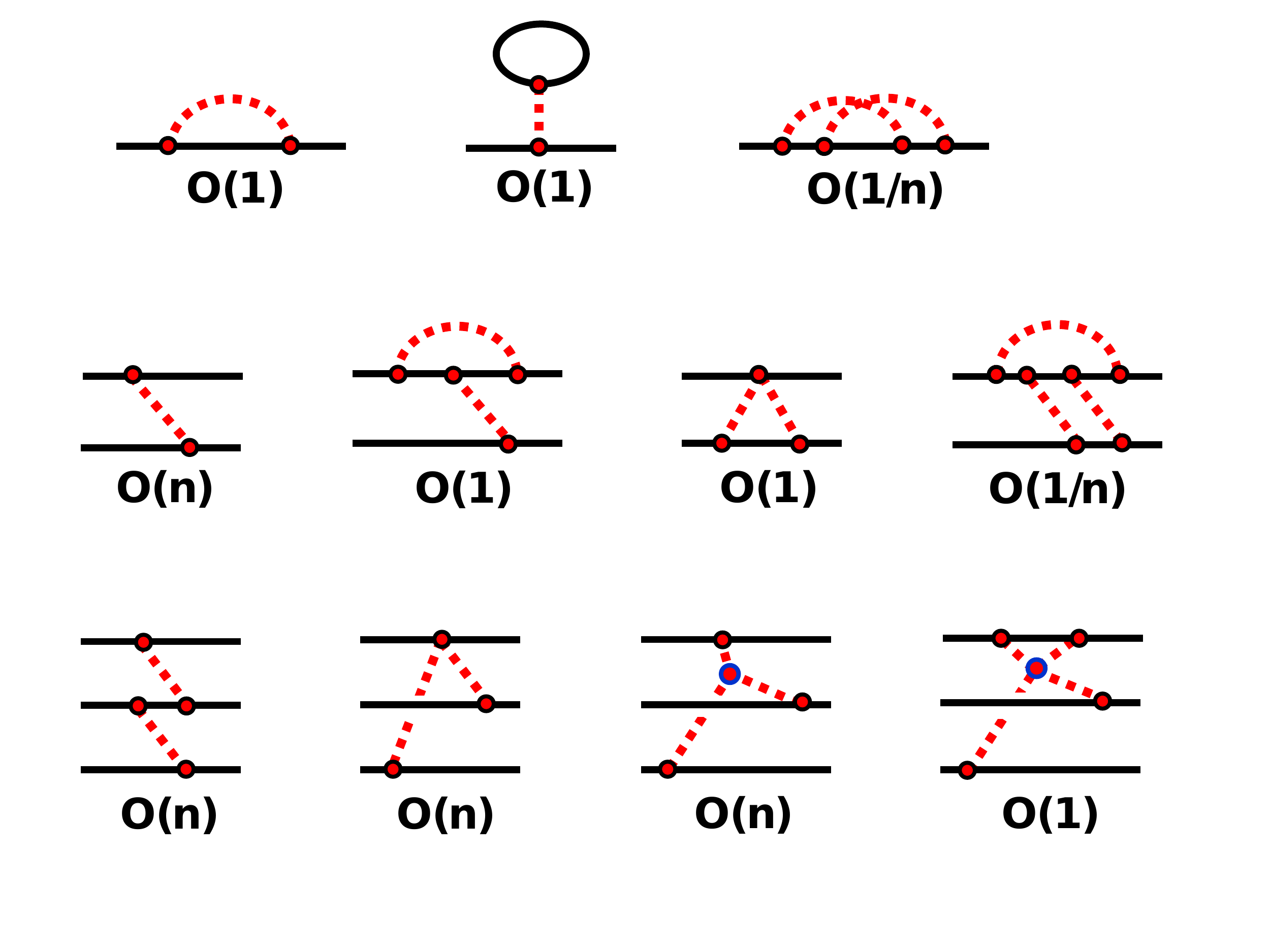}
		\caption{\label{fig:multiL-graphs} ``Untethered" graphs contributing to ${\cal K}_n(2,
			1)$ for a scalar field (compare eqtn. (\ref{eq:propDefn}) above). The top row shows contribution for a single untethered matter line; there are no contribution $\sim O(n)$. The second row shows contributions for two untethered lines; the only graph $\sim O(n)$ is the first one. The third row shows three contributions $\sim O(n)$, and one contribution $\sim O(1)$; there are many other contributions $\sim O(1), O(1/n)$, etc).   }
	\end{figure}
	

	\subsubsection{Structure of Diagrams}
	\label{Ssec:CWL-K2-struct}

	Consider Fig. \ref{fig:multiL-graphs}, which categorizes a representative sample of untethered graphs for ${\cal K}(2,1)$. Note first that none of the standard self-energy graphs, familiar from conventional quantum gravity, contribute at all to ${\cal K}(2,1)$. Three of these self-energy graphs are shown in the top line of the figure. The first of these makes clear what is happening; the contribution of this graph is killed by the graviton loop. By adding more gravitons, we simply lower the order in $n$ still further; adding matter insertions into the gravitons, or between them, does not help here. Nor do tadpole self-energy insertions help either, since they are $\sim O(1)$ (NB: such tadpoles only exist if the matter lines represent fields - they do not exist if these lines represent particle paths).
	
	In graphs with a pair of matter lines (the second row of Fig. \ref{fig:multiL-graphs}), there is only one contribution $\sim O(n)$ (this contribution, and indeed all matter lines, can be decorated with tadpoles). The other 3 graphs illustrate the same principle, that any loops containing gravitons will kill the contribution of the graph. Notice that in the 4th graph, one factor of $1/n$ comes from the graviton self-energy graph, and the other comes from the loop integration involving the two graviton lines linking the matter lines.
	
	The third row contains three graphs $\sim O(n)$, which all therefore contribute to ${\cal K}(2,1)$. They survive precisely because they contain no loops nor single matter line self-interactions - only interactions between the three different matter lines are included. The last graph in this row is $\sim O(1)$, with one graviton-containing loop. To see that this graph is $\sim O(1)$, note that $V=1$, $I=4$, and there are 3 separate `replica sums" over the 3 different matter lines; thus we get $\sim O(n^{1-4+3}) = n^0$.  
	
	If we now go to step (iii) given above, and tie together the ends of these untethered graphs to make diagrams for ${\cal K}(2,1)$, we see that these results are not changed. A systematic study \cite{jordan-largeN} of all contributions to ${\cal K}(2,1)$, incorporating an arbitrary of matter lines, shows that to all orders in $G_N$, the only graphs that survive to give a contribution to ${\cal K}(2,1)$ in the $n \rightarrow \infty$ limit are graphs with no loops involving gravitons. There is however no prohibition on matter loops in which no internal integration over gravitons appears. 
	
	We now need to understand how to interpret all of these results physically - the next 3 sections address this question.

	
	\section{Some Exact Results}
	\label{Ssec:CWL-exact}
	

	In this section we obtain some exact results. We first analyze, in sections 5.A and 5.B, the behaviour of $\mathbb{Q}[J]$, $\mathbb{W}[J]$ and ${\cal K}(2,1)$ at large $N$, ie., containing a very large number $N$ of CWL-coupled paths. Remarkably, as $N \rightarrow \infty$, so that infinitely many paths interact with one another, the leading term gives the exact result---the theory has an intrinsic ``large-$N$'' limit. Without any graphical analysis, we then find that (i) CWL theory yields Einstein's equation of motion for the metric field, with particular matrix elements of $T_{\mu\nu}$ as a source; and (ii) that the matter dynamics is quantum-mechanical, but with CWL correlations, mediated by gravity, between the matter paths. 
	
	Finally, in section 5.C, we expand about flat space, and find the form of ${\cal K}(2,1)$ in this weak field limit. This result is used in the next section to discuss 2-path experiments.

	\subsection{Large $N$ Analysis for $\mathbb{Q}[J]$}
	\label{sec:class-eikQ}
	
	Let us return to the level-$n$ generating functional ${\cal Q}_n$. All of the $n$ matter integrals are identical, and we can formally evaluate them to obtain
	\begin{align}\label{eq:integratedleveln}
		{\cal Q}_{n}[J]&=\int\mathcal{D}g\,e^{inS_{G}[g]}\bigg(e^{iW_{0}[J|g]}\bigg)^{n} \nonumber \\
		&=\int\mathcal{D}g\,e^{in(S_{G}[g]+ W_{0}[J|g])},
	\end{align}
	where we've omitted the superscript $(n)$ on the metric field $g$, since are only considering a single specific tower - the $n$-th tower. As before, $W_{0}[J|g]$ is the connected generating functional for conventional QFT on a fixed background metric $g$. Thus, for a scalar field, $W_{0}[J|g] = -i \log {\cal Z}_{\phi}[g|J]$, with ${\cal Z}_{\phi}[g|J] = \int\mathcal{D}\phi\,e^{i(S_{\phi}[\phi,g]+\int J \phi)}$ (compare eqtn. (\ref{Z-gj})).

	We can now formally evaluate eq.~\eqref{eq:integratedleveln} using the stationary-phase method. We expand the metric $g$ about a stationary point $\bar{g}_{J}$ satisfying
	\begin{equation}\label{eq:semiclassicalEE}
		\bigg(\frac{\delta S_{G}[g]}{\delta g}+\frac{\delta W_{0}[J|g]}{\delta g}\bigg)\bigg|_{g=\bar{g}_{J}}=0.
	\end{equation}
	where we emphasize that $J(x) \neq 0$ in general, so that $\bar{g}_{J}$ is different from its $J=0$ value \cite{existence-comment}.

	The quantity is $\delta_{g}W_{0}[J|g]$ related to the stress-tensor for the matter, 
	\begin{align}
		\frac{\delta W_{0}[J|g]}{\delta g^{\mu\nu}(x)}&=\;\frac{-i}{Z[J|g]}\int \mathcal{D}\phi\bigg(\frac{i \delta S_{\phi}[\phi,g]}{\delta g^{\mu\nu}(x)}\bigg)\,e^{i(S_{\phi}[\phi,g]+\int J\phi)} \nonumber \\
		&=\; -\frac{1}{2} \langle \, T_{\mu\nu}(x|g) \rangle_{J}
		\label{dWdg-T}
	\end{align}
	where $\langle \, T_{\mu\nu}[x|g] \, \rangle_{J}$ is the stress-energy at point $x$, again when there is an external current $J$ coupled to the matter system. 
	
	It is important to emphasize here that for $J\neq 0$, $\langle \, T_{\mu\nu}[x|g] \, \rangle_{J}$ is {\it not a conventional expectation value}, since $J$ generally takes on different values before and after the insertion of the stress tensor. In fact, as we will see, $\langle \, T_{\mu\nu}[x|g] \, \rangle_{J}$ is in general complex unless $J=0$.
	
	Since we also know that $\delta_{g}S_{G}$ is proportional to the Einstein tensor $G_{\mu\nu}$, according to
	\begin{equation} 
		\frac{\delta}{\delta g^{\mu\nu}(x)}S_{G}[g] \;\;=\;\;  {1 \over 16\pi G_N}\;  G_{\mu\nu}(x)
		\label{dSdg-GG}
	\end{equation}
	we then have what looks like a semiclassical form of Einstein's equation of motion, but now in the field $J(x)$, viz.,
	\begin{equation}
		\label{eq:EE}
		G_{\mu\nu}(\bar{g}_{J}(x)) \;=\; 8\pi G_{N}\langle \, T_{\mu\nu}[x|\bar{g}_{J}]\, \rangle_{J}
	\end{equation}
	where 
	\begin{equation}
		G^{\mu\nu}(\bar{g}_J(x)) \;=\; R^{\mu\nu}(\bar{g}_J(x)) - R(\bar{g}_J(x)) \; \bar{g}^{\mu\nu}_J(x)
		\label{G-mn-xJ}
	\end{equation}
	for the Einstein tensor, and $\bar{g}_{J}$ is here the solution to Einstein's equation of motion, in the presence of quantum fields that are themselves sourced by $J(x)$.
	
	We should note at this point the subtle issue of boundary data for the Einstein equation of motion. In flat spacetime QFT one avoids fixing boundary data by implementing small imaginary time rotations in the path integral, which effectively constructs a vacuum-vacuum transition amplitude. in quantum gravity, however, the validity of the Euclidean continuation is a lot less clear \cite{baldazzi}, and the ``vacuum state'' is not known in general. 
	
	To make progress here we will again assume that the defining functional integral is a representation of a perturbative series for fluctuations about a solution to the vacuum Einstein equation. For all calculations in the present paper - which is primarily concerned with weak-field scenarios, relevant to lab-based experiments - this will be assumed to be flat spacetime. The omission of boundary data in eqtn. (\ref{eq:integratedleveln}), along with an $i\epsilon$-prescription, then represents a vacuum-vacuum transition for metric fluctuations about flat spacetime. When solving eqtn. (\ref{eq:semiclassicalEE}) one should then implement past boundary conditions describing asymptotically flat spacetime devoid of incoming gravitational radiation - this is actually always done implicitly when one chooses an  $i\epsilon$-prescription \cite{donoghue2019}.
	
	Let us now write $g=\bar{g}_{J}+n^{-\frac{1}{2}}h$, and expand the effective action in powers of $h$ about the stationary-phase solution (thereby bringing out the behaviour as a function of $n$, while still leaving $h$ dimensionless). Thus we write
	\begin{widetext}
		\begin{align}
			{\cal Q}_{n}[J]=e^{in(S_{G}[\bar{g}_{J}]+W_{0}[J|\bar{g}_{J}])}\int\mathcal{D}h\,\exp\bigg[i\sum_{m=2}^{\infty}\frac{n^{1-m/2}}{m!}\frac{\delta^{m}}{\delta g^{a_{1}}...\delta g^{a_{m}}}(S_{G}[g]+W_{0}[J|g])\big|_{g=\bar{g}_{J}}\times h^{a_{1}}... h^{a_{m}}\bigg].
			\label{Q-exp}
		\end{align}
	\end{widetext}
	where as before we use the ``DeWitt'' notation for tensor indices and spacetime coordinates. We've also omitted a factor of $n$ raised to a power coming from the Jacobian of the integration variable change, because this factor will not be linear in $n$ after taking the logarithm of ${\cal Q}_{n}$.
	
	We can now see that the classical prefactor in (\ref{Q-exp}) is actually the exact result. The term quadratic in $h^{a}$ in the expansion in
	(\ref{Q-exp}) is proportional to $n^{0}$, and all higher vertices are proportional to $n$ to a negative power. We may thus write the level-$n$ generating functional as
	\begin{equation}
		\label{eq:stationaryphase}
		{\cal Q}_{n}[J]\;=\;e^{in(S_{G}[\bar{g}_{J}]+W_{0}[J|\bar{g}_{J})] \;+\; \mathcal{O}(n^{0})},
	\end{equation}
	and, referring back to eqtns. (\ref{Wn-exp}) and (\ref{eq:freeenergy2}), we conclude that the exponent in this equation is actually exact.
	
	We thus arrive at a key result. After taking the product over $n$ and letting $N \rightarrow \infty$, we see that the full CWL generating functional can be written as 
	\begin{equation}
		\label{eq:CWLgenfunc}
		\mathbb{Q}[J]\;=\; e^{i(S_{G}[\bar{g}_{J}]+W_{0}[J|\bar{g}_{J}])},
	\end{equation}
	where, again, $W_{0}[J|g]=-i\log {\cal Z}_{\phi}[J|g]$ is the conventional connected generating functional for a scalar field on a background metric $g$, and $\bar{g}_{J}$ self-consistently solves the full semi-classical Einstein equation, eqtn.~(\ref{eq:EE}). The corresponding result for $\mathbb{W}[J]$ is just
	\begin{equation}
		\label{eq:CWLgenW}
		\mathbb{W}[J] \;=\; S_{G}[\bar{g}_{J}] + W_{0}[J|\bar{g}_{J}].
	\end{equation}
	
	This result can also be written in the form
	\begin{equation}
		\mathbb{Q}[J]\;=\;e^{iS_{G}[\bar{g}_{J}]}\int \mathcal{D}\phi\,e^{i(S_{\phi}[\phi,\bar{g}_{J}]+\int J\phi)}.
	\end{equation}
	
	We see that the `path replicas' have been effectively integrated out, leaving behind a single functional integral for the matter field propagating on a metric $\bar{g}_{J}$ which is self-consistently determined from eqtn.~(\ref{eq:EE}).
	
	In the next section we will discuss the interpretation of this remarkable result. Before doing so, we turn to the propagator ${\cal K}(2,1)$.

	\subsection{Large $N$ Analysis for ${\cal K}(2,1)$}
	\label{sec:class-eikQ}

	Starting from our key result (\ref{eq:propDefn}) for ${\cal K}(2,1)$, we wish again to do an expansion about the stationary phase saddle point. We first note that in the absence of any gravitational dynamics (so that we work on a fixed background $g_{0}$), the conventional propagator for a scalar field between configurations $\Phi_1(x)$ and $\Phi_2(x)$ is just
	\begin{equation}
		K_0(\Phi_2, \Phi_1 |g_{0}) \;=\;  \int_{\Phi_1}^{\Phi_2} {\cal D}\phi \; e^{iS_{\phi}[\phi, g_{0}]},
		\label{K-phi-12}
	\end{equation}
	(compare eqtn. (\ref{K2-QFT-int-g})). We write this as
	\begin{equation}
		K_0(\Phi_2, \Phi_1 |g_{0}) \;=\; e^{i\psi_{0}(\Phi_{2},\Phi_{1}|g_{0})}
		\label{K-phi-12'}
	\end{equation}
	
	When then switch on the gravitational dynamics by integrating over the metric. The conventional propagator, now between configurations $(\Phi_1(x), \mathfrak{h}^{ab}_1)$ and $(\Phi_2(x), \mathfrak{h}^{ab}_2)$, is given precisely by eqtn. (\ref{K2-QFT-int-g}), which again we write as $K(2,1)$ in abbreviated notation. On the other hand we will write the full CWL propagator as
	\begin{equation}
		{\cal K}(2,1) \; \equiv \; {\cal K}(\Phi_2, \mathfrak{h}^{ab}_2; \Phi_1, \mathfrak{h}^{ab}_1)
		\;\; \rightarrow\;\;  e^{i\Psi(2,1)}
	\end{equation}
	where the phase $\Psi(2,1) \equiv \Psi(\Phi_2, \mathfrak{h}^{ab}_2; \Phi_1, \mathfrak{h}^{ab}_1)$ has as its arguments both the matter and metric configurations on the hypersurfaces $\Sigma_1$ and $\Sigma_2$.

	Now, in the same way as before, we expand the phase $\Psi(2,1)$  directly in terms of the tower contributions ${\cal K}_n(2,1)$ to the propagator (recall eqtn. (\ref{eq:propDefn})), as
	\begin{align}\label{eq:Psi}
		\Psi(2,1) \;=\; -i\lim_{N\rightarrow\infty}\bigg[\alpha_{N}\sum_{n=1}^{N}\log {\cal K}_{n}(2,1)\bigg],
	\end{align}
	
	Again, it suffices to have a stationary phase result for ${\cal K}_n(2,1)$. As before, we can then write the level-\textit{n} propagator in the form
	\begin{equation}
		\label{eq:levelnprop}
		{\cal K}_{n}(2,1)=\int^2_1\mathcal{D}g\,e^{in(S_{G}[g]+\psi_{0}(2,1|g))}.
	\end{equation}
	where $\int^2_1 {\cal D}g$ refers to metric propagation between $\mathfrak{h}^{ab}_1$ and  $\mathfrak{h}^{ab}_2$, and we have suppressed the Faddeev-Popov determinant in this equation.

	We can now find the exact result for ${\cal K}(2,1)$. Since $\alpha_{N} \sim  N^{-2}$, we need the log of ${\cal K}_{n}$ to give a quantity linear in $n$.  Then $\sum_{n}^{N}\log {\cal K}_{n}$ yields a factor proportional to $\sum_{n=1}^{N}n=\alpha_{N}^{-1}\sim N^{2}$. Thus in evaluating the path integral for ${\cal K}_n(2,1)$ we need only retain the part scaling as $e^{\mathcal{O}(n)}$, and we get
	\begin{equation}
		{\cal K}_{n}(2,1) \;=\; e^{in (S_{G}[\bar{g}_{21}]+\psi_{0}(2,1|\bar{g}_{21}))\;+\; \mathcal{O}(n^{0})},
		\label{Kn-eik}
	\end{equation}
	where $\bar{g}_{21}$ is the metric satisfying the conditional stationary phase requirement
	\begin{equation}
		\label{eq:semiclassprop}
		\frac{\delta}{\delta g}\bigg(S_{G}[g] +\psi_{0}(2,1|g)\bigg)\bigg|_{g=\bar{g}_{21}}=0.
	\end{equation}
	ie., it is the solution to this differential equation with the metric $\bar{g}(x)$ subject to the boundary condition that the induced metrics on $\Sigma_1$ and $\Sigma_2$ are $\mathfrak{h}^{ab}_1$ and $\mathfrak{h}^{ab}_2$.
	
	Substituting (\ref{Kn-eik}) into eqtn. (\ref{eq:Psi}), and taking the limit $N \rightarrow\infty$, we obtain 
	\begin{equation}
		\mathcal{K}(2,1)=e^{i(S_{G}[\bar{g}_{21}] +\psi_{0}(2,1|\bar{g}_{21}))},
		\label{K-eik}
	\end{equation}
	up to an overall normalization. This is our key result for the CWL propagator. We see it has the same semi-classical form as the generating functional; and again, this result is exact. 
	
	Equation (\ref{eq:semiclassprop}) plays a role analogous to (\ref{dWdg-T}) and (\ref{eq:EE}) above, but must be understood somewhat differently. Let us look first at the 2nd term; this is 
	\begin{align}
		\frac{\delta}{\delta g^{\mu\nu}(x)}\psi_{0}(2,1|g)\; &=\; -i \, \frac{\delta}{\delta g^{\mu\nu}(x)}\log K_{0}(2,1|g) \nonumber \\
		&=\; -i \,\frac{\int_1^2\mathcal{D} \phi \,e^{iS_{\phi}[\phi|g]}i\frac{\delta S_{\phi}[\phi,g]}{\delta g^{\mu\nu}(x)}}{\int_1^2\mathcal{D}\phi\,e^{iS_{\phi}[\phi,g]}} \nonumber \\
		&=\; -\tfrac{1}{2}\frac{\langle \Phi_2|T_{\mu\nu}[x|g]|\Phi_1\rangle}{\langle \Phi_2|\Phi_1\rangle}
		\label{T21-derive}
	\end{align}
	which we think of as a ``conditional stress-energy", ie., the stress energy $T_{\mu\nu}(x)$, subject to the condition that $\phi(x)$ propagates between $\Phi_1$ on $\Sigma_1$ and $\Phi_2$ on $\Sigma_2$ on a background metric $g$. It is essentially a {\it matrix element} of $T_{\mu\nu}(x)$ between the states $|\Phi_1 \rangle$ and $|\Phi_2 \rangle$.
	
	Henceforth we will write this quantity as
	\begin{equation}
		\frac{\langle \Phi_2|T_{\mu\nu}[x|g]|\Phi_1\rangle}{\langle \Phi_2|\Phi_1\rangle} \;\; \equiv \;\; \mathbb{\chi}^{\mathbb{T}}_{\mu\nu}(2,1|x,g)
		\label{2-Tmn-1}
	\end{equation}
	It is clear from its definition that in general it is not real but {\it complex}.

	Consider now the 1st term in (\ref{eq:semiclassprop}).  Using (\ref{dSdg-GG}) above, we then have
	\begin{equation}
		G_{\mu\nu}(\bar{g}_{21}(x))\;\;=\;\; 8\pi G_N \, \mathbb{\chi}^{\mathbb{T}}_{\mu\nu}(2,1|x,\bar{g}_{21}) 
		\label{KG21-KT21}
	\end{equation}
	This equation is completely analogous to the Einstein equation of motion (\ref{eq:EE}), however, since $\mathbb{\chi}^{\mathbb{T}}_{\mu\nu}(2,1|x,g)$ is generally complex, so too is $G_{\mu\nu}(\bar{g}_{21}(x))$.

	The solution of this equation yields $\bar{g}_{21}$. It is obviously very non-linear, with the usual classical non-linearity already inherent in the Einstein tensor, plus the further non-linearity introduced by the back-reaction of the quantum matter. We study the weak-field limit in the next sub-section. 
	
	As in the previous section, we can write this result slightly more explicitly as
	\begin{equation}\label{eq:K-eik2}
		\mathcal{K}(2,1)=e^{iS_{G}[\bar{g}_{21}]}\,\int_{\Phi_{1}}^{\Phi_{2}}\mathcal{D}\phi\,e^{iS_{\phi}[\phi,\bar{g}_{21}]}.
	\end{equation}
	Again one finds an effective theory in terms of a single set of paths for the matter field, wherein the matter propagates on a background metric which is solved for self-consistently from eqtn. (\ref{KG21-KT21}).
	
	To conclude: we see that both the connected generating functional $\mathbb{W}[J]$ and the propagator ${\cal K}(2,1)$ are given exactly by the ``semiclassical'' results in (\ref{eq:CWLgenfunc}) and (\ref{K-eik}, \ref{eq:K-eik2}) respectively. Clearly one can derive similar results for other field theoretical quantities in CWL theory.

	\vspace{2pt}

	\subsection{Form of the weak-gravity CWL propagator}
	\label{sec:weakG-K}

	We begin from our non-perturbative result (\ref{K-eik}) for ${\cal K}(2,1)$, in which the metric $\bar{g}_{21}$ satisfies eqtn. (\ref{KG21-KT21}). We wish to perform a weak-field analysis, writing $(\bar{g}_{21})_{\mu\nu}=\eta_{\mu\nu}+h_{\mu\nu}$, where $\eta_{\mu\nu}$ represents flat spacetime and $|h_{\mu\nu}|$ is small (we assume that we can ignore or otherwise subtract off the effect of other fields coming from the rest of the apparatus, the lab, etc.). We will see that, even in weak field, both $G_{\mu\nu}(\bar{g}_{21}(x))$ and  $\mathbb{\chi}^{\mathbb{T}}_{\mu\nu}(2,1|x)$ have imaginary parts.
	
	Since the flat spacetime metric is a solution to the vacuum Einstein equation and has vanishing action, we immediately have that
	\begin{equation}
		S_{G}[\eta]=\frac{\delta}{\delta g_{\mu\nu}(x)}S_{G}[g]\bigg|_{g=\eta}=0
		\label{saddleP}
	\end{equation}
	
	In this section we will expand out the compact DeWitt notation, to be explicit about spacetime indices, coordinates, and integrations. We will also our assume our system to be a particle with coordinate $q$ propagating between spacetime points $x_1$ and $x_2$; in section 7 we will briefly discuss the case of a real mass having finite spatial extent. We also assume that the particle is propagating between two time slices $x^{0}=t_{1}$ and $x^{0}=t_{2}$.
	
	We begin by expanding, in powers of $h_{\mu\nu}$, the total phase $\Psi(2,1) = S_{G}[\bar{g}_{21}]+\psi_{0}(x_{2},x_{1}|\bar{g}_{21})$ which appears in the exponent of ${\cal K}(2,1)$. We will then  insert the solution to get the linearized version of the propagator Einstein equation. Expanding the phase argument in ${\cal K}(2,1)$ we have
	\begin{widetext}
		\begin{align}
			\label{eq:perturbative-expandedphase}
			\mathcal{K}(x_{2},x_{1})\;\;=\;\;& e^{iS_{G}[\eta]+i\psi_{0}(x_{2},x_{1}|\eta)} \; \exp\Bigg[i\int_1^2 d^{4}y\frac{\delta}{\delta g_{\mu\nu}(y)}\bigg(S_{G}[g]+\psi_{0}[g]\bigg)\bigg|_{g=\eta} h_{\mu\nu}(y)\Bigg] \nonumber \\
			& \qquad \times\exp\Bigg[\frac{i}{2}\int_1^2 d^{4}y\int_1^2 d^{4}y' \frac{\delta}{\delta g_{\mu\nu}(y)}\frac{\delta}{\delta g_{\sigma\rho}(y')}\bigg(S_{G}[g]+\psi_{0}[g]\bigg)\bigg|_{g=\eta} h_{\mu\nu}(y) \, h_{\sigma\rho}(y')\Bigg]\times\exp\bigg[\mathcal{O}(h^{3})\bigg]
		\end{align}
		where we are integrating over the spacetime region bounded by the time slices $y^{0}=t_{1}$ and $y^{0}=t_{2}$.
		
		This expression can be simplified considerably.  First, we use (\ref{saddleP}) to eliminate several terms. Then, from the linearized Einstein equation, it will be obvious that $\frac{\delta}{\delta g}\psi_{0}[g]\big|_{g=\eta}=\mathcal{O}(h)$, so that can drop the matter term in the second line of \ref{eq:perturbative-expandedphase}, as it gives a result $\sim \mathcal{O}(h^{3})$. The resulting CWL propagator for a system with weak gravitational fields is then
		\begin{align}
			\mathcal{K}(x_{2},x_{1}) \;\;=\;\; & e^{i\psi_{0}(x_{2},x_{1}|\eta)} \exp\Bigg[i\int_1^2 d^{4}y\frac{\delta \psi_{0}[g]}{\delta g_{\mu\nu}(y)}\bigg|_{g=\eta}\times h_{\mu\nu}(y)\Bigg] \nonumber \\
			& \qquad \times\exp\Bigg[\frac{i}{2}\int_1^2 d^{4}y\int_1^2 d^{4}y' \frac{\delta^{2}S_{G}[g]}{\delta g_{\mu\nu}(y)\delta g_{\sigma\rho}(y')}\bigg|_{g=\eta}  h_{\mu\nu}(y) \; h_{\sigma\rho}(y')\Bigg]  \; \exp\bigg[\mathcal{O}(h^{3})\bigg]
			\label{K21-h2}
		\end{align}
	\end{widetext}
	where the prefactor in this expression is just the flat spacetime propagator for the particle in the absence of gravity (compare eqtn. (\ref{K-phi-12'}), ie., 
	\begin{equation}
		e^{i\psi_{0}(x_{2},x_{1}|\eta)} \; = \;  K_{0}(2,1|\eta) \; \equiv \; K_{0}(2,1)
		\label{psi-Ko}
	\end{equation}

	The expression (\ref{K21-h2}) simplifies one step further if one inserts into it a formal expression for the linearized semiclassical Einstein equation, which we write as
	\begin{equation}
		\Bigg(\int d^{4}y\frac{\delta^{2}S_{G}[g]}{\delta g_{\mu\nu}(x)\delta g_{\sigma\rho}(y)}h_{\sigma\rho}(y)+\frac{\delta \psi_{0}[g]}{\delta g_{\mu\nu}(x)}\Bigg)\Bigg|_{g=\eta}=0,
	\end{equation}
	
	This then gives the required result for the propagator ${\cal K}(2,1)$ of the particle in CWL theory, in this linearized approximation, as
	\begin{equation}
		\label{eq:perturbative-CWLpropagator}
		\mathcal{K}(2,1)  =  K_{0}(2,1)\; e^{i \Theta_{21}}
	\end{equation}
	where the linearized phase $\Theta_{21}$ is
	\begin{equation}  
		\Theta_{21} \;=\;  \frac{1}{2}\int_1^2 d^{4}y\frac{\delta \psi_{0}[g]}{\delta g_{\mu\nu}(y)}\bigg|_{g=\eta} h_{\mu\nu}(y)
		\label{chi-21}
	\end{equation}
	and we have dropped terms $\sim \mathcal{O}(h^{3})$ in this phase. 
	
	All that remains is to explicitly solve the linearized semiclassical Einstein equation. This calculation is standard in classical gravity \cite{MTW73}; the quantum discussion here assumes a Faddeev-Popov gauge-fixing procedure \cite{FP67}, and we will fix the gauge here to be harmonic, so that
	the linearized Einstein tensor is 
	\begin{equation}
		G^{(1)}_{\mu\nu}(\eta+h)=\frac{1}{2}\partial^{2}\bar{h}^{\mu\nu}.
	\end{equation}
	with $\bar{h}_{\mu\nu}=h_{\mu\nu}-\frac{1}{2}\eta_{\mu\nu}h$. Notice that strictly speaking the field $\bar{h}_{\mu\nu}(x)$ also depends on the endpoints $x_1$ and $x_2$ in  ${\cal K}(2,1)$ and $\mathbb{\chi}^{\mathbb{T}}_{\mu\nu}(2,1|x)$. To avoid clutter we suppress the indices $1,2$ in $\bar{h}_{\mu\nu}(x)$. 
	
	Linearizing the matter side of the Einstein equation fixes the source as equal to the flat-spacetime stress tensor, so that (\ref{KG21-KT21}) becomes
	\begin{equation}
		\label{eq:perturbative-linearizedEinstein}
		\partial^{2}\bar{h}_{\mu\nu}(x) \;=\; 16\pi G_N \, \mathbb{\chi}^{\mathbb{T}}_{\mu\nu}(2,1|x)
	\end{equation}
	where $\mathbb{\chi}^{\mathbb{T}}_{\mu\nu}(2,1|x)$ is given for a particle by 
	\begin{equation}
		\label{eq:perturbative-stressenergy}
		\mathbb{\chi}^{\mathbb{T}}_{\mu\nu}(2,1|x)   \;=\;\frac{\int_{x_{1}}^{x_{2}}\mathcal{D}q\,e^{iS[q]}T_{\mu\nu}(x)}{\int_{x_{1}}^{x_{2}}\mathcal{D}q\,e^{iS[q]}}.
	\end{equation}

	Inverting the differential operator in (\ref{eq:perturbative-linearizedEinstein}), we get the retarded flat spacetime Green's function for $h_{\mu\nu}(x)$ as
	\begin{equation}
		{\cal G}_o(x,y) \;=\; \frac{\delta\big((x^{0}-y^{0})-|\vec{x}-\vec{y}|\big)}{|\vec{x}-\vec{y}|}
		\label{Do-xy}
	\end{equation}
	yielding the solution
	\begin{equation}
		h_{\mu\nu}(x) \;=\; -4 G_N \int d^{4}y\, {\cal G}_o(x,y) \; \mathbb{\chi}^{\mathbb{T}}_{\mu\nu}(2,1|x)  
	\end{equation}
	
	Inserting this solution into (\ref{eq:perturbative-CWLpropagator}), and using (\ref{chi-21}), we obtain the final expression for the weak field CWL propagator in the form of eqtn. (\ref{eq:perturbative-CWLpropagator}), with the phase $\Theta_{21}$ given by 
	\begin{align}
		\Theta_{21} \;=\;& 
		G_N \int d^{4}y\int d^{4}y' \nonumber \\
		&  \qquad \times  \mathbb{\chi}^{\mathbb{T}}_{\mu\nu}(2,1|y) \, {\cal G}_o(y,y') \, 
		\mathbb{\chi}^{\mathbb{T}}_{\mu\nu}(2,1|y') 
		\label{theta-21}
	\end{align}
	
	This expression is valid for any particle trajectory.  For a slow-moving particle (as for any lab experiment involving massive objects) we go to the non-relativistic limit. Then $T_{00}$ dominates $T_{\mu\nu}$, and it moreover is not changing appreciably on relativistic time scales. We can then simplify the phase to
	\begin{align}
		\label{eq:perturbative-nonrelprop}
		\Theta_{21} \;\rightarrow \;\;\; & 
		\frac{1}{2}G_N \int_{t_{1}}^{t_{2}} dt\int d^{3}r \, d^{3}r' \frac{1}{|{\bf r}(t) - {\bf r'}(t)|} \nonumber \\
		& \qquad \times \mathbb{\chi}^{\mathbb{T}}_{00}(2,1|{\bf r}, t) 
		\, \mathbb{\chi}^{\mathbb{T}}_{00}(2,1|{\bf r'}, t) 
	\end{align}
	involving simple 3-space integrations over ${\bf r}$ and ${\bf r'}$, along with integration between the 2 time slices. 
	
	Now, in spite of appearances to the contrary, the expressions in (\ref{theta-21}) and (\ref{eq:perturbative-nonrelprop}) are {\bf not the same} as the standard result for the lowest order self-energy correction to the propagator in quantum gravity. This is because these expressions are written in terms of $\mathbb{\chi}^{\mathbb{T}}_{\mu\nu}(2,1|y)$ rather than $T_{\mu\nu}(y)$. This will become very clear in the next section.

	To compute (\ref{eq:perturbative-nonrelprop}), we simply need to compute the two standard quantum mechanics quantities
	\begin{equation}
		\int_{x_{1}}^{x_{2}}\mathcal{D}q\,e^{iS[q]}\hspace{10pt}\textrm{and}\hspace{10pt}\int_{x_{1}}^{x_{2}}\mathcal{D}q\,e^{iS[q]}T_{00}(x|q),
	\end{equation}
	taken along the path $q$ followed by the system, and then assemble the results to get $\mathcal{K}(x_{2},x_{1})$. We do this in the next section for the 2-path system. 
	
	Before proceeding to a specific application of these results, let us first comment on the limitations of this weak-field linearized approximation. We notice that the source $\chi^{\mathbb{T}}$ in eqtn. (\ref{eq:perturbative-stressenergy}) no longer depends on the dynamical metric. We can contrast this with the full source in eqtns. (\ref{T21-derive}, \ref{KG21-KT21}), where the matter propagates on a metric which is solved for self-consistently; in the linearized approximation, the matter path-integrals are instead evaluated in flat spacetime. 
	
	This is of course completely analogous to the situation in classical gravity when one linearizes Einstein's equation; truncating the expansion to linear order will cause the matter to source a gravitational field, but it will not respond to this field. As in classical linearized gravity, the linearized approximation discussed here will fail when it is no longer consistent to ignore the back-reaction of the gravitational field onto the propagating matter. We will discuss this further when interpreting the 2-path results in the next section.

	
	\section{2-path experiment}
	\label{sec:2-path}
	
\begin{figure}
		\includegraphics[width=3.2in]{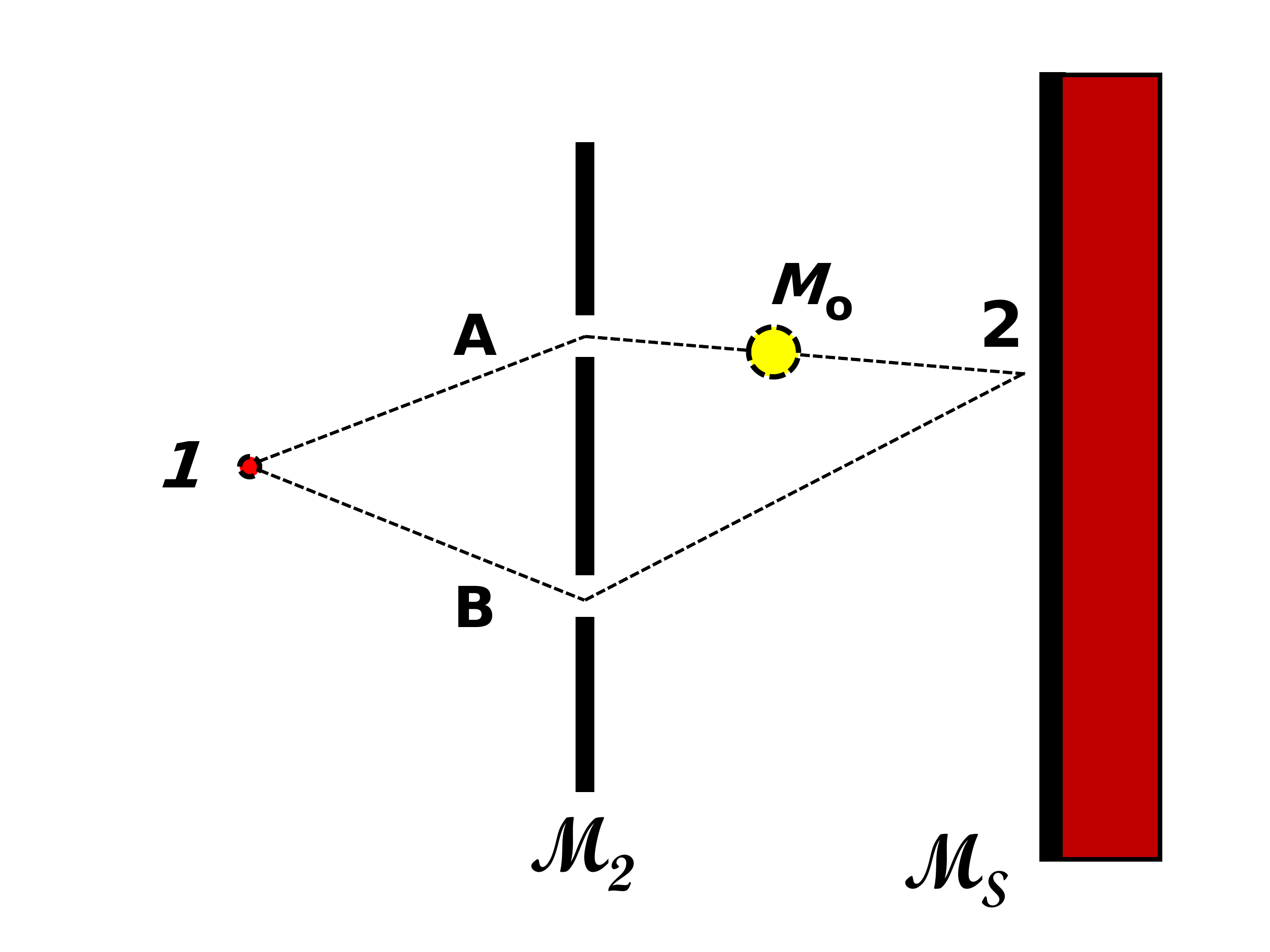}
		
		\caption{\label{fig:2slit}  A schematic 2-slit experiment. A mass $M_o$, beginning from point 1, passes through a 2-slit system ${\cal M}_2$, and is then incident on a screen ${\cal M}_S$ at point 2. The 2 paths are labelled by A and B. One can also introduce a test mass which interacts gravitationally with $M_o$ (see text for details).}
	\end{figure}

	
	The standard 2-path set-up is shown in Fig. \ref{fig:2slit}. This is the same thought experiment as that considered by Feynman \cite{feynmanGR} and Kibble \cite{kibble2} in their original discussions of low-energy quantum gravity. In this low-energy context, the 2-path set-up has been discussed repeatedly over the years \cite{feynmanGR,kibble2,page81,unruh84,baym09,mari17,belenchia18}; and analogous real experiments have been the topic of much discussion \cite{QGrav-Expt}. 
	
	A proper treatment of the 2-path system includes the dynamics of the slit system ${\cal M}_2$, itself of mass $M_2$, along with the screen system ${\cal M}_S$ with mass $M_S$, to get the correct coupling of all the masses to $g^{\mu\nu}(x)$. We ignore these details here; it is easy to calculate their effects for a specific geometry, at least in weak field gravity \cite{colbyPRL,colbyBMV}.
	
	One can also perform ``which path" measurements on this system, designed to probe the position of $M_o$. To do this one can, eg., introduce a `test mass' $\bar{m}$ (as in a Cavendish experiment) to monitor the position of $M_o$. Any real measurement is of course more likely to use optical probes to determine the path followed by $M_o$.

	In Fig. \ref{fig:2slit} the {\it single} path shown passing through a given slit actually represents the set of {\it all} paths for the mass leaving point 1, and then passing through this slit on its way to point 2. One can unambiguously separate the 2 different classes of path by introducing surfaces of ``final crossing'' of the paths \cite{auerbach84}. The set of all paths labeled by A is then defined as the set of all paths originating at the source 1 and terminating on ${\cal M}_S$ at point 2, whose last passage through the slit system ${\cal M}_2$ is through slit A.
	
	In what follows we assume the paths contributing to the propagator are `channeled' by a 2-path potential, and so cluster very strongly about the 2 relevant paths - this will be the case anyway, even in conventional QM, for large masses \cite{baym09}. This obviates the need to employ the formal techniques described in ref. \cite{auerbach84}.

	
	To actually do a 2-path experiment is very difficult for a large mass, because of the strong environmental decoherence effects then acting on $M_o$ (the largest mass for which 2-path experiments have been done so far \cite{arndt19} is $\sim 34,000~D$, where $1~D \equiv 1~$Dalton is the atomic mass unit). However here we will only be interested in what theory predicts in the absence of environmental decoherence.
	
	In this section we do three things. First, in section 6.A, we calculate the propagator for the 2-path system in conventional quantum gravity. In section 6.B we find the result for CWL theory; and finally, in section 6.C, we see what is predicted by semiclassical gravity. In all 3 cases we work in the weak-field regime, ie., where the gravitational field sourced by $M_o$ is weak (which will be the case in any experiment). The differences between the 3 results are very illuminating.

	
	\subsection{Propagator in Conventional Quantum Gravity}
	\label{sec:2-path}
	
	
	In a conventional QM analysis of the 2-path system, one has the choice between evaluating the propagator $K(2,1)$ for the system alone, or including the test mass $\bar{m}$ as well (or some other measurement system in its place). In this latter case, one can either:
	
	(i) treat the test mass $\bar{m}$ as a quantum system, so that its coordinate entangles with the position of $M_o$, putting the pair of systems in a state which we can write schematically as 
	\begin{equation}
		\Psi = \tfrac{1}{\sqrt{\alpha^2 + \beta^2}} ( \alpha |A_{M_o} A_{\bar{m}} \rangle + \beta |B_{M_o} B_{\bar{m}} \rangle)
		\label{2path-QM} 
	\end{equation}
	or alternatively
	
	(ii) treat the test mass $\bar{m}$ as a classical system. In this case, in standard QM the test mass acts as a measuring device, and over some time period the coordinate state of the mass $M_o$ is supposed to `collapse' onto one or other of the paths A or B.  
	
	As noted in the introduction, discussion of the measurement process in CWL theory is rather lengthy. Thus in what follows we will largely ignore the measurement apparatus, and simply calculate the propagator.

	\subsubsection{Long-wavelength Calculation}
	\label{sec:longWave1}
	
	In the 2-path system, as just discussed, we assume that the paths for the particle cluster around one or other of 2 paths A and B. There are thus two `semiclassical' paths $q^{(\alpha)}$, with $\alpha = \{ A,B \}$ labelling these paths; and there will be fluctuations around these paths which we will assume small. 
	
	If we completely ignore all gravitational fields, the `bare' flat space propagator $K_0(2,1)^{(\alpha)}$ (cf. eqtn. (\ref{psi-Ko}) along each of these paths can be written
	\begin{equation} 
		K_0^{(\alpha)}(2,1) \;=\; \Omega_o^{(\alpha)} \, e^{i S^0_{21}[q^{(\alpha)} |\eta]}
		\label{Ko-alpha}
	\end{equation}
	where the prefactors $\Omega_o^{(\alpha)}$ are van Vleck fluctuation determinants representing fluctuations around the paths $q^{(\alpha)}$. Then QM predicts that
	\begin{equation}
		K_0(2,1) \;=\; \sum_{\alpha}^{A,B} \Omega_o^{(\alpha)} \, e^{i S^0_{21}[q^{(\alpha)} |\eta]}
		\label{2slit-QM}
	\end{equation}
	
	This expression can be simplified if we suppose that the small oscillation frequencies in (\ref{Ko-alpha}) are the same for each path, ie., that $\Omega_o^{(\alpha)}\rightarrow \Omega_o$. Defining sum and difference actions as
	\begin{align}
		\bar{S}^0_{21} \;\equiv\;  &\tfrac{1}{2}\Big(S^0_{21}[q^{(A)}]+S^0_{21}[q^{(B)}]\Big) \nonumber \\
		\Delta S_{21} \; \equiv\;  &\tfrac{1}{2} \Big( S^0_{21}[q^{(A)}]-S^0_{21}[q^{(B)}] \Big)
		\label{sumdiff}
	\end{align}
	we have
	\begin{equation}
		K_0(2,1) \;=\; 2 \, \Omega_o \, e^{i \bar{S}^0_{21}} \, \cos (\Delta S_{21})
		\label{Ko-cos}
	\end{equation}
	
	In what follows we will often assume this simplification, which  would be fairly accurately obeyed in many 2-path experiments.

	We assume that the field deviation $h^{\mu\nu}(x)$ is small (we will return to this assumption below). We also assume that $h^{\mu\nu}(x)$ will vary slowly, on a spatial scale of order the 2-path system size, and a timescale comparable to the system traversal time for the particle. These scales are $\gg$ than the wavelength and inverse frequency of the particle, even for a microscopic particle like an electron, unless it is moving at very low velocity. For a more massive system the difference is huge \cite{baym09}. 
	
	This suggests we use a long wavelength eikonal approximation to represent the QM weak field propagator $K_{0}^{(\alpha)}(2,1|h)$ along the paths $q^{(\alpha)}$. For low energies, this can be done using standard methods developed for both relativistic and non-relativistic systems \cite{fradkin,fried,khvesh}. For the leading order we write
	\begin{equation}
		\label{eq:applications-eikonalprop}
		K_{0}^{(\alpha)}(2,1|h) \;\approx \; e^{-\frac{i}{2}\int_1^2 d^{4}x \, h^{\mu\nu}(x)\,  T_{\mu\nu}^{(\alpha)}(x)} \; K_{0}^{(\alpha)}(2,1)
	\end{equation}
	where $K_0^{(\alpha)}(2,1)$ is given by (\ref{Ko-alpha}) above, and  $T_{\mu\nu}^{(\alpha)}(x|q^{(\alpha)})$ is just the stress tensor at point $x$ when the particle follows the $\alpha$-th path $q^{(\alpha)}$ between the endpoints, ie., 
	\begin{equation}
		T^{(\alpha)}_{\mu\nu}(x|q^{(\alpha)}) \;=\;  M_o \int ds \, u^{(\alpha)}_{\mu}(s) u^{(\alpha)}_{\nu}(s)\, \delta^{(4)}(x - q^{(\alpha)}(s) ) \qquad
		\label{eq:Stress}
	\end{equation}
	in which $u^{(\alpha)}_{\mu}(s) \equiv dq^{(\alpha)}_{\mu}(s)/ds$ is the 4-velocity for the particle.
	
	If this lowest-order eikonal form is accurate, we can then write the weak field 2-path QM propagator as
	\begin{equation}
		K_{0}(2,1|h) \;=\; \sum_{\alpha}^{A,B} e^{-i\int_1^2 d^{4}x \, h^{\mu\nu}(x)\,  T_{\mu\nu}^{(\alpha)}(x)} \, K_{0}^{(\alpha)}(2,1)
		\label{K-eik-AB}
	\end{equation}
	using the shorthand $T_{\mu\nu}^{(\alpha)} \equiv T^{(\alpha)}_{\mu\nu}(q^{(\alpha)}|x)$. 
	
	To calculate $K_0(2,1)$ in conventional quantum gravity, we then integrate over the fluctuation field $h^{\mu\nu}(x)$. Since these fluctuations are small, we can simply write
	\begin{equation}
		K(2,1) \;=\; \int {\cal D}h e^{{1 \over 2} \int d^4x \partial_{\mu}h \partial^{\mu}h } K_0(2,1|h)
		\label{Kh-K}
	\end{equation}
	with $K_0(2,1|h)$ taking the 2-path form just given in eqtn. (\ref{K-eik-AB}).

	\subsubsection{Result for Propagator}
	\label{sec:K-CWL-2p1}
	
	The path integration over $h^{\mu\nu}(x)$ in (\ref{Kh-K}) is independent of the sum over the pair of paths in {\ref{K-eik-AB}). Carrying out the functional integration, and defining the flat-space graviton propagator as

		
		\begin{figure}
			\includegraphics[width=3.2in]{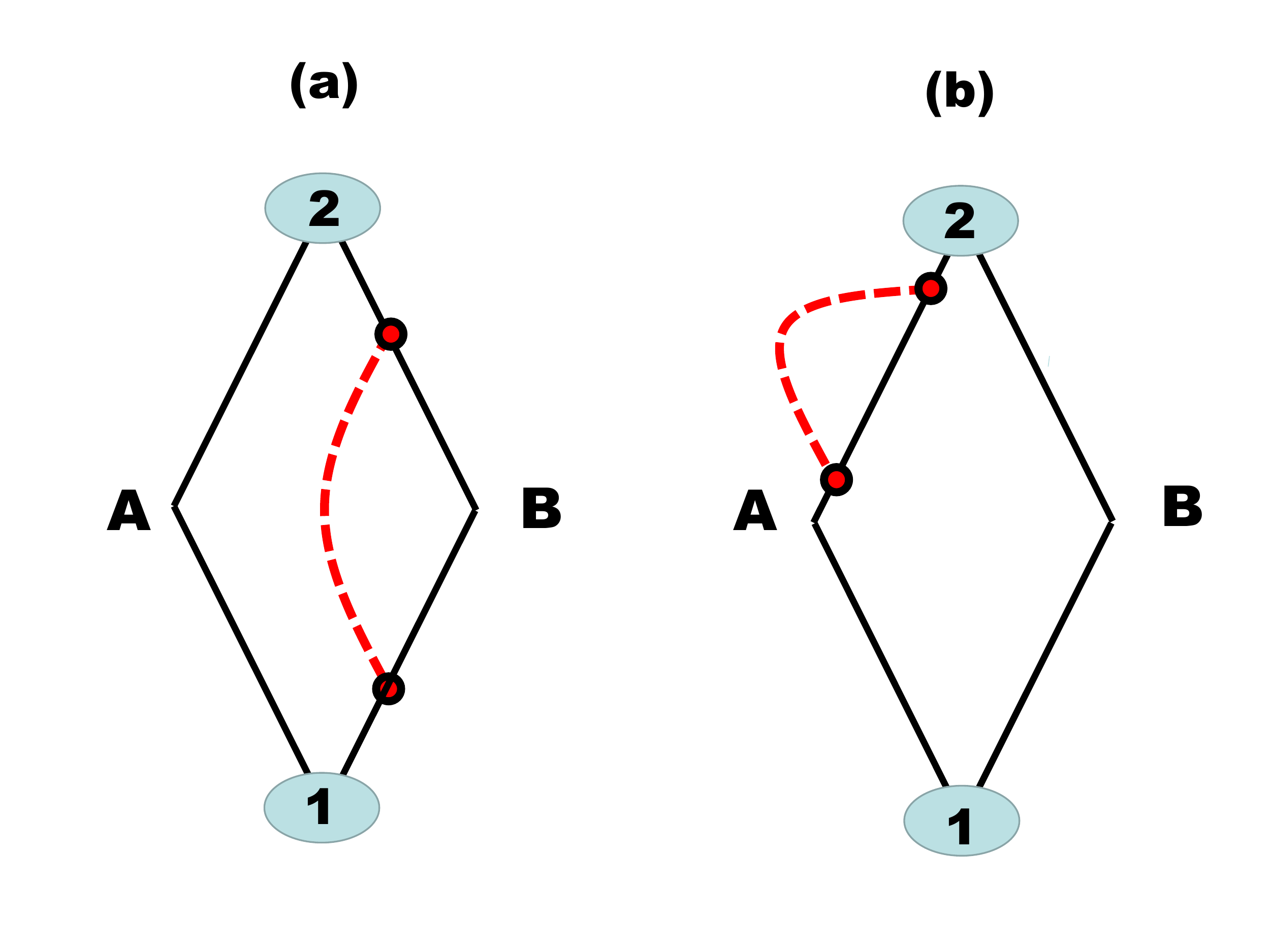}
			
			\caption{\label{fig:2slit-QG}  The 2 lowest-order diagrams contributing to the propagator $K(2,1)$ in equation (\ref{K-ePsi}), calculated in a long-wavelength weak-field treatment of conventional quantum gravity. There are no interactions between separate paths. }
		\end{figure}
		

		\begin{align}
			D_{\mu\nu\lambda\sigma}(x,x') \;=\;& {\cal G}_o(x,x') \nonumber \\
			&\;\; \times \, [\eta_{\mu\lambda} \eta_{\nu\sigma} + \eta_{\mu\sigma} \eta_{\nu\lambda} - \eta_{\mu\nu} \eta_{\lambda\sigma}] \; 
			\label{graviton}
		\end{align}
		where ${\cal G}_o(x,x')$ was defined in eqtn. (\ref{Do-xy}), we then get the result for the particle propagator as a simple sum over paths:
		\begin{widetext}
			\begin{equation}
				K(2,1) \;=\;  \sum_{\alpha}^{A,B} K_{0}^{(\alpha)}(2,1) \; e^{{i \over 2} \int d^4x \int d^4x' \, T_{\mu\nu}(x|q^{(\alpha)})\,  D^{\mu\nu\lambda\sigma}(x,x')\, T_{\lambda\sigma}(x'|q^{(\alpha)})   }
				\label{K-ePsi}
			\end{equation}
		\end{widetext}
		
		This conventional result just involves a self-energy correction to each path, which we represent in the usual way by the sum of the 2 Feynman diagrams shown in Fig. \ref{fig:2slit-QG}. These diagrams represent the lowest-order terms coming from the exponent in (\ref{K-ePsi}). There are no diagrams corresponding to the CWL inter-path correlations in Fig. \ref{fig:2path-CWL}).

		If we ignore the very small imaginary part of $\int T^{(\alpha)}D T^{(\alpha)}$, and again assume that $\Omega_o^{(\alpha)}\rightarrow \Omega_o$, we simply end up with a renormalized version of $K_0(2,1)$ for the propagator, as 
		\begin{equation}
			K_0(2,1) \;=\; 2 \, \Omega_o \, e^{i\bar{S}^{(R)}_{21}} \, \cos (\Delta S^{(R)}_{21})
			\label{Ko-cosR}
		\end{equation}
		where 
		\begin{align}
			\bar{S}^{(R)}_{21} \;=\; \bar{S}^0_{21} +  &\tfrac{1}{2} \int  (T_A D T_A + T_BDT_B)  \nonumber \\
			\Delta S^{(R)}_{21} \;=\;     \Delta S_{21} + &\tfrac{1}{2}          
			\int  (T_A D T_A - T_BDT_B)         
			\label{sumdiffR}
		\end{align}
		and where $\int T_{\alpha} D T_{\alpha}$ refers to the integral in the exponent of (\ref{K-ePsi}) for a particle moving on the $\alpha$-th path between 1 and 2.

		\subsection{CWL Propagator for the 2-path System}
		\label{sec:KCWL-2path}

		Turning now to CWL theory, we will proceed as follows. We first find an expression for the conditional stress energy 
		$\mathbb{\chi}^{\mathbb{T}}_{\mu\nu}(2,1|x)$, as defined in eqtns. (\ref{2-Tmn-1}, \ref{eq:perturbative-stressenergy}), in the weak field regime. Then, to get ${\cal K}(2,1)$, we substitute this result into eqtn. (\ref{eq:perturbative-nonrelprop}) for the phase in ${\cal K}(2,1)$.

		\subsubsection{Long-wavelength Calculation for ${\cal K}(2,1)$}
		\label{sec:longWaveK21}
		
		Recall from eqtn. (\ref{T21-derive}) that 
		\begin{align}
			2\frac{\delta \psi_{0}(2,1|g)}{\delta g^{\mu\nu}(x)} \;=\; -
			\mathbb{\chi}^{\mathbb{T}}_{\mu\nu}(2,1|x,g)
			\label{dPsidg-T21}
		\end{align}
		where, as before, $\psi_{0}(2,1|g)$ is the phase for the particle in a fixed background. In the weak-field regime we have
		\begin{equation}
			\mathbb{\chi}^{\mathbb{T}}_{\mu\nu}(2,1|x)
			\;=\; -2\frac{\delta \psi_{0}(2,1|h)}{\delta h^{\mu\nu}(x)} \Bigg|_{h=0}
			\label{dPsidh-T21}
		\end{equation}

		We may now write a long-wavelength result for the conditional stress-energy propagator $\mathbb{\chi}^{\mathbb{T}}_{\mu\nu}(2,1|x)$, starting from eqtn. (\ref{dPsidh-T21}). Taking the differential of (\ref{K-eik-AB}) with respect to $h^{\mu\nu}$, we get
		\begin{align}
			\label{eq:twopathstressenergy1}
			\mathbb{\chi}^{\mathbb{T}}_{\mu\nu}(2,1|x)
			\;=\; \frac{T^{(A)}_{\mu\nu}(x)K_{0}^{(A)}(2,1)+T^{(B)}_{\mu\nu}(x) K_{0}^{(B)}(2,1)}{K_{0}^{(A)}(2,1)+K_{0}^{(B)}(2,1)}
		\end{align}
		in which the numerator $K_0 = K_0^A + K_0^B$ normalizes the propagator (cf. eqtn. (\ref{eq:perturbative-stressenergy})). 
		
		This expression can be evaluated straightforwardly, and reduces to
		\begin{widetext}
			\begin{equation}
				\label{eq:twopathstressenergy2}
				\mathbb{\chi}^{\mathbb{T}}_{\mu\nu}(2,1|x)
				\;\;=\;\; {1 \over 2} \left[ \big(T_{\mu\nu}^{(A)}(x)+T_{\mu\nu}^{(B)}(x) \big) \;\;+\;\; i \, \big(T_{\mu\nu}^{(A)}(x)-T_{\mu\nu}^{(B)}(x) \big) \, \tan\left(\Delta S_{21} \right) \right]
			\end{equation}
		\end{widetext}
		which is complex. As we have seen, a complex $\mathbb{\chi}^{\mathbb{T}}_{\mu\nu}(2,1|x)$ implies a complex $G_{\mu\nu}(\bar{g}_{21}(x))$ (cf. eqtn. (\ref{KG21-KT21})).  In the absence of any phase information here (ie., no phase interference between the paths) the imaginary part of $\mathbb{\chi}^{\mathbb{T}}_{\mu\nu}(2,1|x)$ is zero.

		Continuing on, we insert (\ref{eq:twopathstressenergy2}) into (\ref{eq:perturbative-nonrelprop}) to find the final form of the 2-path propagator ${\cal K}(2,1)$. The prefactor $K_0(2,1)$ is as before (cf. equation (\ref{Ko-cos})); for the phase term we simplify the notation and write 
		$T_{00}^{(\alpha)} ({\bf r}, t) \; \rightarrow \; T_{\alpha}$ and 
		$T_{00}^{(\alpha)} ({\bf r'}, t) \; \rightarrow \; T'_{\alpha}$
		respectively. Then, inserting our result for $\chi^{\mathbb{T}}(2,1|x)$ into eqtn. (\ref{eq:perturbative-nonrelprop}), we have
		\begin{widetext}
			\begin{align}
				\Theta_{21} \;\;=\;\; {G_N \over 4} \int_{t_{1}}^{t_{2}} dt\int \frac{d^{3}r d^{3}r'}{|{\bf r} - {\bf r'}|} \Bigg\{ \bigg[ (T_A T_A' + T_B T_B') (1 - \tan^2 (\Delta S_{21})) &\;+\; 2 \frac{T_A T_B'}{\cos^{2}(\Delta S_{21})} \bigg]  \nonumber \\ & \;+\; 2i \, \big(T_A T_A' - T_B T_B' \big) \, \tan (\Delta S_{21}) \Bigg\}
				\label{Theta-AB}
			\end{align}
			ie., this phase is complex. Obviously we can absorb this imaginary part of the phase into the prefactor, and write
			\begin{equation}
				\mathcal{K}(2,1)  \;=\;  A(2,1)\; e^{i \Phi_{21}}
				\label{Ko-Ao}
			\end{equation}
			where we have
			\begin{equation}
				A(2,1) \;\;=\;\; 2\Omega_o \cos(\Delta S_{21})\;  \exp \Bigg\{ - {G_N \over 2} \int_{t_{1}}^{t_{2}} dt\int \frac{d^{3}r d^{3}r'}{|{\bf r} - {\bf r'}|} \;  \big(T_A T_A' - T_B T_B' \big) \, \tan (\Delta S_{21}) \Bigg\} 
				\label{Ao-Phi}
			\end{equation} 
			\begin{equation}
				\Phi_{21} \;\; =\;\; \bar{S}^{o}_{21}+{G_N \over 4} \int_{t_{1}}^{t_{2}} dt\int \frac{d^{3}r d^{3}r'}{|{\bf r} - {\bf r'}|}\; \bigg[ (T_A T_A' + T_B T_B') \, (1 - \tan^2 (\Delta S_{21})) \;+\;  \frac{(T_A T_B' + T_A' T_B)}{\cos^{2}(\Delta S_{21})} \bigg]
				\label{Phi-21}
			\end{equation}
		\end{widetext}
		for the renormalized prefactor and phase respectively, and with $K_0(2,1)$ given by eqtn. (\ref{Ko-cos}). 
		
		Before interpreting these results, note that they depend on 2 approximations, both of which are questionable, viz.,
		
		(i) The approximation of a point particle used here breaks down for any extended mass. As we discuss in more detail in section 7, even for objects of nanometre size the effective interaction between CWL paths is no longer of singular $1/|{\bf r} - {\bf r'}|$ form as $|{\bf r} - {\bf r'}| \rightarrow 0$; for objects exceeding $\sim O(10^2)$ nm in size it is quite different. 
		
		(ii) The assumption of weak fields. As we shall see immediately below, this can fail. In this paper we will not try to go beyond this approximation, although the techniques developed by Fradkin \cite{fradkin,fried,khvesh} could be used to do so.

		\subsubsection{Interpretation of CWL Results}
		\label{sec:2path-interp}
		
		The different terms in (\ref{Ao-Phi}) and (\ref{Phi-21}) come either from `self-interaction' of the mass along the same set of paths, or from interactions across paths, ie., between a path along A and another along B. These 2 contributions are shown at lowest order in $G_N$ in Fig. \ref{fig:2path-2int}. Self-interactions along a specific path (cf. Fig. \ref{fig:2path-2int}(a)) renormalize the action along this path; the renormalization of the prefactor from $K_0(2,1)$ to $A(2,1)$ involves such a term.  
		
		More interesting is the effect of the attractive cross-interactions between paths A and B, which we will re-interpret in the next subsection in the context of `path-bunching', caused by the mutual attraction of paths \cite{stamp15,stamp12}. These cross-terms are examples what we showed in Fig. \ref{fig:2path-CWL}(b) at the beginning of this paper. 
		
		Formally, the point is that while we are exponentiating the classical action in eqtns. (\ref{eq:CWLgenfunc}) and (\ref{K-eik}), so that the spacetime metric is just the classical solution to the Einstein equation, the matter term in these equations still represents the full quantum-mechanical matter path integrals (compare, eg., eqtns. (\ref{K-phi-12}) and (\ref{K-phi-12'})).
		
		Thus, if we denote by $\langle \bar{g} \rangle_{AB}$ the particular solution for the metric to our 2-path problem, and then substitute $\langle \bar{g} \rangle_{AB}$ back into our expression (\ref{K-eik}) for the CWL propagator, we get a result for the full CWL propagator of schematic form
		\begin{equation}
			\mathcal{K}(2,1)=e^{iS_{G}[\langle \bar{g} \rangle_{AB}]} \, \sum_{\alpha}^{A,B}  e^{iS_M[q^{(\alpha)}| \langle \bar{g} \rangle_{AB}]}
		\end{equation}
		in which the gravitational term is just the path integral for the classical action $S_{G}[\langle \bar{g} \rangle_{AB}]$, integrated along the classical path in configuration for a metric field $\langle \bar{g} \rangle_{AB}$ sourced by both matter paths; and the matter term sums over the 2 matter paths, in the presence of the same background metric field $\langle \bar{g} \rangle_{AB}$. Thus the matter is still propagating quantum-mechanically.

		
		\begin{figure}
			\includegraphics[width=3.2in]{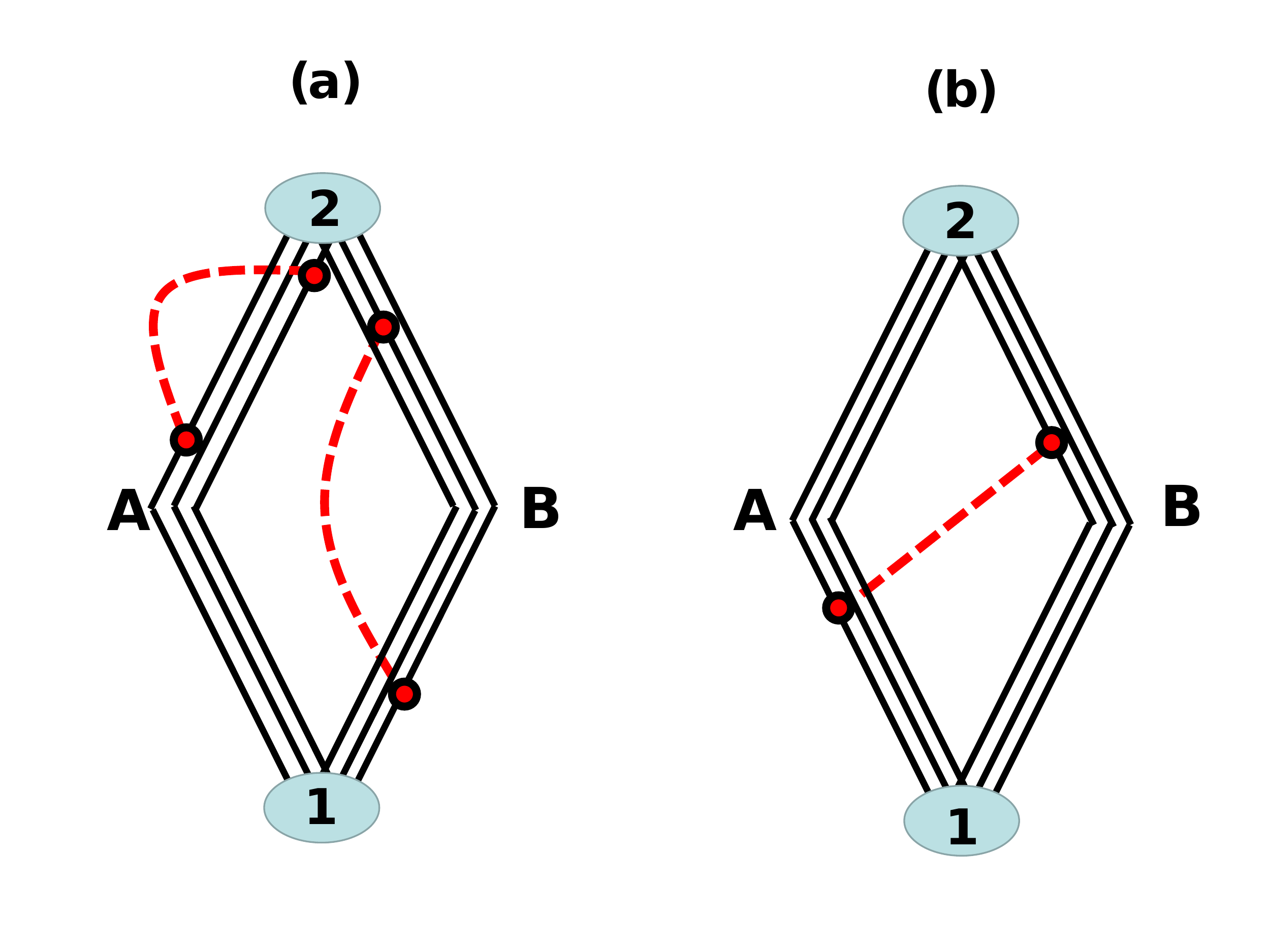}
			
			\vspace{-4mm}
			
			\caption{\label{fig:2path-2int}  Graphical representation of the two different kinds of terms in eqtns. (\ref{Ao-Phi}) and (\ref{Phi-21}) for ${\cal K}(2,1)$. In (a) we show `self-energy' contributions involving pairings like $T_AT'_A$ or $T_BT_B'$. In (b) we show inter-path pairings like $T_AT_B'$, or $T_A'T_B$, which we refer to as `path-bunching' terms. In both (a) and (b), 3-path contributions to ${\cal K}(2,1)$ are shown. }
		\end{figure}


		Let us now look in more detail at our results (\ref{Ko-Ao})-(\ref{Phi-21}) for ${\cal K}(2,1)$. Consider first their dependence on the relative phase $\Delta S_{21}$. We note that the imaginary part of $\mathbb{\chi}^{\mathbb{T}}_{\mu\nu}(2,1|x)$ vanishes when $\Delta S_{21} =2n\pi$ for integer $n$---precisely when the two paths interfere constructively. For this case of constructive interference, $\mathbb{\chi}^{\mathbb{T}}_{\mu\nu}(2,1|x)$ is then a simple average over the two paths. 
		
		We see that for constructive interference, the prefactor $A(2,1)$ in ${\cal K}(2,1)$ is unrenormalized; however this is not true of the phase $\Phi_{21}$, which contains both intra-path and inter-path contributions. The last term in $\Phi_{21}$ in eqtn. (\ref{Phi-21}), when $\Delta S_{21} =2n\pi$, is precisely the Newtonian interaction between paths considered in previous discussions of path-bunching (see next section). 
		
		If we move away from the constructive interference regime, so that $\Delta S\neq 2n\pi$, several things happen. First, the renormalization of the prefactor enters. In principle it can suppress the propagator, but we notice that it is proportional to the difference between the gravitational self-energies for the two paths, which is zero for a symmetric 2-path system. If this difference is non-zero, then eqtn. (\ref{Ao-Phi}) predicts that the renormalization will drive $A(2,1)$ rapidly to zero as one approaches the destructive interference regime around $\Delta S_{21} = \pm\pi$, because of the factor $\tan (\Delta S_{21})$. This will happen much faster than would happen without the renormalization.
		
		Turning now to the phase $\Phi_{21}$, the `path-bunching' term grows like $\sec^2 (\Delta S_{21})$, and ultimately diverges when $\Delta S_{21} = \pm\pi$, ie., the phase becomes singular.  The other self-energy term in $\Phi_{21}$ is now modified by the $-\tan^{2}$ term; when $\Delta S_{21}=\pm\pi/2$ this term switches from positive to negative, also diverging when $\Delta S_{21} =\pm\pi$. 
		
		These singular effects are interesting, as they appear to signal specific locations in which our approximation scheme breaks down. As mentioned in the previous section, by linearizing the semiclassical Einstein equation we have forgone the self-consistency of the full metric solution $\bar{g}_{21}$. Near the ``bright fringes'', ie. where $\Delta S_{21}\approx 2 n\pi$, the CWL result simply describes the Newtonian interaction between paths A and B. Since this is a small interaction, the linear approximation is still valid.  Near the ``dark fringes'' though, ie. where $\Delta S_{21}\approx (2n+1)\pi$, we apparently have an effective CWL interaction which is arbitrarily strong. It is clear then, that the linear approximation is failing near the dark fringes, since the result is no longer self-consistent. 
		
		We can start to anticipate what is happening here. In a self-consistent calculation we must allow the matter to respond to the gravitational field it sources. Near the bright fringes the effective gravitational interaction is relatively weak, so we expect the classical paths considered above to remain approximately correct.  Near the dark fringes though, the effective gravitational interaction must significantly alter the dynamics of the particle. Remarkably, this then indicates that CWL `path-bunching' must become relevant near the locations of dark fringes.
		
		We expect that a self-consistent calculation will ensure that the conditional stress energy does not diverge as the endpoints $(1,2)$ are varied. Note that a realistic calculation will also involve an extended mass rather than the simple particle approximation used here.  Looking at the expression (\ref{eq:twopathstressenergy1}), we might then anticipate that in a proper CWL calculation, we will see the prevention of total destructive interference at the locations of the dark fringes. We leave this for another paper. 
		
		To summarize: in CWL theory, the mutual attraction of the paths causes a breakdown of the usual 2-slit interference result. The CWL interactions can lead to divergent corrections of the conventional result, which will need to be dealt with by a full self-consistent calculation.

		\subsection{Comparison with Semiclassical gravity}
		\label{sec:2path-semiGR}
		
		Semiclassical gravity has a long history \cite{moller62,rosenfeld63}, which has been repeatedly reviewed \cite{davies82,singh89,kiefer07,verdaguer20,sudarsky}. In this theory, one writes the semiclassical equation of motion as
		\begin{equation}
			\label{eq:EE'}
			G_{\mu\nu}(x|\bar{g}) \;=\; 8\pi G_{N} \langle \, T_{\mu\nu}[x|\bar{g}]\, \rangle
		\end{equation}
		which is the same equation (\ref{eq:EE}) as we found for $G_{\mu\nu}(x)$ in CWL theory, in the special case that $J=0$.
		
		The literature describing the predictions of semiclassical theory appears to be quite confusing. In the original papers of Kibble \cite{kibble2}, Page and Geilker \cite{page81}, and others, it was argued that a semiclassical analysis of the 2-path experiment leads to an obvious violation of QM. Thus, suppose the mass $M_o$ is in a symmetric superposition of states paths A and B. It has then been claimed (see, eg., ref. \cite{kibble2}), that $\langle \, T_{\mu\nu}[x|\bar{g}]\, \rangle$ will source a field which is apparently generated by the average of the 2 paths, ie., by a source mid-way between the 2 paths.

		If one then employs a `test mass' $\bar{m}$ (as in a Cavendish experiment) to monitor the position of $M_o$, via the gravitational interaction between $M_o$ and $\bar{m}$, then according to this argument, semiclassical theory predicts that it will detect $M_o$ at this mid-point.

		This result is not entirely clear to us. If a particle which is simultaneously following paths ${\bf r}_A(t)$ and ${\bf r}_B(t)$, it will be in a state 
		\begin{equation}
			|\psi(t) \rangle \;\sim\;  \tfrac{1}{\sqrt{2}}[\delta({\bf r} - {\bf r}_A(t)) + \delta({\bf r} - {\bf r}_B(t))]
			\label{psi-2state}
		\end{equation} 
		
		Then one has
		\begin{align}
			\langle T_{00}(x) \rangle \;&=\; {\langle \psi |T_{00}(x)| \psi \rangle \over \langle \psi|\psi \rangle} \nonumber \\
			& =\; \tfrac{m}{2} [\delta({\bf r} - {\bf r}_A(t)) + \delta({\bf r} - {\bf r}_B(t))]
			\label{expT-psi}
		\end{align}
		for the expectation value of $T_{00}(x)$. 
		
		On the other hand the argument just given indicates that in semiclassical gravity one should instead have
		\begin{equation}
			\langle T_{00}(x) \rangle \;=\; 
			m \,\delta\left({\bf r} - \tfrac{1}{2}[{\bf r}_A(t) + {\bf r}_B(t)] \right)
			\label{expT-psi2}
		\end{equation}
		
		To clarify this question, let us expand the semiclassical equation (\ref{eq:EE'}) for the particle in state $|\psi \rangle$ as 
		\begin{align}
			\label{eq:EE2}
			G_{\mu\nu}(g({\bf r}, t)) \;=&\; 8\pi G_{N} \langle \psi | e^{i \hat{H} t} T_{\mu\nu}({\bf r}) e^{-i \hat{H}t} | \psi \rangle \nonumber \\
			= &\; 8\pi G_{N} \langle \psi(t) |  T_{\mu\nu}({\bf r})  | \psi(t) \rangle
		\end{align}
		which, using (\ref{psi-2state}) for $|\psi(t) \rangle$, gives
		\begin{equation}
			G_{\mu\nu}(g({\bf r},t)) \;=\; 4\pi G_N \sum_{\alpha}^{A,B} \langle {\bf r}_{\alpha} | T_{\mu\nu}({\bf r}) | {\bf r}_{\alpha} \rangle
			\label{EE-semicl}
		\end{equation}
		and this result is shown in Fig. \ref{fig:2slit-semiC}(a).

		
		\begin{figure}
			\includegraphics[width=3.2in]{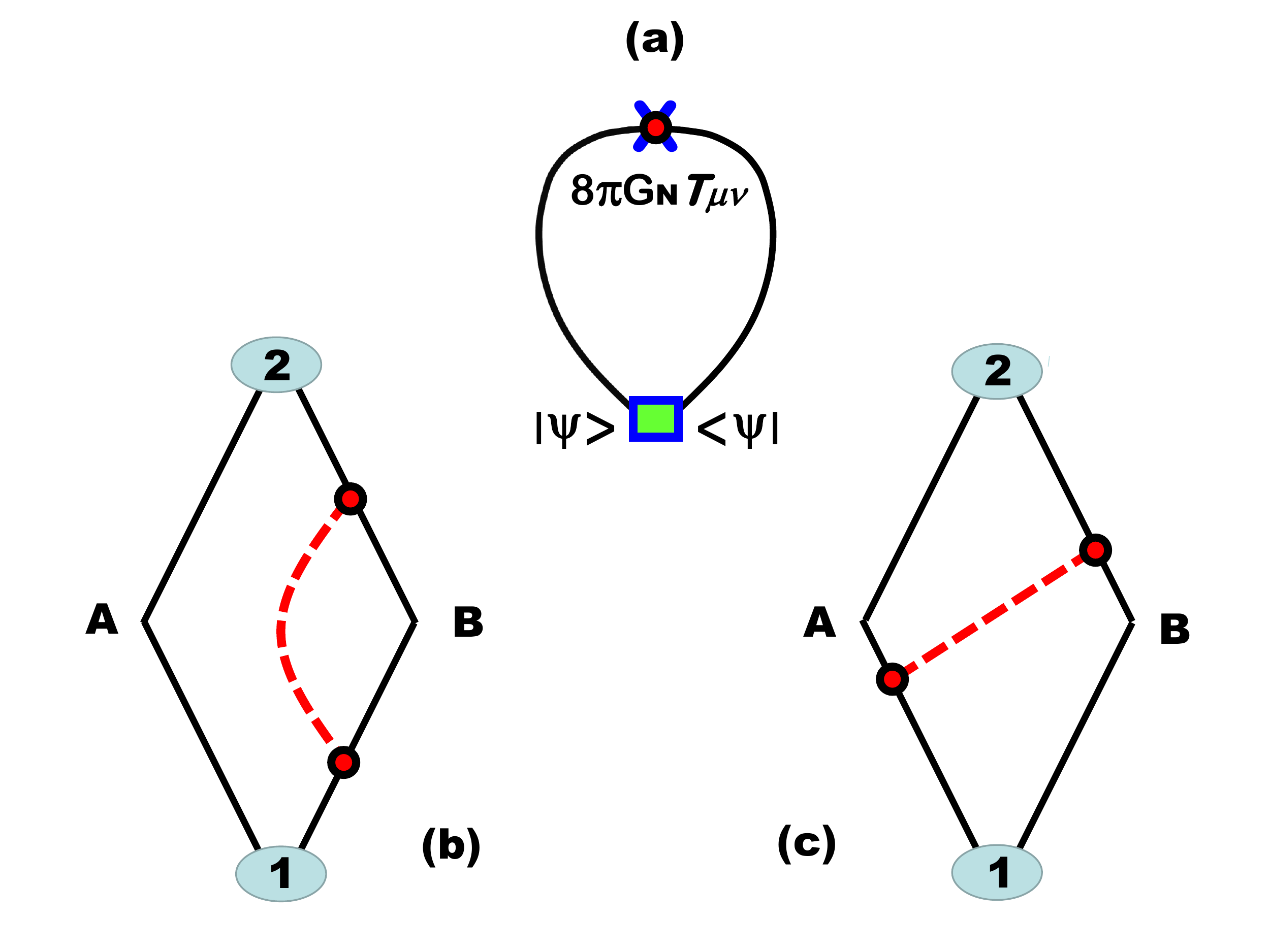}
			
			\vspace{-4mm}
			
			\caption{\label{fig:2slit-semiC}  Results in semiclassical gravity theory. The graphical representation of eqtn (\ref{EE-semicl}) is shown in (a). In (b) and (c) the two contributions to the semiclassical propagator $K_{sc}(2,1)$ are shown for the 2-path system; in (b) we see the intra-path term proportional to $T_BT_B'$ in the phase $\Phi_{21}$ in eqtn. (\ref{Phi-21-sc}), and (c) shows one of the inter-path cross-terms in this eqtn.   }
		\end{figure}


		To proceed further we again introduce the eikonal expansion of the weak field deviation $h^{\mu\nu}(x)$; proceeding as before we obtain the semiclassical propagator in the form $K_{sc}(2,1) = A^{(sc)}(2,1)e^{i\Phi_{21}}$, where the prefactor has the unrenormalized form
		\begin{equation}
			A^{(sc)}(2,1) \;=\; 2\Omega_{o}\cos(\Delta S_{21})
			\label{Ao-Phi'}
		\end{equation}
		and the phase is now
		\begin{align}
			\Phi^{(sc)}_{21} \; =&\; \bar{S}^{o}_{21} + {G_N \over 4} \int_{t_{1}}^{t_{2}} dt\int \frac{d^{3}r d^{3}r'}{|{\bf r} - {\bf r'}|} \nonumber \\ &\; \times \bigg[ (T_A T_A' + T_B T_B') \;+\;  (T_A T_B' + T_A' T_B) \bigg]
			\label{Phi-21-sc}
		\end{align}
		when written out in full. 
		
		This result for the semiclassical propagator is clearly different from both the conventional result in eqtns. (\ref{K-ePsi})-(\ref{sumdiffR}), and the CWL result in eqtns. (\ref{Ko-Ao})-(\ref{Phi-21}). 
		
		Notice that we can get exactly the same result for $K_{sc}(2,1)$ by noting that in semiclassical theory, we only expect the mean stress-energy to be involved (compare eqtn. (\ref{eq:EE'})), and so we naturally guess that $\mathbb{\chi}^{\mathbb{T}}_{\mu\nu}(2,1|x)$ will have the form
		\begin{equation}
			\mathbb{\chi}^{\mathbb{T}}_{\mu\nu}(2,1|x) \;=\; \tfrac{1}{2}  \big(T_{\mu\nu}^{(A)}(x)+T_{\mu\nu}^{(B)}(x) \big)
			\label{T21-semiGR}
		\end{equation} 
		
		If we now substitute this into (\ref{eq:perturbative-nonrelprop}) we again get back (\ref{Ao-Phi'}}) and (\ref{Phi-21-sc}) for $K^{(sc)}(2,1)$.
	
	In this long-wavelength, weak field approximation we can depict these semiclassical results diagrammatically (Fig. \ref{fig:2slit-semiC}(b) and (c)).  The Hartree pairing of terms in the phase $\Phi_{21}$ in (\ref{Phi-21-sc}), in the form $(T_A + T_B)(T_A' +T_B')$, is what we would expect from a Schrodinger-Newton analysis in the non-relativistic regime. One gets not only self-interactions along each path, but also interactions between paths. 
	
	To summarize: one finds inter-path interactions in both semiclassical and CWL theory. The difference between the results for the 2 theories comes entirely from the imaginary part of $\mathbb{\chi}^{\mathbb{T}}_{\mu\nu}(2,1|x)$ in eqtn. (\ref{eq:twopathstressenergy2}), which is absent from the semiclassical result.

	
	\section{The Propagator ${\cal K}(2,1)$ in $\ell_P^2$ Approximation}
	\label{sec:CWL-K-lP2}
	

	As we have just seen, a key feature in CWL theory is the cross-correlation between paths. We would like to better understand how this works. In this section, we drop the restriction to 2-path system, and now look at the lowest-order graphs in an expansion in powers of $G_N$ (ie., in $\ell_P^2$), for ${\cal K}(2,1)$. A preliminary analysis of ${\cal K}(2,1)$ to order $\ell_P^2$ was given in a previous paper \cite{stamp15}. Here we justify the previous work, in section 7.A, by showing that at order $\ell_P^2$, only one graph survives after we take the CWL product over $N$, the same graph that was analyzed \cite{stamp15} in the earlier work. 
	
	We then give, in section 7.B, a more detailed treatment of the physics emerging in this approximation, in the non-relativistic regime relevant to experiment, and show what kind of dynamics emerges. Finally, in section 7.C, we discuss what we might expect to happen in a more realistic calculation, where a dissipative coupling to the background environment is included, and where we go beyond the ``$\ell_P^2$ approximation" used here. 
	
	We emphasize before starting that CWL results obtained in the $\ell_P^2$ approximation are mainly of methodological interest. They allow simple calculations, which allow one to explore the physics of path-bunching, and estimate the relevant energy and length scales in the problem. They can also be related to calculations done in semiclassical theory using, eg., the Schrodinger-Newton approximation \cite{penrose-SchN,yanbei-SchN}. However they have very obvious limitations \cite{stamp15}, which we will reiterate in this section.

	\subsection{Evaluation of Graphs}
	\label{sec:K-lP2-graph}
	
	In a previous paper \cite{CWL2} we derived all the terms appearing up to $\sim O(\ell_P^2)$ in the generating functional $\mathbb{Q}$ and the correlation functions. We now extend this analysis to the propagator ${\cal K}(2,1)$, to the same order.

	As before, we collect all the $n$ matter field paths in the $n$-th tower into one big ``vector field" $\Phi_n \equiv \{ \phi_i^{(n)}(x) \}$, so that $S[\Phi]=\sum_{i=1}^{n}S_{\phi}[\phi_{i}^{(n)}]$, and use the contracted DeWitt-style notation in which $a,b,c$ label all internal indices (including tower and replicated path indices), and subscripts denote functional derivatives around a background field $g_o$. Thus, eg., $S_a \equiv \delta S/\delta g_{a}|_{g = g_o}$ and the 2nd derivative $I_{ab}$ is the inverse of the graviton propagator, ie., $I_{ac}D^{cb}=\delta^b_a$. We will also use the 3-graviton vertex $I_{bcd} = \delta^3 I/\delta g_b \delta g_c \delta g_c|_{g = g_o}$. Note that in this section we will be more explicit about gauge-breaking terms, ie. rather than $S_{G}[g]$ we use $I[g]$ as defined in eqtn.~(\ref{S-EH}). We will omit the Faddeev-Popov ghost terms because they ultimately do not contribute---since graviton loops all vanish, so too do ghost loops.

\begin{figure}
		\includegraphics[width=3.2in]{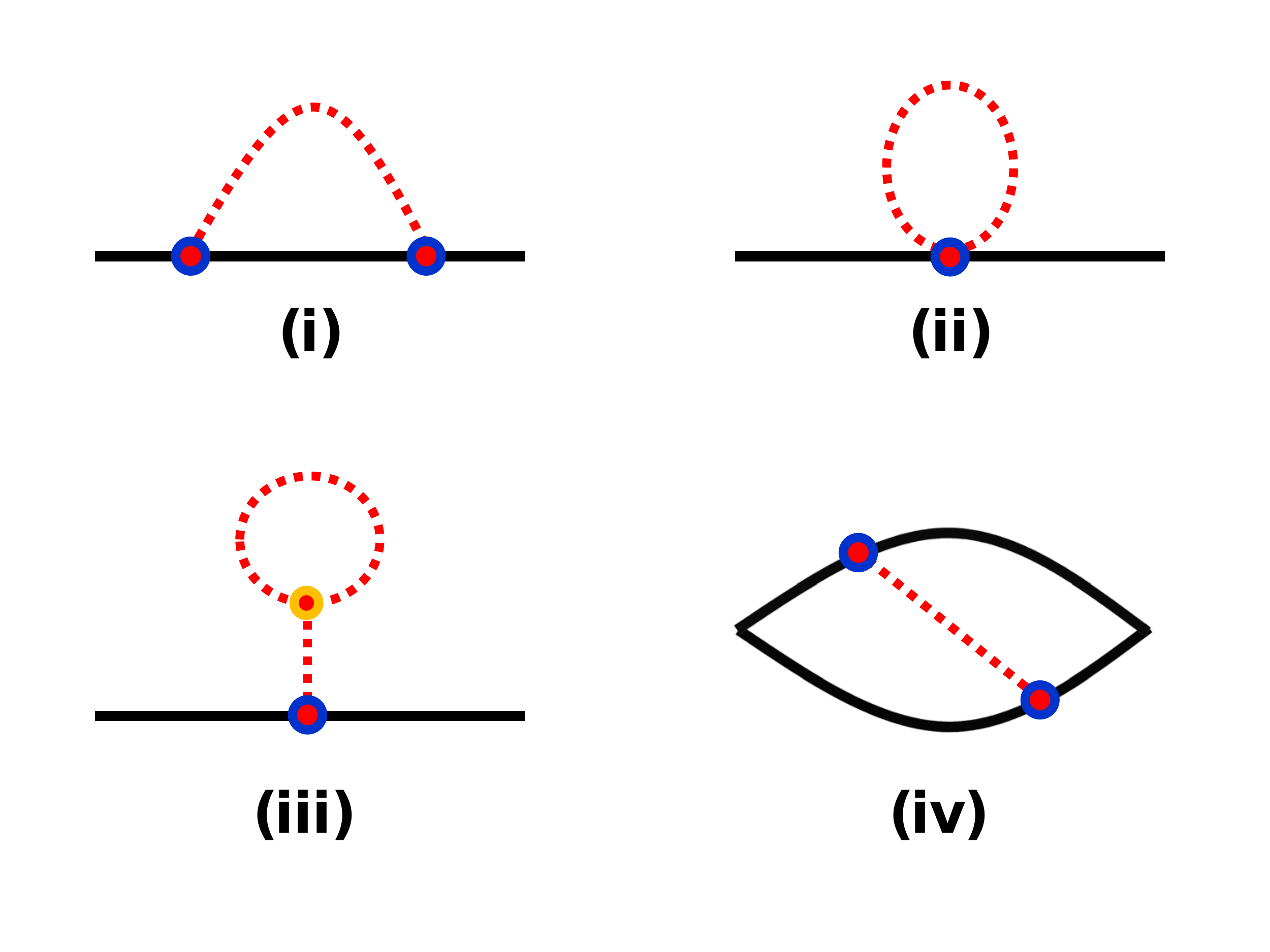}
		
		\vspace{-4mm}
		
		\caption{\label{fig:CWL-selfE}  Graphical representation of the four terms in eqtns. (\ref{K-CWL-1}) and (\ref{K-CWL-2}). In (i), (ii) and (ii) we have the graphs corresponding to the 1st, 2nd, and 3rd terms in eqtn. (\ref{K-CWL-2}). In (iv) we have the CWL graph corresponding to the last term in eqtn. (\ref{K-CWL-2}), in which one sums over two different sets of paths.  }
	\end{figure}

	Let us now expand the propagator up to $O(l_P^2)$, precisely as was done for ${\cal Q}_n$ and for $\mathbb{Q}$ in ref \cite{CWL2}. Note that, as in \cite{CWL2}, we assume below that the metric fluctuations propagate between vacuum states, and that these fluctuations have already been integrated out.  This leaves an effective action for the matter propagator in terms of graviton correlators/vertices. The terms are shown graphically in Fig. \ref{fig:CWL-selfE}); for the $n$-path contribution ${\cal K}_n$ one gets
	\begin{widetext}
		\begin{equation}
			{\cal K}_{n}(2,1)\;\;=\;\; \int_{\Phi_{1}}^{\Phi_{2}}\mathcal{D}\Phi_n\,e^{iS[\Phi_n]}
			\left( 1-\frac{\ell_{P}^{2}}{2n}D^{ab}\bigg(iS_{a}[\Phi_n]S_{b}[\Phi_n]
			+S_{ab}[\Phi_n]-S_{a}[\Phi_n]I_{bcd}D^{cd}\bigg)\right) \;+\; O(\ell_P^4)
			\label{K-CWL-1}
		\end{equation}
		which when expanded out in terms of the different configurations $\phi_{k}^{(n)}$ takes the form
		\begin{align}
			{\cal K}_{n}(2,1) \;=\;\int_{\Phi_{1}}^{\Phi_{2}}\mathcal{D}\Phi_n\,e^{iS[\Phi_n]}\bigg[ 1&-\frac{\ell_{P}^{2}}{2n} D^{ab}  \sum_{k=1}^{n}\left(iS_{a}[\phi_{k}^{(n)}]S_{b}[\phi_{k}^{(n)}]+S_{ab}[\phi_{k}^{(n)}]-S_{a}[\phi_{k}^{(n)}]I_{bcd}D^{cd}\right) \nonumber \\
			&-i\frac{\ell_{P}^{2}}{2n}D^{ab}\sum_{k\neq k'=1}^{n}S_{a}[\phi_{k}^{(n)}]S_{b}[\phi_{k'}^{(n)}]\bigg] \;+\; O(\ell_P^4)
			\label{K-CWL-2}
		\end{align}
	\end{widetext}
	in which the cross-terms in the last term (ie., the interaction between $\phi_{k}^{(n)}$ and $\phi_{k'}^{(n)}$) are written explicitly. This, as we will see presently, is the CWL term, ie., the term that does not exist in conventional quantum gravity at order $\ell_P^2$, and which leads to path-bunching. The 4 terms in (\ref{K-CWL-2}) are shown in Fig. \ref{fig:CWL-selfE}).

	
	Now, since each of the different paths in the sums in (\ref{K-CWL-2}) is indistinguishable from the others, we can easily evaluate these sums. The result is conveniently expressed in the form
	\begin{equation}
		\label{3Terms}
		{\cal K}_{n}=K_{0}^{n}+\ell_{P}^{2} (A K_{0}^{n-1}+(n-1) B K_{0}^{n-2}) + O(\ell_P^4),
	\end{equation}
	where the ``single path" CWL contribution $A = A(2,1)$ (ie., the term arising from a single sum over paths) is
	\begin{eqnarray}
		A &=& \frac{D^{ab}}{2}\int^{\Phi_{2}}_{\Phi_{1}}\mathcal{D}\phi\,e^{iS_M[\phi]} \,\big(I_{bcd}D^{cd}S_{a}[\phi] \nonumber \\
		&& \qquad\qquad\qquad\qquad -S_{ab}[\phi]-iS_{a}[\phi]S_{b}[\phi] \big) \;\;\;
	\end{eqnarray}
	and the two-path CWL contribution $B = B(2,1)$ is
	\begin{eqnarray}
		B &=& -i\frac{D^{ab}}{2}\int^{\Phi_{2}}_{\Phi_{1}}\mathcal{D}\phi\int^{\Phi_{2}}_{\Phi_{1}}\mathcal{D}\phi'  \nonumber \\
		&& \qquad\qquad  \times \; e^{iS_M[\phi]+iS_M[\phi']} \; S_{a}[\phi]S_{b}[\phi'] \qquad
	\end{eqnarray}
	with a gravitational interaction mediated by $D^{ab}$ between pairs of paths $\phi$ and $\phi'$.
	
	We can now reorganize eqtn. (\ref{3Terms}), by defining new correlators as follows:
	\begin{equation}
		\mathbb{A} \;=\; (AK_0 - B)/K_0^2 \; ;  \qquad\qquad
		\mathbb{B} \;=\; B/K_0^2
		\label{AB}
	\end{equation}
	where we note that $\mathbb{A} \equiv \mathbb{A}(2,1)$ etc. We immediately see where this $\ell_P^2$ approximation fails - when $K_0 \rightarrow 0$, then $\mathbb{A}$ and $\mathbb{B}$ are no longer small. This is precisely the same failure of the linearized theory that we saw in the last section, near the `dark fringes' of the 2-path system. 
	
	Assuming $K_{o}$ is not too close to zero, we then have, to order $O(\ell_P^2)$, 
	\begin{align}
		{\cal K}_{n}&\;=\; K_{0}^{n}\left(1+\ell_{P}^{2}\mathbb{A}+n\ell_{P}^{2}\mathbb{B} \right) + O(\ell_P^4) \nonumber \\
		&\;\sim\;  \left[K_{0}\left(1+\ell_{P}^{2}\mathbb{B}\right)\right]^{n}\left(1+\ell_{P}^{2}\mathbb{A}\right).
	\end{align}

	This result for ${\cal K}_n$ is in a form suitable to do the product over $n$, to get a result for the full CWL propagator up to order $\ell_P^4$. We find
	\begin{equation}
		{\cal K} \;=\;\lim\limits_{N\rightarrow\infty}\left[\prod_{n=1}^{N}
		\left[K_{0}\left(1+\ell_{P}^{2}\mathbb{B}\right)\right]^{n}
		\left(1+\ell_{P}^{2}\mathbb{A}\right)\right]^{\alpha_N}
		\label{K-alpha2}
	\end{equation}
	with the result that we simply have
	\begin{equation}
		{\cal K}(2,1) \;=\; K_{0}(2,1) \, \left(1+\ell_{P}^{2}\mathbb{B}(2,1)\right)
		\label{K-alpha2'}
	\end{equation}
	in which the term $\mathbb{A}(2,1)$, which refers to those terms in the propagator that do {\it not} involve CWL terms, has disappeared! Only the contribution from the 4th graph in Fig. \ref{fig:CWL-selfE}, ie., the path-bunching term, has survived.
	
	We have thus found that the propagator to lowest order perturbation theory becomes
	\begin{widetext}
		\begin{align}
			\label{K-fin}
			{\cal K}(2,1)\; &=\; K_{0}(2,1)-i K^{-1}_{0}(2,1) \; \ell_{P}^{2}\frac{D^{ab}}{2}\int^{\Phi_2}_{\Phi_1}
			\mathcal{D}\phi\int^{\Phi_1}_{\Phi_1}\mathcal{D}\phi'
			\,e^{iS[\phi]+iS[\phi']}S_{a}[\phi]S_{b}[\phi']  \;\;+ \;\; O(\ell_P^4)  \nonumber \\
			&=\;K^{-1}_{0}(2,1)\int^{\Phi_2}_{\Phi_1}\mathcal{D}\phi\int^{\Phi_2}_{\Phi_1}\mathcal{D}\phi'\,e^{iS[\phi]+iS[\phi']}\left(1-i \ell_{P}^{2}\frac{D^{ab}}{2}S_{a}[\phi]S_{b}[\phi']\right) \;\;+ \;\; O(\ell_P^4)
		\end{align}
		where the path bunching term in the effective action is, to this order in $\ell_P^2$, is given by
		\begin{align}
			S_{CWL}[\phi,\phi'] \;&=\; -\ell_{P}^{2}\frac{D^{ab}}{2}S_{a}[\phi]S_{b}[\phi'] \nonumber \\
			&=\; -{\ell_{P}^{2} \over 8} \int d^4x \int d^4x'D^{\mu\nu\alpha\beta}(x-x') T_{\mu\nu}(\phi(x))T_{\alpha\beta}(\phi'(x'))
			\label{S-CWL1}
		\end{align}
		where $D^{\mu\nu\alpha\beta}(x,x')$ is the graviton propagator (again defined with respect to the background field $g_0$), and we rewrite $S_s$ in terms of the stress-energy, using $2T_a = S_a$.
		
		Occasionally we will rewrite the result (\ref{K-fin}) in the exponentiated form
		\begin{equation}
			{\cal K}(2,1) \;=\;\sim \; K^{-1}_{0}(2,1)\int^{\Phi_2}_{\Phi_1}\mathcal{D}\phi\int^{\Phi_2}_{\Phi_1}\mathcal{D}\phi'\,e^{i(S[\phi]+S[\phi'])} \; e^{iS_{CWL}[\phi,\phi']}   \;\;+ \;\; O(\ell_P^4)
			\label{K-fin2}
		\end{equation}
	\end{widetext}
	but for the same reasons as given above, this form is only valid if $|S_{CWL}[\phi,\phi']| \ll 1$. 
	
	It is easy to see that we would have found find precisely the same results as above if we had done the calculation in the unscaled version of the theory. In both calculations the extra factor of $n$, coming from the double sum over replicas in the path-bunching term, singles out this term, and the other 3 terms are eliminated.

\begin{figure}
		\includegraphics[width=3.2in]{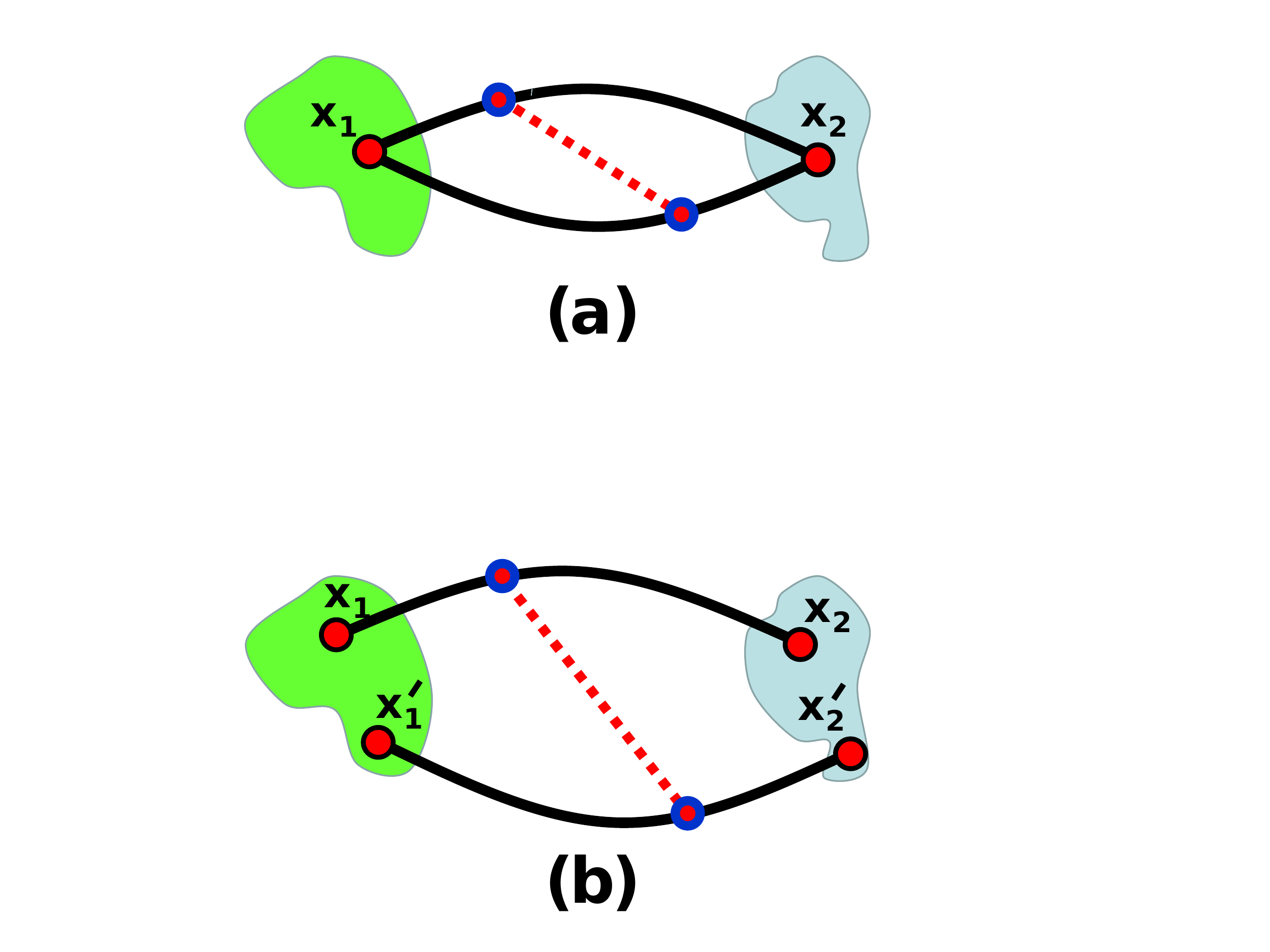}
		
		\caption{\label{fig:CWL-int} Graphical representation of two possible ways of writing the path integration for the lowest-order CWL contribution to ${\cal K}(\beta.\alpha)$ in eqtn. (\ref{K2-CWL-ab}). In (a) the end-points for the two paths are the same, and in (b) they are different; the latter corresponds to eqtn. (\ref{D-int-defn}), and is the correct prescription. The graviton is shown as a hatched line.
		}
		
	\end{figure}

	The foregoing calculation is trivially modified to deal with the propagator between general states defined by `wave-functions' $\psi_{\alpha}(x)$ and $\psi_{\beta}(x)$ (for a particle).  The relativistic CWL propagator ${\cal K}(\beta \alpha)$ becomes
	\begin{eqnarray}
		{\cal K}(\beta \alpha) & \sim & K^{-1}_{0}(\beta \alpha)\int^{\beta}_{\alpha}\mathcal{D} q
		\int^{\beta}_{\alpha}\mathcal{D}q'
		\,e^{i(S[q]+S[q'])}  \nonumber \\
		&& \qquad\qquad\qquad\qquad \times \;\; e^{iS_{CWL}[q,q']}
		\label{K2-CWL-ab}
	\end{eqnarray}
	where the arguments surrounding eqtns. (\ref{pathI-CWL-q})-(\ref{K-phi-ab}) tell us how to treat the path integrations $\int^{\beta}_{\alpha}\mathcal{D}q$ and $\int^{\beta}_{\alpha}\mathcal{D}q'$; one has
	\begin{eqnarray}
		\int^{\beta}_{\alpha}\mathcal{D}q &=& \int d^4 x_1 d^4x_2 \;\langle \beta | x_2 \rangle \, \langle x_1 | \alpha \rangle \, \int^{x_2}_{x_1} {\cal D}x    \\
		\int^{\beta}_{\alpha}\mathcal{D}q' &=& \int d^4x'_1 d^4x'_2 \;\langle \beta | x'_2 \rangle \, \langle x'_1 | \alpha \rangle \, \int^{x'_2}_{x'_1} {\cal D}x'  \;\;\;\;
		\label{D-int-defn}
	\end{eqnarray}
	as shown in Fig. \ref{fig:CWL-int} (b). 
	
	It will also be obvious how one generalizes these considerations to, eg., a scalar field propagating between different wave functionals (recall the discussion in section 4.A).

	
	\subsection{Non-Relativistic Regime}
	\label{sec:K-lP2-psi}
	
	To get some intuition for these results, it is helpful to go to the non-relativistic regime - the one that will be relevant for future lab experiments. We summarize the results for a single particle, and then go on to discuss what happens if one deals more realistically with an extended mass coupled to its environment.

	\subsubsection{Particle Dynamics}
	\label{sec:K-lP2-psi}
	
	To be specific, we begin again with the simple case of a single particle of mass $M_o$, moving along some path in spacetime. In this case the path-bunching term in ${\cal K}(2,1)$ is
	\begin{align}
		S_{CWL}[q,q'] \;&=\; -{\ell_{P}^{2} \over 2} \int d^4x \int d^4x' \nonumber \\
		& \qquad \times D^{\mu\nu\alpha\beta}(x,x') T_{\mu\nu}(q, x)T_{\alpha\beta}(q',x')
		\label{S-CWL-X}
	\end{align}
	where $T_{\mu\nu}(x|q)$ is again the stress-energy for a particle following trajectory $q(s)$.
	
	If we then go to the limit where the particle is moving slowly, with velocity $v \ll c$, and spacetime is flat, we get the very simple result
	\begin{eqnarray}
		\lim_{v \, \ll \, c} \; S_{CWL}[q,q'] &=&  S_{CWL}[{\bf r},{\bf r'}] \nonumber \\
		&=& \frac{1}{2}\int_{t_{1}}^{t_{2}} dt \frac{GM_o^{2}}{|{\bf r}(t)-{\bf r'}(t)|}
		\label{CWL-nonRP}
	\end{eqnarray}
	where ${\bf r}(t)$ is the spatial coordinate of the particle. We then just have a Newtonian interaction between the 2 paths in the CWL propagator. 
	
	We have previously noted some of the effects of this Newtonian term on particle propagation in CWL theory (compare ref. \cite{stamp15}, section 5.2.3, and ref. \cite{stamp12}). It is useful to describe things here in more detail. Quite generally we can say that 
	
	\vspace{2mm}
	
\begin{figure}
		\includegraphics[width=3.2in]{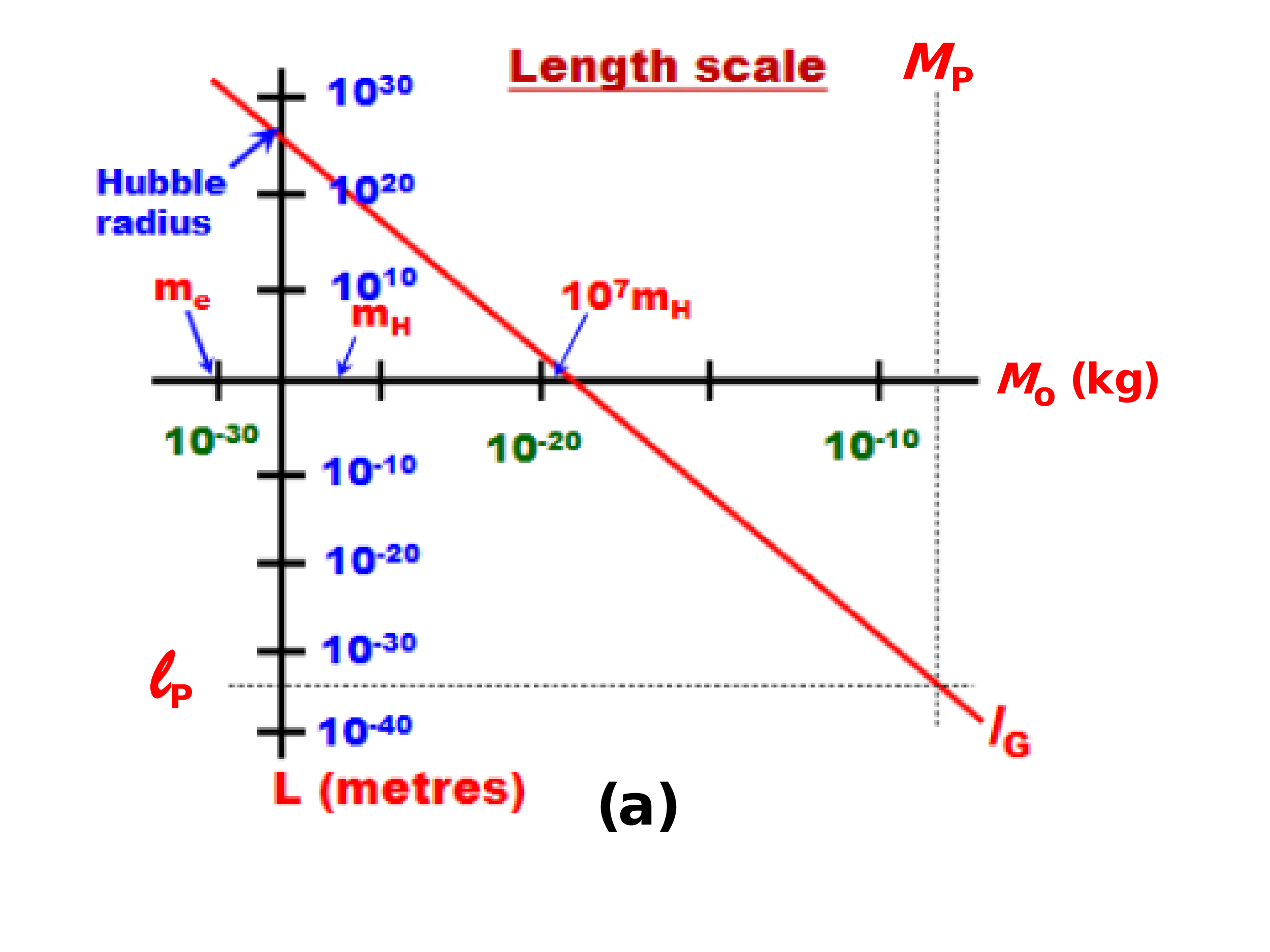}
		\includegraphics[width=3.2in]{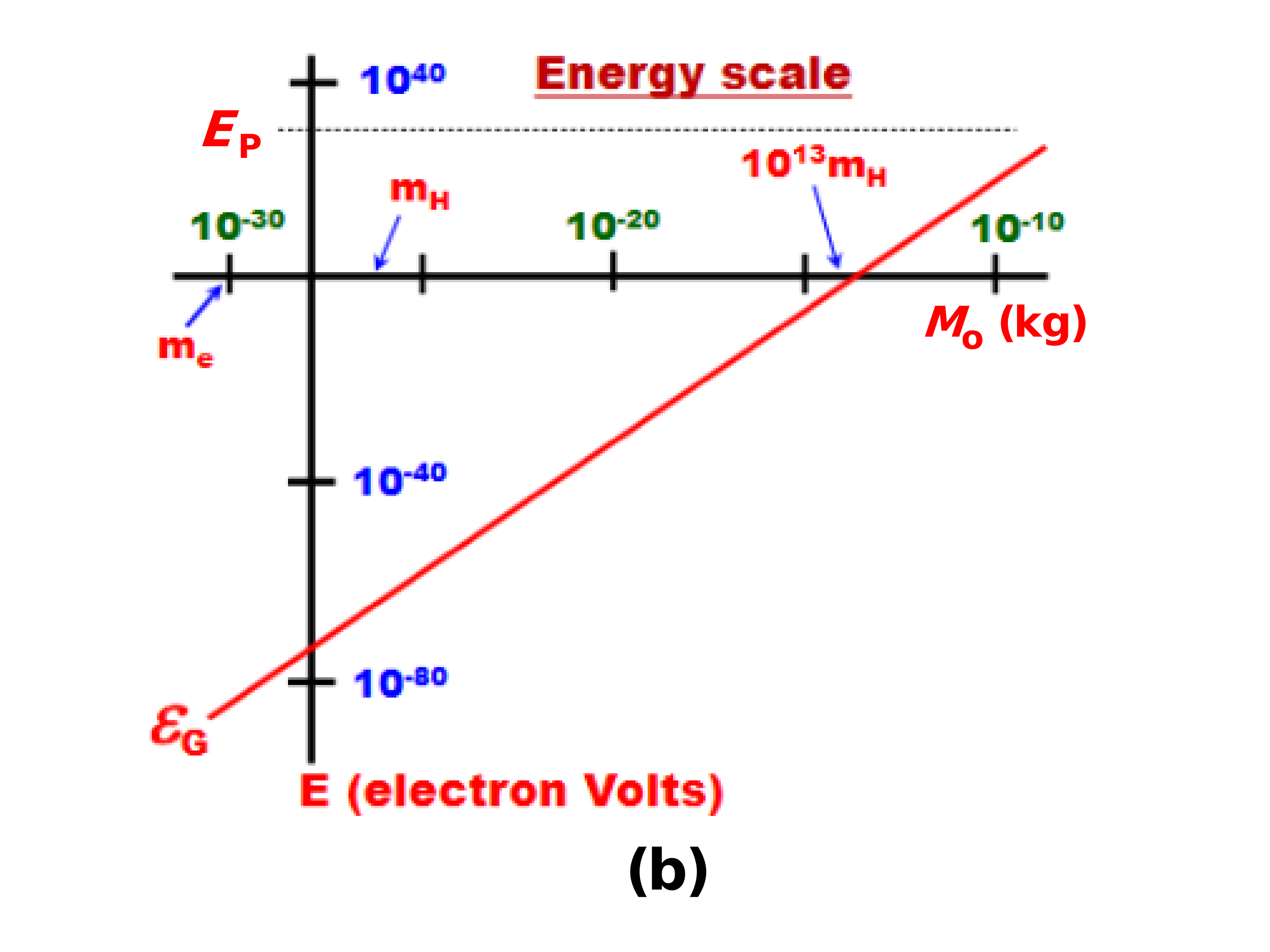}
		\caption{\label{fig:lP2-scale}  The length and energy scales which emerge in the $\ell_P^2$ approximation for the dynamics of a single free particle in CWL theory, according to equation (\ref{lP2-scale}). In (a) we plot the length scale $\ell_G(M_o)$, and in (b) we plot the energy scale $\epsilon_G(M_o)$.}
	\end{figure}

	(i) There are two key scales inherent in the attractive Newtonian potential in (\ref{CWL-nonRP}), viz., 
	\begin{align}
		\ell_G(M_{o}) &= (M_P/M_o)^3 \ell_P  \nonumber \\ 
		\epsilon_G(M_{o}) & = (M_o/M_P)^5 E_P
		\label{lP2-scale}
	\end{align}
	The length scale $\ell_G(M_o)$ is the analogue of the Bohr radius for this potential, and the energy scale $\epsilon_G(M_o)$ is the analogue of the Coulomb binding energy (ionization energy). Here $M_P$, $\ell_P$ and $E_P$ are the Planck mass, length, and energy respectively (see the 1st paragraph of this paper). In Fig. \ref{fig:lP2-scale} we show these scales graphically for a wide range of masses. 
	
	\vspace{2mm}
	
	(ii) Any external potential $V({\bf r})$ acting on the mass $M_{o}$ can upset the effects of the Newtonian attraction between paths. Roughly speaking, if $\ell_G(M_o) |\nabla V({\bf r})| \geq \epsilon_G(M_o)$, then the Newtonian attraction will be destabilized. 
	
	\vspace{2mm}
	
	(iii) Both $\ell_G(M_o)$ and $\epsilon_G(M_o)$ are extremely rapid functions of $M_o$. To get a feel for the numbers it is useful to look at some examples; three will suffice:
	
	For an electron, $\ell_G(M_o) \sim 3.6 \times 10^6~R_H$, where $R_H$ is the Hubble radius, and $\epsilon_G(M_o) \sim 1.4 \times 10^{-84}~eV$; 
	
	For an object like a vaccinia virus, of linear dimension $3 \times 10^{-7}$m and mass $10^{-17}$kg (ie., $6 \times 10^9$D), one has $\ell_G(M_o) \sim 1.7 \times 10^{-7}$m, and $\epsilon_G(M_o) \sim 2.6 \times 10^{-19}~eV$;
	
	For an object like a Dunaliella Salina alga, with linear dimension $10~\mu$m and mass $1.5 \times 10^{-13}$kg (ie., $9 \times 10^{13}$D, or $7 \times 10^{-6}M_p$), one has $\ell_G(M_o) \sim 4.9 \times 10^{-20}$m, and $\epsilon_G(M_o) \sim 200~eV$.
	
	\vspace{2mm} 
	
	From these numbers it is clear that the point-particle model used to calculate ${\cal K}(2,1)$ in the $\ell_P^2$ approximation, to give (\ref{K-fin}) or (\ref{K-fin2}), is extremely accurate for an electron - where however it gives utterly negligible corrections to standard QM. For the vaccinia virus $\epsilon_G(M_o)$ is still fantastically small, so $S_{CWL}[q,q']$ is also very small, and the $\ell_P^2$ approximation is still valid, as is QM. However $\ell_G(M_o)$ is by then smaller than the virus, and at this point one expects the point-particle approximation to be breaking down - one then needs to redo the calculation for an extended body.

	
	Finally, for the Dunaliella alga, it is clear that both the $\ell_P^2$ approximation and the point-particle approximation have broken down irretrievably - the CWL interaction energy now being $\sim 200$ eV -  and we need to do completely change the calculation. Even at this point we are still far below the Planck mass - we see clearly that CWL effects become prominent already for masses $\ll M_P$ \cite{stamp15,stamp12,carney19}.
	
	In the context of CWL theory, in this $\ell_P^2$ approximation, it is clear that if we are examining the behaviour of a particle at length scales $L \gg \ell_G(M_o)$, then the particle will look as though it is a point particle - at low energies the paths will seem so closely bound as to behave like a single path. On the other hand if $L \ll \ell_G(M_o)$, the opposite is true; the 2 separate paths are clearly visible at length scale $L$. 
	
	This is as far as we can go in the $\ell_P^2$ approximation for point particles. We now turn to a brief description of what one can do to go beyond these calculations.

	\subsubsection{Extended Body coupled to an Environment}
	\label{sec:extMass}
	
	In the following, for completeness, we describe qualitatively how one can go beyond the point-particle model - a full derivation of these results appears elsewhere (see ref. \cite{jordanPhD,jordan-largeN}). One can extend the calculations in 3 different ways. Within the $\ell_P^2$ approximation one can (a) generalize to an extended mass, and (b) add a dissipative coupling to an environment. Finally (c) one can go to higher orders in $\ell_P^2$. We look at these in turn. 
	
	\vspace{2mm}
	
	{\bf (a) Extended Mass}: In the point mass $\ell_P^2$ approximation, even when $\epsilon_G(M_o) \sim 10^{-13}~kg$, more than 5 orders of magnitude below the Planck mass, we still have $\ell_G(M_o) \sim 3 \times 10^{-19}~m$, already far less then the typical size ($\sim 10^{-5}~m$) of an object with this mass. Clearly, one has to do calculations for an extended mass to get realistic results. 
	
	To study this problem in an $\ell_P^2$ approximation, one describes the mass as a nanoscopic or mesoscopic body of some shape, assembled from a set of particles distributed either in some crystalline array, or as in an amorphous solid \cite{jordan-largeN}). The mass is concentrated almost entirely in the atomic nuclei, and one must take account of the fluctuations of these nuclei around their equilibrium positions (which at low temperature $T$ are zero-point in nature, of amplitude $\xi_o \sim 1-5 \times 10^{-11}$m). The result can be entirely characterized in terms of the phonon spectrum of the solid, the sample shape, and $T$. 
	
	One finds that when the size $L$ of the extended 
	mass $\gg \ell_G(M_o)$, then the non-relativistic $1/|{\bf r}(t)-{\bf r'}(t)|$ inter-path CWL potential in (\ref{CWL-nonRP}) is replaced by a very different low-$T$ interaction with 2 potential wells, one of range $\sim L$, the other, 
	inside the first, of range $\sim \xi_o$. At low energies, this latter `zero point' potential well has a low-energy harmonic form. 
	
	As an example, one can consider a solid made entirely from a total of $N_o = M_o/m$ ions of mass $m$ particles \cite{jordan-largeN}. Then the oscillation frequency $\omega_{eff}$ in the zero-point harmonic well which now binds the 2 paths is $\omega^2_{eff} = (2^{1/2}G_Nm/3 \pi^{1/2} \xi_o^3)$, a result also found using the Schrodinger-Newton equation \cite{bassi}.  Thus $\omega_{eff}$ is independent both of the shape of the extended mass, and of its mass - it depends only on microscopic details of the object. 
	
	We can evaluate this for a crystalline $Si O_2$ system (quartz); one finds an oscillation period $t_o = 2\pi/\omega_{eff} \sim 16$ secs. Thus the relative oscillatory motion of pairs of paths, in the $\ell_P^2$ approximation, is rather slow. 
	
	\vspace{2mm}
	
	{\bf (b) Dissipative Effects}: Just as radiative coupling to a photonic bath is required for decay of an orbit in QED, the effect of dissipative coupling to the environment will facilitate the path-bunching process. To treat this process in the $\ell_P^2$ approximation, one calculates the dynamics of the reduced density matrix for the matter degrees of freedom, once the environmental modes are integrated out.  
	
	Dissipative (and decohering) effects are typically described by coupling the system to an `oscillator bath' \cite{feynV63,CL83}, which describes delocalized environmental modes (phonons, photons, electronic quasiparticles, etc.), or to a 'spin bath' \cite{PS00} which describes localized modes (solid-state defects, nuclear and paramagnetic spins, etc.). One then integrates out these modes to derive an influence functional for the matter dynamics, in the presence of CWL interactions. 
	
	One simple conclusion emerges in the regime of low dissipation, which can modelled for many systems of relevance here \cite{CL83,adhikari} in terms of a simple friction coefficient $\eta$. In the $\ell_P^2$ approximation, one then sees pairs of paths spiraling into each other on a timescale $\tau_{PB} \sim  Q/\omega_{eff}$, where $Q = M_o \omega_{eff}/\eta$ is the quality factor associated with the frictional damping. Thus if $Q \gg 1$, the `path-bunching' time $\tau_{PB}$ can be extremely long. 
	
	Results like this are preliminary - they neglect the effect of multi-path CWL correlations (discussed immediately below). Nevertheless they suggest that when $Q \gg 1$ (as for the mirrors in LIGO-type experiments) it may take a long time for the classical path-bunched dynamics to emerge, even for mirrors with mass  $\gg M_p$. 
	
	\vspace{2mm}
	
	{\bf (c) Multiple Paths and the Classical Limit}: To truly characterize path-bunching in ${\cal K}(2,1)$, we clearly need to incorporate higher-order terms in $\ell_P$, in which graphs containing 3 or more matter lines interact. The following remarks should be viewed as preliminary. 
	
	Note first that the same sort of path-bunching will take place amongst $n$-tuples of lines for ${\cal K}(2,1)$; and again, it will be influenced by coupling to an environment. One can then ask what happens once this path-bunching has taken place.
	
\begin{figure}
		\includegraphics[width=3.2in]{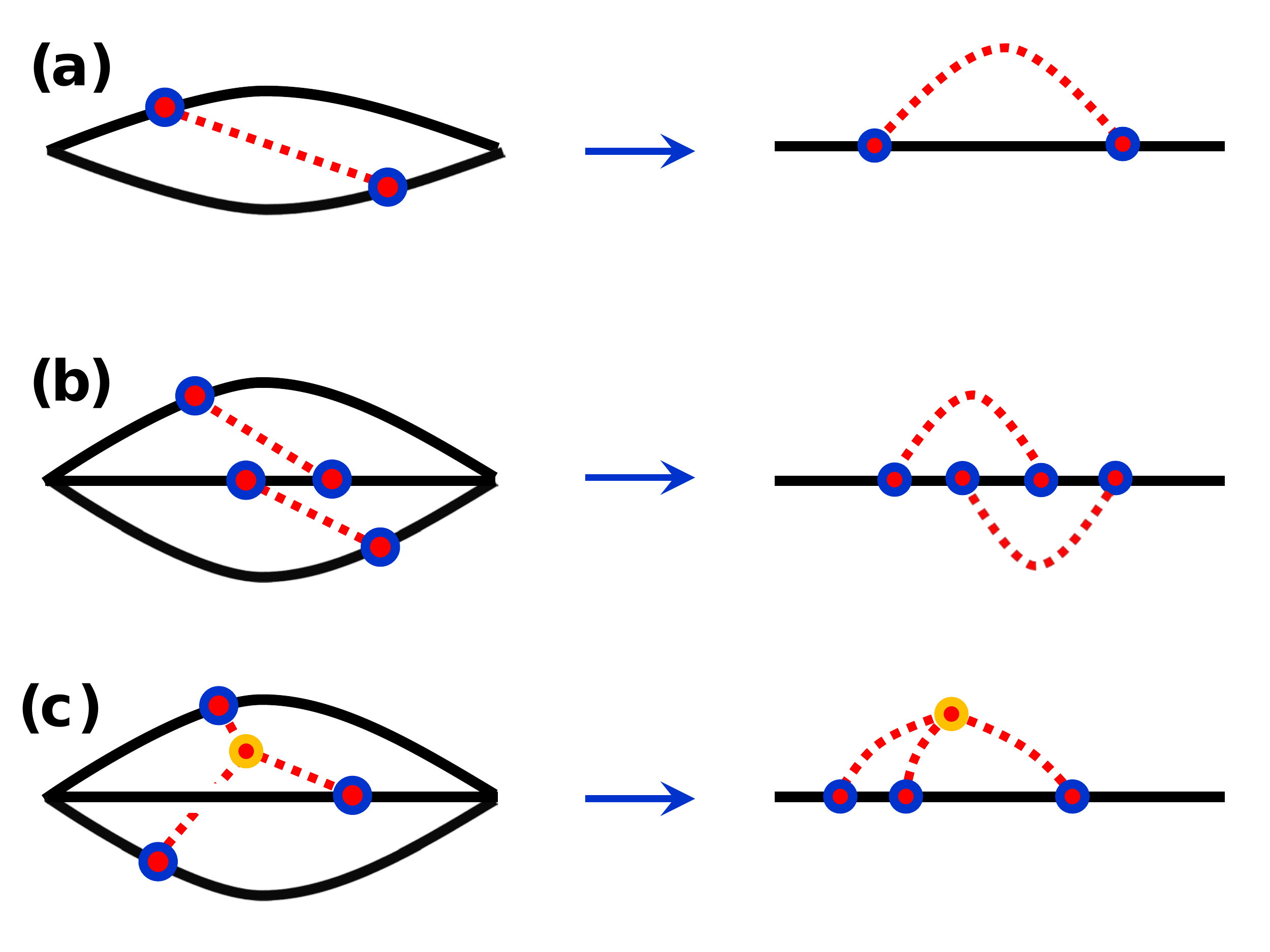}
		
		\caption{\label{fig:CWL-collapse}  Diagrammatic representation of the transition to classical behaviour for sufficiently massive matter lines.  On the left-hand side we show 3 different diagrams for the propagator ${\cal K}(2,1)$. On the right-hand side we show what happens to each diagram in the massive limit, when path-bunching causes all matter lines to collapse to a single matter line, with the result that graviton loops appear.}
	\end{figure}

	Notice first that in the non-relativistic regime, for a set of $n$ matter lines, the same energy and length scales emerge as in the $\ell_P^2$ approximation (for $n$ lines, the coupling between each is $\propto 1/n$). Suppose we now deal with a particle of mass $M_o$. In Fig. \ref{fig:CWL-collapse}, we show what we expect to happen to several graphs for ${\cal K}(2,1)$, as path-bunching occurs. 
	
	Suppose now that path binding has occurred over a length scale $L \ll$ any experimental length. Then, for all practical purposes, the matter lines all collapse onto each other, so that the graviton lines now `fold back' onto the single `composite' matter line that is left.  These graviton lines are still necessarily on-shell, so we retain only the classical gravity contributions to the loop diagrams.
	
	Consider first Fig. \ref{fig:CWL-collapse}(a). As we just saw, this is the only graph $\sim O(\ell_P^2)$ for ${\cal K}(2,1)$ in CWL theory. Once the 2 matter lines have path-bunched, we get the rainbow graph shown in Fig. \ref{fig:CWL-collapse}(a) at right. This graph is simply the lowest order contribution to the classical self-energy, in which the perturbation of the metric caused by a mass reacts back on the mass. If the mass is accelerating, then we get a contribution  $\sim O(\ell_P^2)$ to the radiation damping and radiation reaction in the classical theory.
	
	The graphs in Fig. \ref{fig:CWL-collapse}(b) and (c) show the same features. The left-hand graphs are permitted by the CWL graphical rules; after path-bunching, the right-hand side gives further contributions to the classical self-energy of the mass.

	
	Note that this result allows us to address the 2 paradoxes noted at the end of section 3, regarding the absence of loops containing gravitons in CWL theory. We see that in the `classical regime', defined here as the regime in which path-bunching has occurred, these graviton loop contributions are restored, along with classical self-energy and radiation reaction terms.    
	
	Clearly one then needs to show that the CWL graphs, at arbitrary order in $\ell_P^2$, collapse precisely to those of the same order in classical GR expanded in powers of $\ell_P^2$. We examine this question elsewhere.
	
	This concludes our brief survey of results whose full description is beyond the scope of this paper. We see that although the $\ell_P^2$ approximation cannot be relied on for any quantitative predictions, it can give a good qualitative idea of some of the physics.

	
	\section{Conclusions}
	\label{sec:Concl}
	
	
	In this paper we have given an extended discussion of the low-energy (ie., $E \ll E_p$) behaviour of CWL theory. We have focussed on the behaviour of the connected generating functional $\mathbb{W}[J]$ and the matter propagator ${\cal K}(2,1)$. A combination of perturbative (in $G_N$) and non-perturbative large $N$ analyses leads to exact results for these 2 functions. We also give results for the weak field approximation to CWL theory, where it can be linearized.  
	
	A key result of this work is that the matter field moves in a background metric field whose dynamics is determined by the matter field, in a way superficially reminiscent of (but not the same as) semi-classical quantum gravity. To see in some detail how CWL theory works, we give extended analyses of both the 2-path experiment and lowest-order perturbation theory; the results show clearly how CWL predictions differ from both conventional low-energy quantum gravity and from semiclassical quantum gravity.  
	
	The key distinguishing feature of CWL theory is the way in which different paths in all path integrals are coupled to each other via gravity - this causes `path-bunching' of the matter paths, in a way which explicitly violates the usual superposition principle in quantum mechanics. The formulation of the theory in terms of Feynman paths is essential. 
	
	We note that no new interactions or constants of nature are introduced; nor any fields apart from traditional metric and matter fields. Thus no {\it ex cathedra} noise fields or classical fields are involved, and in fact the theory is entirely quantum-mechanical in that all fields are quantized in a universe defined by the dynamics of the quantized metric field $g^{\mu\nu}(x)$. The difference with conventional QM or QFT is in the dynamical rules, and a key consequence of these rules is that for large masses, the dynamics of $g^{\mu\nu}(x)$ is classical, and governed by Einstein's equation. 
	
	The resulting theory realizes the idea discussed by Kibble \cite{kibble1,kibble2}, viz., that QM and QFT are transformed into non-linear theories, violating QM, by the coupling to gravity. As we have discussed elsewhere \cite{BCS18,CWL2}, CWL theory appears to be a consistent theory; expansions in $G_N$ and $\hbar$ are consistent, the theory has a consistent classical limit, and it obeys all Ward identities. In this paper we have added to this work by finding exqact results for the dynamics. Thus the consistency problems, which have bedevilled earlier non-linear theories, are circumvented. 
	
	It is clear that in CWL theory, measurements and experiments are just ordinary physical processes; measurements play no central role of the kind found in conventional QM.  The transition to classical behaviour comes for macroscopic system via path-bunching. For a microscopic system ${\cal S}$ it happens once it couples to some macroscopic system ${\cal M}$ which is sufficiently massive and complex that it exhibits path-bunching \cite{stamp15}.
	
	Finally we can ask - what is CWL theory good for? Any consistent theory still has to pass experimental tests to be taken seriously. The present paper has laid the foundation for this. Clearly more detailed analysis of specific real experiments is now required, in which quantitative predictions are made. This will be the subject of several future papers, in which we work out these predictions for a solid object of arbitrary composition and shape, for various experimental designs.

	
	\section{Acknowledgements}
	\label{sec:Ackn}
	
	
	We are very grateful to both A.O Barvinsky and Y. Chen for extensive conversations regarding this work. We also think M. Aspelmeyer, C. Delisle, R. Penrose, W.G. Unruh, and B. Whaley for discussions, and Y. Chen and K.S. Thorne for partial support. The work was supported in Vancouver by the National Sciences and Engineering Research Council of Canada, and in Caltech by the Simons Foundation (Award 568762) and the National Science Foundation (Award PHY-1733907). J.W.-G. was also supported by a Burke fellowship in Caltech and a NSERC PGS-D award in Vancouver.

	\appendix

	
	\section{Generating Functional for conventional theory}
	\label{sec:AppA}
	
	
	We derive here the results from section 2 which relate the generating functional to propagators; we do this for ordinary QM and for scalar field theory.

	\subsection{QM from a generating functional}
	\label{Ssec:pathI-QM}
	
	We begin with a non-relativistic particle in an external current ${\bf j}(t)$; the generating functional is then 
	\begin{equation}
		{\cal Z}_o[{\bf j}] \;=\; \oint {\cal D} {\bf r}(t) \; e^{i(S_o[{\bf r}] + \int dt \; {\bf j}\cdot {\bf r})}
		\label{Zo-QM'}
	\end{equation}
	(cf eqtn. (\ref{Zo-QM})). Recall that in the main text we were interested in evaluating a function 
	\begin{equation}
		\aleph(2,1) = \int d{\bf j}_1  d{\bf j}_2 \; e^{-i ({\bf j}_1 \cdot {\bf x}_1 + {\bf j}_2 \cdot {\bf x}_2)}  {\cal Z}_o[{\bf j}_1, {\bf j}_2]
		\label{Ko-Zo'}
	\end{equation}
	with cuts at times $t_1, t_2$, in which ${\cal Z}_o[{\bf j}_1, {\bf j}_2] \equiv {\cal Z}_o[{\bf j}_1 \delta(t-t_1) + {\bf j}_2 \delta(t-t_2)]$ (compare eqtns. (\ref{Zo-j1j2}) and (\ref{Ko-Zo})).

	
	\begin{figure}
		\vspace{7mm}
		
		\includegraphics[width=3.2in]{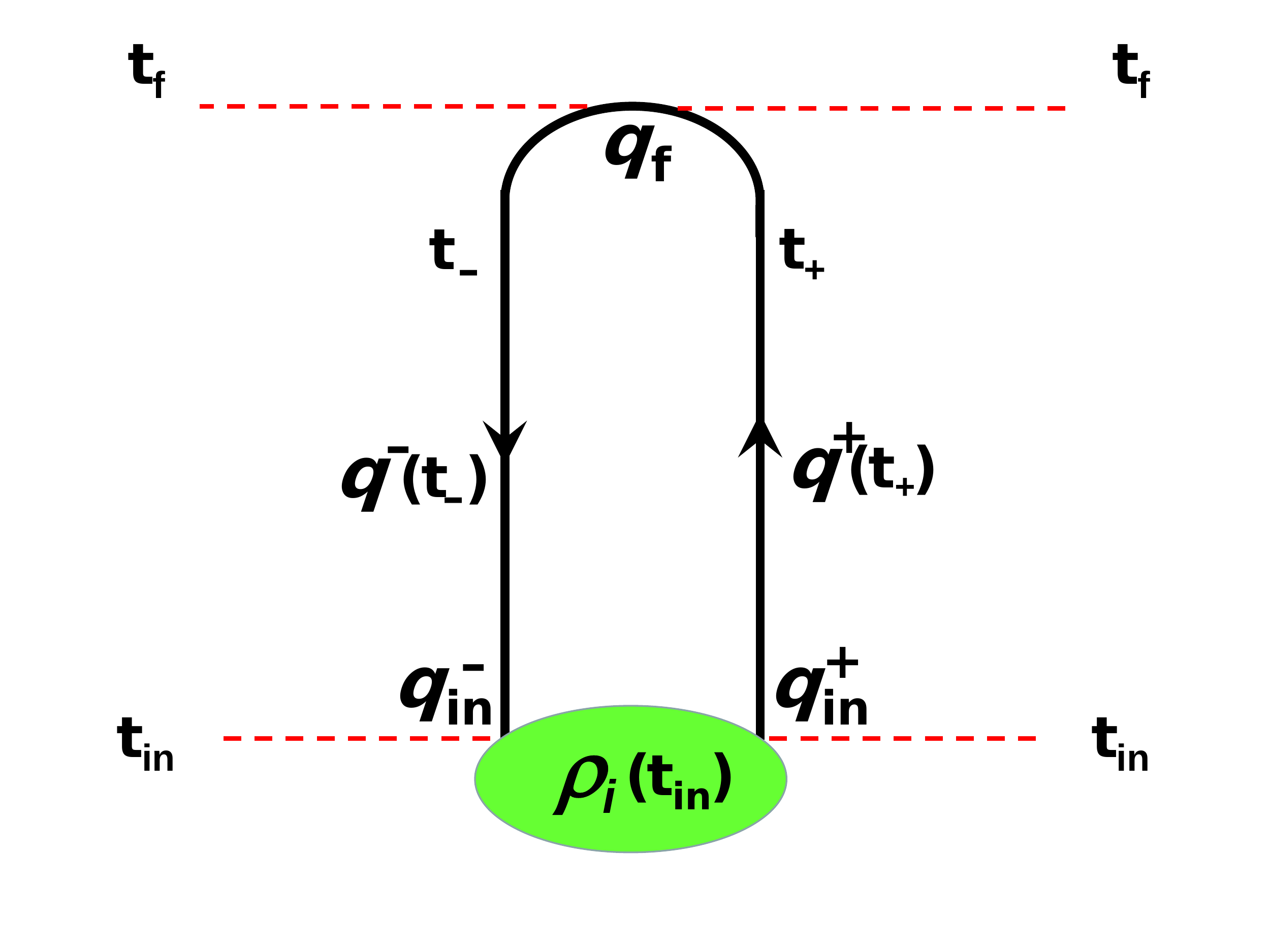}
		\vspace{-7mm}
		\caption{\label{fig:ringP-rho}  The ring path integral written in Keldysh form. The path begins and ends at times $t_{in}$, with $t_{in} \rightarrow - \infty$. It proceeds up via the path ${\bf q}^+(t_+)$, a function of time $t_+$, to time $t_f$; we let $t_f \rightarrow \infty$. It then proceeds back down via path ${\bf q}^-(t_-)$,a function of time $t_-$. The contribution around the closed imaginary time contour at $t_{in}$ (compare Fig. \ref{fig:ring+K1}) gives the density matrix $\rho_{i} ({\bf q}_{in}^+, {\bf q}_{in}^- \,; t_{in})$.}
	\end{figure}
	

	Let us write the ring path integral ${\cal Z}_o[{\bf j}]$ in Keldysh form \cite{keldysh-R}. We define the times $t_{\pm}$ on the upwards/downwards parts of the ring contour respectively (see Fig. \ref{fig:ringP-rho}).  These times extend between $t_{in}$ and $t_f$. We also define particle coordinates ${\bf q}^+(t_+)$ and ${\bf q}^-(t_-)$ on the upwards/downwards paths, with limiting values
	\begin{align}
		{\bf q}_f \;&=\; {\bf q} (t = t_f) \nonumber \\
		{\bf q}^+_{in}  \;&=\;    {\bf q^+} (t_{in})  \nonumber \\
		{\bf q}^-_{in}  \;&=\;   {\bf q^-} (t_{in}) 
		\label{qf-qin}
	\end{align}
	Finally, we let $t_{in} \rightarrow -\infty$, and $t_f \rightarrow \infty$.
	
	The integral around the imaginary time loop at $t_{in} \rightarrow -\infty$ defines the particle thermal density matrix $\rho_{i} ({\bf q}_{in}^+, {\bf q}_{in}^- \,; t_{in}) \equiv \langle {\bf q}_{in}^+ | \hat{\rho}_i (t_{in}) | {\bf q}_{in}^- \rangle $. We can then write the generating functional as 
	\begin{align}
		{\cal Z}_o[{\bf j}] \;&=\; \int d{\bf q}_{in}^+  d{\bf q}_{in}^- \int d{\bf q}_f \nonumber \\
		&\qquad \qquad \times \,  \rho_{i} ({\bf q}_{in}^+, {\bf q}_{in}^-) \; \mathbb{G} ({\bf q}_{in}^+, {\bf q}_{in}^-; {\bf q}_f |\, {\bf j}\,)
		\label{Zj-rho-in}
	\end{align}
	where $\mathbb{G} ({\bf q}_{in}^+, {\bf q}_{in}^-; {\bf q}_f |\, {\bf j}\,)$ describes the integration around the rest of the ring, and is written as
	\begin{widetext}
		\begin{equation}
			\mathbb{G} ({\bf q}_{in}^+, {\bf q}_{in}^-; {\bf q}_f |\, {\bf j}\,) \;=\; \int_{{\bf q}_{in}^+}^{{\bf q}_f}  {\cal D} {\bf q}^+ \; e^{i(S_o[{\bf q}^+] \,+\, \int dt_+ \; {\bf j}(t_+)\cdot {\bf q}^+(t_+))}     \;  \int_{{\bf q}_{in}^-}^{{\bf q}_f} {\cal D} {\bf q}^- \;  e^{i(S_o[{\bf q}^-] \,+\, \int dt_- \; {\bf j}(t_-)\cdot {\bf q}^-(t_-))}
			\label{bbG-int}
		\end{equation}

		We can also write this expression in terms of the Hamiltonian ${\cal H}_o$ of the system, as the trace
		\begin{equation}
			{\cal Z}_o[{\bf j}] \;=\; Tr \left[ \; \hat{\cal T} \{ e^{-i \int_{t_{in}}^{t_f} dt_+ [{\cal H}_o \,+\, {\bf j}(t_+) \cdot \hat{\bf q}(t_+]} \} \;\; \hat{\rho}_{i}(t_{in}) \;\;  \hat{\cal T}^{-1} \{ e^{-i \int_{t_{in}}^{t_f} dt_- [{\cal H}_o \,+\, {\bf j}(t_-) \cdot \hat{\bf q}(t_-]} \} \right]
			\label{Zj-H}
		\end{equation}
		in which $\hat{\cal T}$ is the time ordering operator, and $\hat{\cal T}^{-1}$ its inverse. 
		
		We may now substitute this form directly into eqtn. (\ref{Ko-Zo'}) for $\aleph(2,1)$, to get 
		\begin{align}
			\aleph(2,1) \;&=\; \int d{\bf j}_1  d{\bf j}_2 \;e^{-i ({\bf j}_1 \cdot {\bf x}_1 + {\bf j}_2 \cdot {\bf x}_2)} \; Tr \left[ e^{-i {\cal H}_o (t_f - t_2)} \;  e^{i {\bf j}_2 \cdot {\bf q}} \;  e^{-i {\cal H}_o (t_2 - t_1)} \; e^{i {\bf j}_1 \cdot {\bf q}} \; e^{-i {\cal H}_o (t_1 - t_{in})} \; \hat{\rho}_{i}(t_{in}) \; e^{-i {\cal H}_o (t_f - t_{in})} \right] \nonumber \\
			&= \;  \langle {\bf x}_2 | e^{-i {\cal H}_o (t_2 - t_1)} | {\bf x}_1 \rangle \; \langle {\bf x}_1 | e^{-i H(t_1 - t_{in})} \, \hat{\rho}_{in} \,  e^{i H(t_2 - t_{in})} | {\bf x}_2 \rangle \nonumber \\
			& \equiv \; K_0(2,1) f(2,1)
			\label{aleph-H}
		\end{align}
	\end{widetext}
	with no integration over ${\bf x}_1$ or ${\bf x}_2$. Thus we get the product form for $\aleph(2,1)$ given in eqtn. (\ref{aleph}) of the main text. We can also define various time-ordered Keldysh propagators, starting from here, using standard techniques \cite{time-O}.  
	
	At a temperature $T$ the thermal density operator $\hat{\rho}_i (t_{in}) = \sum_n |n \rangle e^{-\beta \epsilon_n} \langle n |$, where the $\{ \epsilon_n \}$ are the particle eigenstates and $\beta = 1/kT$; one then has the limiting cases
	
	(i) in the infinite temperature limit where $\beta \rightarrow 0$, we get $f(2,1) \rightarrow \langle {\bf x}_1 | e^{-i {\cal H}_o (t_1 - t_2)} | {\bf x}_2 \rangle$, so that $f(2,1) = K_0^*(2,1)$ and $\aleph(2,1) = |K_0(2,1)|^2$;
	
	(ii) in the $T \rightarrow 0$ limit where $\beta \rightarrow \infty$, and $\hat{\rho}_{in} \rightarrow \hat{\rho}_o \equiv |0 \rangle \langle 0 |$, the vacuum state density operator, we get $f(2,1) = \langle {\bf x}_1 | 0 \rangle \langle 0 | {\bf x}_2 \rangle e^{-i \epsilon_0 (t_1 - t_2)}$, ie., the time-dependent vacuum density matrix between states $|{\bf x}_1 \rangle$ and $|{\bf x}_2 \rangle$.

	\subsection{Extension to more complicated cases}
	\label{Ssec:pathI-QFT}

	This technique is easily generalized to cover other kinds of propagator. In particular one has
	
	\vspace{2mm}
	
	(i) {Density matrix}: the propagator ${\cal K}_o(2,2'; 1,1')$ for the density matrix, which in path integral language is written as \cite{feynH65,feynV63}
	\begin{equation}
		\rho(2,2'; t_2)  \;=\; \int d \, 1 \int d \, 1' \; {\cal K}_o(2,2'; 1,1') \rho(1,1'; t_1)
		\label{p-K-p}
	\end{equation}
	where, eg., $\rho(1,1'; t_1) = \langle 1 | \hat{\rho}(t_1) | 1' \rangle$ is the density matrix element between states $|1 \rangle$ and $|1' \rangle$ at time $t_1$. The derivation of the path integral form from ${\cal Z}$ is the same as for the propagator, only now we introduce four cuts, instead of two.

	\vspace{3mm}
	
	(ii) {\it Relativistic particle}: Starting from the generating functional for a relativistic particle, we can apply the same techniques to find the propagator for this particle while propagating on a fixed background metric $g_o$. One gets
	\begin{equation}
		K_0(2,1|g_0) \;=\; \int_{0}^{\infty}ds \int^2_1 {\cal D}X(\tau) \,e^{i\int_{0}^{s}
			d\tau\,\mathcal{L}_0(X|g_{o},j)}
		\label{Ko-21'}
	\end{equation}
	where the action $S_o[X|s,g_o] = \int_{0}^{s} d\tau\,\mathcal{L}_0(X|g_o)$ is a functional of the background field $g_o$ and a function of the proper time $s$.
	
	\vspace{3mm}
	
	(iii) {\it Scalar Field}: Consider a scalar field $\phi$ with action $S[\phi]$, defined on a spacetime in which a hypersurface $\Sigma$ bounds a `bulk' spacetime region ${\cal M}$. The surface $\Sigma$ is divided into spacelike past and future surfaces $\Sigma_1$ and $\Sigma_2$, along with a region $\Sigma_B$ at spatial infinity.
	
	Starting from $Z_{\phi}[J]$, and using the same methods as before (now imposing cuts on $\Sigma_1$ and $\Sigma_2$), we have a propagator between scalar field configurations $\Phi_1(x)$ and $\Phi_2(x)$, localized on $\Sigma_1$ and $\Sigma_2$, given by
	\begin{equation}
		K(2,1) \; \equiv \; K(\Phi_2,\Phi_1)  \;=\;  \int^{\Phi_2}_{\Phi_1} {\cal D}\phi\, e^{iS_{\phi}[\phi]}
		\label{K2-QFT-g0'}
	\end{equation}
	
	Here we have assumed flat spacetime for simplicity. The same development can be carried out for a gauge field theory like QED - for details see refs. \cite{jordan20,jordanPhD}. The derivation of propagators in conventional quantum gravity from the generating functional is described in the main text.

	
	\section{The Regulator Function}
	\label{sec:AppB}
	

	Here we show how one fixes the form of the regulator function $c_n$ introduced in eqtn. (\ref{bbQ-J1'}), to get $c_n = 1$ for all values of $n$. To do this, we will evaluate a typical normalized correlation function for our scalar field system, but this will be done for the case of a finite $J(x)$, instead of the more usual case $J(x) \rightarrow 0$; and we'll do this in the $G_{N}\rightarrow 0$, limit where we require conventional QFT to hold.
	
	Before beginning we simplify the algebra by working in a fixed background field $g_o(x)$, ie., we drop the functional integration over $g(x)$, so that
	\begin{equation}
		{\cal Q}_n[J, g_o] \; \rightarrow \; \left(Z_{\phi}\Big[\,g_o,
		\frac{J}{c_n}\,\Big]\right)^n.  \qquad
		\label{bbQ-Jgo}
	\end{equation}
	Freezing the metric dynamics in this way, about a solution $g_o$ to the vacuum Einstein equation, is the same as taking the $G_{N}\rightarrow0$ limit of the theory.
	
	We now calculate the correlation function ${\cal G}_l(\{ x_k \}|J_o(x))$, which in conventional QFT is given by
	\begin{equation}
		{\cal G}_l(\{ x_k \}|J_o(x))  \;\;=\;\;  \langle\Phi[J_o]|\,\phi(x_{1})...\phi(x_{l})\,|\Phi[J_o]\rangle
		\label{calG-cn}
	\end{equation}
	where $|\Phi[J_o]\rangle$ denotes the vacuum state of the scalar field in the presence of the current $J_o(x)$.
	
	Observe now that if we work this out explicitly, according to the unscaled prescription (\ref{corrCWL-poss}), we find
	\begin{widetext}
		\begin{align}
			{\cal G}_l(\{ x_k \}|J_o(x)) & \;\;=\;\;  \left(\sum_{n=1}^{\infty}\frac{n}{c^{l}_n}\right)^{-1}\frac{(-i\hbar)^{l}\delta^{l}}{\delta J(x_{1})...\delta J(x_{l})}\ln \mathbb{Q}[J]\bigg|_{J=J_o}
			\nonumber \\
			&\;\;=\;\;\left(\sum_{n=1}^{\infty}\frac{n}{c^{l}_n}\right)^{-1}\sum_{n=1}^{\infty}\frac{n}{c^{l}_n}\langle\Phi[J_o/c_n]|\,\phi(x_{1})...\phi(x_{l})\,|\Phi[J_o/c_n]\rangle
			\label{deriv-cn}
		\end{align}
	\end{widetext}
	
	However, we now observe that the result in (\ref{deriv-cn}), with operators sandwiched between states $|\Phi[J_o/c_n]\rangle$, is not in general equal to the initial result in (\ref{calG-cn}), with the same operators sandwiched between vacuum states $|\Phi[J_o]\rangle$. 
	
	In fact the only way we can get consistency is if $c_n = 1$ for all values of $n$. Thus we conclude that
	\begin{equation}
		c_n = 1
		\label{c-n-1}
	\end{equation}
	for all values of $n$.


\end{document}